\documentclass[a4paper,11pt]{article}
\pdfoutput=1
\usepackage{jheppub}
\usepackage{slashed}
\usepackage[T1]{fontenc}
\usepackage{adjustbox}
\usepackage{multirow}
\usepackage{hhline}
\usepackage{tabularx,booktabs,caption}
\usepackage{float}
\usepackage{placeins}
\usepackage{caption}
\usepackage{verbatim}
\usepackage{soul}
\usepackage{color}
\usepackage[utf8]{inputenc}
\usepackage{xcolor}

\newcommand{\ie}{$i.e.,$ }
\newcommand{\eg}{$e.g.,$ }

\newcommand{\m}{\mu}

\newcommand{\vp}{\varphi}
\newcommand{\vrho}{\varrho}
\newcommand{\bvrho}{\bar \varrho}

\usepackage[bb=dsserif, scr=rsfso]{mathalpha}
\usepackage{bm}
\newcommand{\Id}{\mathbb{1}}
\newcommand{\id}{\mathbb{1}}

\newcommand{\Tr}{\mathrm{Tr}}

\newcommand{\nn}{\nonumber{\nonumber}}
\newcommand{\abs}[1]{\left| #1 \right|}
\newcommand{\beq}{\begin{equation}}
\newcommand{\eeq}{\end{equation}}
\newcommand{\beqn}{\begin{eqnarray}}
\newcommand{\eeqn}{\end{eqnarray}}

\newcommand{\coeffa}{\alpha}
\newcommand{\dd}{\mathrm{d}}

\newcommand{\hef}{^4 {\rm He}}
\newcommand{\het}{^3 {\rm He}}
\newcommand{\lis}{^7 {\rm Li}}

\newcommand{\yp}{{Y\rm_P}}

\newcommand{\Tcm}{T_{\rm CM}}
\newcommand{\Neff}{N_{\rm eff}}
\newcommand{\neff}{N_{\rm eff}}
\newcommand{\xin}{\xi_\nu}

\newcommand{\ev}{{ \rm eV}}

\newcommand{\diag}{{ \rm diag}}

\newcommand{\eq}{Eq. }
\newcommand{\cref}{Ref. }
\newcommand{\fig}{Fig. }

\mathchardef\mhyphen="2D 

\title{\boldmath Primordial lepton asymmetries: neutrino transport, spectral distortions and cosmological constraints}

\author[a,b]{Yuan-Zhen Li}
\author[a,b,c,d]{and Jiang-Hao Yu} 

\affiliation[a]{Institute of Theoretical Physics, 
Chinese Academy of Sciences, Beijing 100190, China}

\affiliation[b]{School of Physical Sciences, University of Chinese Academy of Sciences, Beijing 100049, P.\ R.\ China}
\affiliation[c]{School of Fundamental Physics and Mathematical Sciences, Hangzhou Institute for Advanced
Study, UCAS, Hangzhou 310024, China}
\affiliation[d]{International Centre for Theoretical Physics Asia-Pacific, Beijing/Hangzhou, China}

\emailAdd{liyuanzhen@itp.ac.cn}
\emailAdd{jhyu@itp.ac.cn}

\abstract{
The primordial neutrino asymmetry leaves profound imprints on the evolution history of the universe, which can be constrained by cosmological observations, including Big Bang Nucleosynthesis (BBN), Cosmic Microwave Background (CMB), and Large-Scale Structure (LSS). We present comprehensive analysis on implications and constraints of the primordial neutrino asymmetry, based on a precise treatment of neutrino decoupling by solving the complete (anti)neutrino quantum kinetic equations in the Closed-Time-Path formalism. { Assuming the same primordial asymmetry $\xin$ for all neutrino flavors,} the effective number of neutrinos $\Neff$ and (anti)neutrino spectral distortions are calculated, and we find that the non-instantaneous decoupling correction is given by $\delta\Neff = 0.0440 + 0.0102 \, \xin^2$. Then we perform the state-of-the-art calculation for the abundance of light elements including (anti)neutrino spectral distortions, which indicate a positive asymmetry $0.032 \leq \xin \leq 0.052$ from EMPRESS data. The implications of the neutrino asymmetry for the CMB and LSS are studied in detail, and we find that the Baryon Acoustic Oscillations (BAO) are also significantly affected by $\xin$ in addition to the sum of neutrino masses. A combined analysis with EMPRESS BBN, Planck CMB and BOSS BAO data yields a tighter constraint $\xin = 0.024 \pm 0.012$, which provides constraints on UV models capable of producing large asymmetries.
}

\begin{document} 
\captionsetup[figure]{labelfont={bf},labelformat={default},labelsep=period,name={FIG.}}
\maketitle
\flushbottom

\section{Introduction}\label{sec:intro}

The presence of neutrinos, as evidenced by various cosmological observations, is crucial for understanding the evolution of the universe~\cite{Lesgourgues:2013sjj}. 
In the early universe, neutrinos and antineutrinos remain in thermal equilibrium with the Standard Model (SM) thermal bath as long as the interaction rates between (anti)neutrinos and other SM particles exceed the Hubble expansion rate. This equilibrium breaks down when the temperature falls below approximately $1 \, \text{MeV}$, leading to the decoupling of neutrinos from the SM thermal bath. 
Shortly after decoupling, the annihilation of electrons and positrons reheats the SM thermal bath, causing the photon temperature to exceed that of the neutrinos, with $T_\gamma/T_\nu \simeq (11/4)^{1/3}$. 
Subsequent processes in the universe, in particular Big Bang Nucleosynthesis (BBN) and the formation of the Cosmic Microwave Background (CMB) anisotropies and Large Scale Structure (LSS), are both affected by the relic neutrinos, allowing constraints on neutrino properties to be obtained from relevant observations, see \eg \cite{Pitrou:2018cgg,Planck:2018vyg,empress,DESI:2024mwx}.

A precise understanding of the neutrino decoupling process is essential for accurate constraints on neutrino properties. 
Notably, the partial reheating of neutrinos and antineutrinos due to the small overlap between neutrino decoupling and $e^{\pm}$ annihilations results in slightly non-thermal spectra and an increased total neutrino energy density, typically described by the effective number of relativistic species, $\Neff$. 
The values of $\Neff$ and the spectral distortions of neutrinos and antineutrinos are then used as initial conditions to evaluate the impact of relic neutrinos in subsequent processes. 
For the Standard Model case, where asymmetries between neutrinos and antineutrinos are assumed to be negligible, significant progress has been made in detailing the neutrino decoupling process over recent decades~\cite{Hannestad:1995rs,Dolgov:1997mb,Mangano:2001iu,Mangano:2005cc,Grohs:2015tfy,deSalas:2016ztq,Akita:2020szl,Froustey:2019owm,EscuderoAbenza:2020cmq,Froustey:2020mcq,Bennett:2020zkv}.
Including flavor oscillations, matter effects, Finite-Temperature QED (FTQED) corrections up to {$\mathcal{O}(e^3)$, where $e$ is the elementary electric charge}, and full neutrino-electron and neutrino-neutrino collision terms, the current standard prediction yields $\Neff^{\rm SM} = 3.0440 \pm 0.0002$~\cite{Froustey:2020mcq,Bennett:2020zkv,Akita:2020szl}.

On the other hand, inspired by the recent EMPRESS survey on primordial helium abundance~\cite{empress} and its potential to alleviate cosmological tensions~\cite{Barenboim:2016lxv,Yeung:2020zde,Seto:2021tad,Kumar:2022vee,Yeung:2024krv}, the scenario where asymmetries between neutrinos and antineutrinos are not negligible has recently attracted significant attention. 
{  Typically, such neutrino asymmetries can be parameterised with the degeneracy parameters $\xi_{\nu_\alpha} \equiv \mu_{\nu_\alpha}/T_{\nu_\alpha}$, defined as the chemical potential for the flavour $\alpha$ neutrino normalised to its temperature, so that
\begin{equation}
    \label{eq:etaL}
    \eta_\nu \equiv \frac{1}{n_{\gamma}}\sum_{\alpha=e,\mu,\tau}(n_{\nu_\alpha}-n_{\bar{\nu}_\alpha}) = \frac{\pi^2}{33 \zeta(3)} \sum_{\alpha=e,\mu,\tau} \left( \xi_{\nu_\alpha} + \frac{\xi_{\nu_\alpha}^3}{\pi^2} \right) \ ,
\end{equation}
where $n_{\gamma}$ is the photon number density and $n_{\nu_{\alpha}} (n_{\bar\nu_{\alpha}}) $ is the (anti)neutrino number density with flavour $\alpha$. {For the second equality, the SM value for the neutrino-photon temperature ratio $T_{\nu_{i}}/T_{\gamma} = (4/11)^{1/3}$ is assumed.} Beyond thermal equilibrium, an accurate $\eta_\nu$ must be obtained by a precise treatment of the neutrino decoupling process for given $\xi_{\nu_\alpha}$ as initial conditions.    }
Due to the sphaleron processes in the early universe~\cite{Kuzmin:1985mm,Khlebnikov:1988sr,Harvey:1990qw,Dreiner:1992vm}, one might naively expect neutrino asymmetries, or lepton asymmetries, to be of the same order as the baryon asymmetry of the universe, which is strongly constrained by BBN and CMB observations: $\eta_B \equiv (n_B - n_{\bar{B}})/n_\gamma = (6.14 \pm 0.04) \times 10^{-10}$~\cite{Planck:2018vyg}, where $n_B$ and $n_{\bar{B}}$ denote the number densities of baryons and antibaryons, respectively. 
However, substantial lepton asymmetries compared to the baryon asymmetry before the neutrino decoupling epoch can still be generated by various models, see \eg \cite{Casas:1997gx,Dolgov:1989us,Bajc:1997ky,Asaka:2005pn,Asaka:2005an,Pilaftsis:2003gt,Borah:2022uos,Kawasaki:2002hq,Kawasaki:2022hvx}.

The lepton asymmetries present before the neutrino decoupling epoch, henceforth referred to as primordial neutrino asymmetries, have significant implications for the universe's evolution. During the neutrino decoupling epoch, these asymmetries contribute to an increased $\Neff$ and induce specific spectral distortions in neutrinos and antineutrinos~\cite{Grohs:2016cuu}. If the asymmetries vary between different neutrino flavors, neutrinos and antineutrinos can undergo synchronous oscillations~\cite{Bell:1998ds,Pastor:2001iu,Abazajian:2002qx,Wong:2002fa}, leading to a redistribution of asymmetries among the flavors~\cite{Dolgov:2002ab,Pastor:2008ti,Mangano:2010ei,Mangano:2011ip,Castorina:2012md,Barenboim:2016shh,Johns:2016enc,Froustey:2021azz}.
After neutrino decoupling, during the Big Bang Nucleosynthesis (BBN) epoch, primordial neutrino asymmetries primarily affect the final proton-to-neutron ratio and the resulting light element abundances. This effect depends largely on the distribution of electron neutrinos and antineutrinos, while corrections from the increased $\Neff$ are secondary~\cite{Sarkar:1995dd,Iocco:2008va,Pitrou:2018cgg,Serpico:2005bc,Mangano:2011ip,Chu:2006ua,Simha:2008mt,Saviano:2013ktj,Burns:2023sgx}. Additionally, these asymmetries influence cosmic microwave background (CMB) anisotropies and large-scale structure (LSS) through the increased $\Neff$ and modified light element abundances, along with the effects of massive neutrinos. Consequently, observations of BBN, CMB, and LSS can constrain primordial neutrino asymmetries~\cite{Simha:2008mt,Oldengott:2017tzj,Pitrou:2018cgg,Matsumoto:2022tlr,Burns:2022hkq,Escudero:2022okz}. Notably, recent findings from the EMPRESS survey suggest a preference for a positive asymmetry in electron neutrinos~\cite{Matsumoto:2022tlr,Burns:2022hkq,Escudero:2022okz}. Specifically, combining BBN and CMB observations yields $\xi_{\nu_e} = 0.034 \pm 0.014$~\cite{Escudero:2022okz}.

However, the current situation is less satisfactory compared to the Standard Model (SM) scenario due to persistent uncertainties concerning neutrino decoupling. These uncertainties involve neutrino spectral distortions, flavor oscillations, and finite-temperature quantum electrodynamics (FTQED) corrections. Consequently, the effects of primordial neutrino asymmetries on BBN and CMB, as well as the constraints derived from observational data, are generally evaluated using approximations, such as the instantaneous neutrino decoupling approximation and the assumption of massless neutrinos.
To derive precise constraints on primordial neutrino asymmetries from both current and future observations, it is essential to rigorously analyze these asymmetries within a comprehensive framework of neutrino decoupling. 
A recent study~\cite{Froustey:2024mgf} makes an effort towards a rigorous analysis of primordial neutrino asymmetries. 
This study focuses on the implications of neutrino synchronous oscillations on the neutrino flavor equilibration and BBN process, while the corresponding constraints on the primordial neutrino asymmetries are obtained by combining the BBN measurements and the constraints on the parameter set ($\Neff, Y_P, \omega_b$) from CMB + BAO experiments.
However, the resulting (anti)neutrino spectral distortions and their implications for the BBN and CMB are not discussed in~\cite{Froustey:2024mgf}. 
Furthermore, the implications of primordial neutrino asymmetries for both the CMB angular power spectrum and the BAO are not addressed in~\cite{Froustey:2024mgf}, which according to our results are actually important for the accurate determination of the cosmological constraints.

In this work, we focus on evaluating the neutrino decoupling process and its subsequent effects on BBN, CMB, and LSS in the presence of primordial neutrino asymmetries. 
{  Since the synchronous oscillations before neutrino decoupling will redistribute the neutrino asymmetries between different flavors, the resulting neutrino asymmetries of different flavors are usually of the same order, so that $\xi_{\nu_e} \simeq \xi_{\nu_\mu} \simeq \xi_{\nu_\tau}$~\cite{Dolgov:2002ab,Pastor:2008ti,Mangano:2010ei,Mangano:2011ip,Castorina:2012md,Barenboim:2016shh,Johns:2016enc,Froustey:2021azz}. Therefore, a complete flavor equilibration with $\xi_{\nu_e} = \xi_{\nu_\mu} = \xi_{\nu_\tau} = \xin$ is usually assumed in the study of the cosmological constraints on neutrino asymmetries~\cite{Simha:2008mt,Oldengott:2017tzj,Pitrou:2018cgg,Matsumoto:2022tlr,Burns:2022hkq,Escudero:2022okz}. 
The framework we have developed in this work can also be straightforwardly applied to studies of the general flavour equilibration process. However, a detailed illustration is beyond the scope of this paper. Therefore, we will also assume complete flavor equilibration in this paper and leave further studies of general flavor equilibration to future work.
Relevant previous discussions of the exact flavor equilibration process and its cosmological implications can also be found in, \eg \cite{Pastor:2008ti,Mangano:2011ip,Froustey:2024mgf}.}
To accurately describe the evolution of neutrino density matrices during decoupling, including flavor oscillations and full collision effects, we employ the complete set of neutrino quantum kinetic equations (QKEs). These equations are derived using several approaches, such as perturbative expansions of the density matrix~\cite{Sigl:1993ctk}, relativistic generalizations of the Bogoliubov-Born-Green-Kirkwood-Yvon (BBGKY) equations~\cite{Volpe:2013uxl,Volpe:2015rla,Froustey:2020mcq}, and the Closed-Time Path (CTP) formalism from non-equilibrium quantum field theory~\cite{Vlasenko:2013fja,Blaschke:2016xxt,Kainulainen:2023ocv,Kainulainen:2024fdg}.
The main improvements of our study are as follows:
\begin{itemize}
    \item We re-derive the most general form of the QKEs for neutrinos and antineutrinos using the CTP formalism, incorporating primordial neutrino asymmetry.
    \item We numerically solve the QKEs for the neutrino decoupling process, accounting for flavor oscillations, matter interactions, FTQED corrections up to $\mathcal{O}(e^3)$, and complete neutrino-electron and neutrino-neutrino collision terms, matching the accuracy of state-of-the-art Standard Model (SM) predictions~\cite{Froustey:2020mcq,Bennett:2020zkv}\footnote{Additional QED corrections, such as those to interaction vertices~\cite{Cielo:2023bqp,Jackson:2023zkl,Drewes:2024wbw}, remain under debate for the SM case and are not included in this study.}.
    \item We use the resulting density matrices to study the impacts of primordial neutrino asymmetry on BBN, CMB, and LSS, with a particular emphasis on (anti)neutrino spectral distortions by comparing these results with those obtained using the Fermi-Dirac distribution for neutrinos.
    \item Recognizing that Baryon Acoustic Oscillations (BAO) are also influenced by primordial neutrino asymmetry, we derive constraints on primordial neutrino asymmetry by combining traditional BBN and CMB datasets with BAO observations for the first time.
\end{itemize}

\begin{figure}\centering
\includegraphics[width=1.0\textwidth]{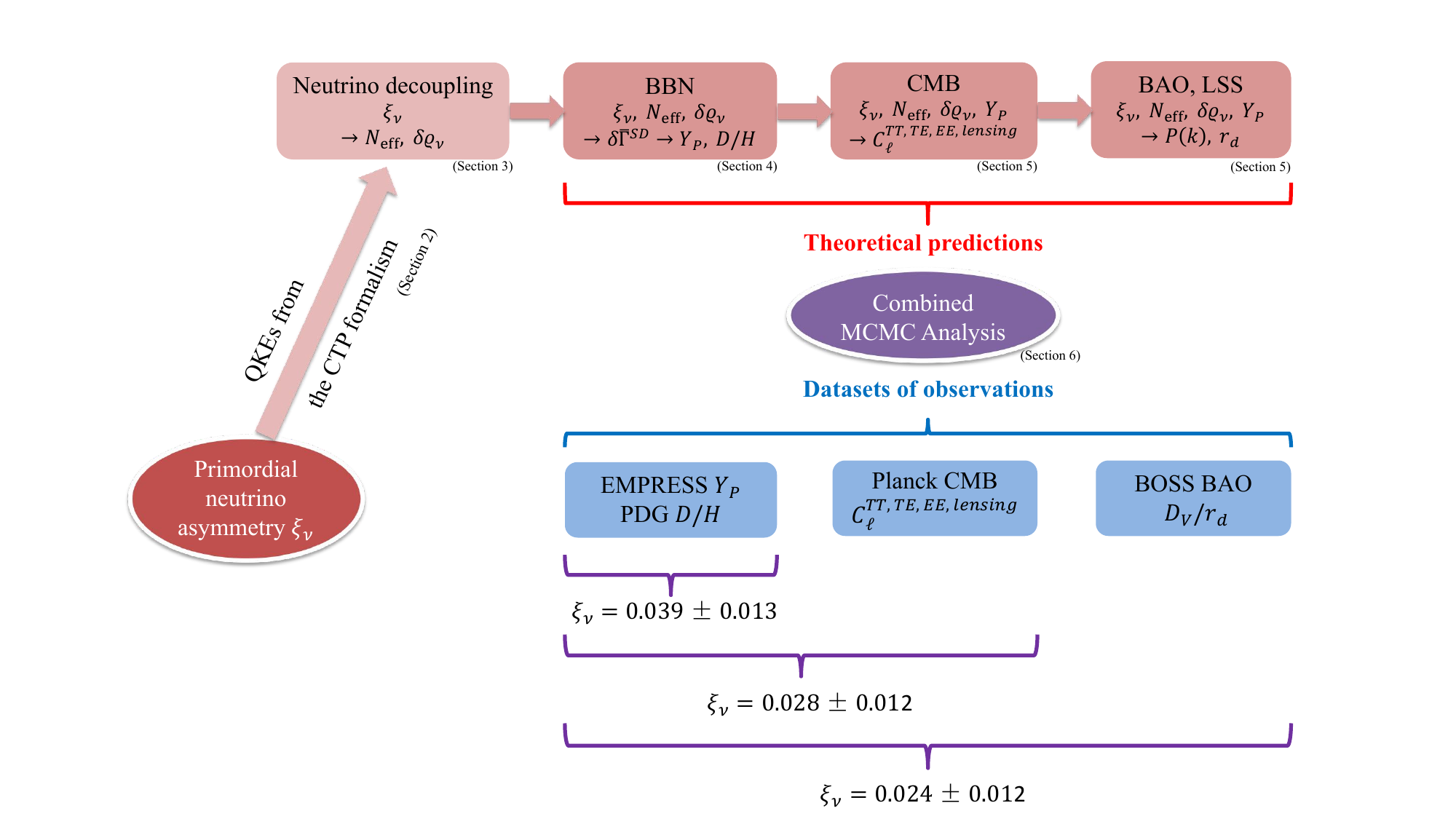}
\caption{
A schematic illustration of the paper's structure. Section~\ref{sec:qkes} derives the QKEs for neutrinos using the CTP formalism. These QKEs are then applied to study neutrino decoupling in the presence of primordial neutrino asymmetry $\xin$ in Section~\ref{sec: decoupling}. Based on this precise treatment of neutrino decoupling, Sections~\ref{sec:bbn} and~\ref{sec:cmb} explore the effects of primordial neutrino asymmetry on BBN, the CMB, and LSS. 
For each section, the primary input and output physical quantities dependent on $\xin$ are labeled in each box.
Finally, in Section~\ref{sec:mcmc}, we obtain cosmological constraints on $\xin$ through Markov Chain Monte Carlo (MCMC) analyses, combining theoretical predictions with observational data from BBN, CMB, and LSS.
}
\label{fg:guide}
\end{figure}

The key physical results of this study are:
\begin{itemize}
    \item We numerically evaluate $\neff$ and the (anti)neutrino spectral distortions, incorporating primordial neutrino asymmetries with state-of-the-art accuracy. 
    For primordial neutrino asymmetries with complete flavor equilibration before neutrino decoupling, \ie $\xi_{\nu_e} = \xi_{\nu_\mu} = \xi_{\nu_\tau} = \xin$, we find that $\neff$ can be expressed as $ \neff = \neff^{\rm SM} + \delta \neff^{\xin, \rm ther} + 0.0102 \, \xin^2$, where \(\neff^{\rm SM} = 3.0440 \pm 0.002\) represents the SM prediction~\cite{Froustey:2020mcq,Bennett:2020zkv}. The term \(\delta \neff^{\xin, \rm ther}\) accounts for the contribution from primordial neutrino asymmetry under the instantaneous decoupling approximation and is given by $\delta \neff^{\xin, \rm ther} \simeq 3 \left( \frac{30}{7 \pi^2} \xi_{\nu}^2 + \frac{15}{7 \pi^4} \xi_{\nu}^4 \right)$. The final term represents the additional correction from non-instantaneous decoupling effects in the presence of non-zero primordial neutrino asymmetry.
    \item We provide state-of-the-art predictions for the abundances of light elements produced by BBN. Using the EMPRESS measurement of helium abundance, $Y_P |_{\rm EMPRESS} = 0.2370^{+0.0034}_{-0.0033}$, we immediately indicate a positive primordial neutrino asymmetry, $0.032 \leq \xin \leq 0.052$, with fixed baryon abundance $\omega_b$. We also find that (anti)neutrino spectral distortions are crucial for precise BBN predictions, especially for the abundance of Helium-4, $Y_P$.
    \item For constraints on primordial neutrino asymmetry, combining EMPRESS BBN and Planck CMB data yields $\xin = 0.028 \pm 0.012$, while incorporating BOSS BAO data results in $\xin = 0.024 \pm 0.012$. Compared to the previous result obtained in Ref.~\cite{Escudero:2022okz}, $\xin = 0.034 \pm 0.014$, our results show an approximately 18\% and 29\% reduction in the central value and a slight decrease in uncertainty, respectively.
    \item We also discuss the impacts of treating both the sum of neutrino masses $\sum m_\nu$ and the primordial neutrino asymmetry $\xin$ as free parameters on the resulting constraints for each. Finally, we demonstrate how the constraints on $\xin$ can be converted into constraints on ultraviolet (UV) model parameters, using a specific Q-ball decay model capable of generating large primordial neutrino asymmetries.
\end{itemize}

This paper is organized as follows.
In Section~\ref{sec:qkes}, we re-derive the QKEs for Dirac and Majorana neutrinos using the CTP formalism. 
Section~\ref{sec: decoupling} presents the QKEs in the context of the early universe and provides corresponding numerical results, considering primordial neutrino asymmetries.
The impact of primordial neutrino asymmetries on the BBN epoch is discussed in Section~\ref{sec:bbn}, while their effects on both the CMB and LSS are analyzed in Section~\ref{sec:cmb}. 
In Section~\ref{sec:mcmc}, we present results from a joint MCMC analysis of neutrino primordial asymmetries and neutrino masses, utilizing datasets from BBN, CMB, and BAO observations. 
Finally, we conclude in Section~\ref{sec:conclusion}. 
To guide the reader, a schematic illustration of this work is presented in Fig.~\ref{fg:guide}.
Throughout this manuscript, we use natural units where $\hbar = c = k_B = 1$.

\section{Quantum Kinetic Equations for Dirac and Majorana Neutrinos}\label{sec:qkes}

In this section, we derive the quantum kinetic equations (QKEs) for both Dirac and Majorana neutrinos using the closed-time-path (CTP) formalism of non-equilibrium quantum field theory. The QKEs generalize the traditional Boltzmann equation by governing the evolution of density matrices and incorporating both the phase-space distribution functions for the eigenstates and the quantum coherence between them. These equations have also been derived through various approaches and approximations in the literature (see \eg~\cite{Sigl:1993ctk,Volpe:2013uxl,Volpe:2015rla,Froustey:2020mcq}).

\subsection{The Closed-Time-Path formalism}

The Closed-Time-Path (CTP) formalism, also known as the in-in formalism, provides a general framework for non-equilibrium quantum field theory (QFT). This approach is necessary because, in non-equilibrium situations, the final state of the system is not known a priori. The CTP formalism employs a closed time contour to handle such cases.
Specifically, the path integral in this formalism is defined over a closed time path that starts at an initial time $t_0$, extends to infinity on the positive branch, and then returns to $t_0$ on the negative branch. This closed time path allows for the consistent treatment of quantum fields in non-equilibrium scenarios by incorporating both forward and backward evolution in time.
Accordingly, depending on the location of the time arguments on the $\pm$ branches, there are four kinds of real-time Green functions, which we denote as
\begin{subequations}
\begin{align}
\left(S_{ab} \right)^> (x,y) & \equiv \langle \nu_a (x_-) \bar{\nu}_b (y_+) \rangle ~, \\
\left(S_{ab} \right)^< (x,y) & \equiv \langle \bar{\nu}_b (y_-) \nu_a (x_+) \rangle ~, \\
\left(S_{ab} \right)^{\rm T} (x,y) & \equiv \langle \mathcal{T} \left( \nu_a (x_+) \bar{\nu}_b (y_+) \right) \rangle ~, \\
\left(S_{ab} \right)^{\rm \bar T} (x,y) & \equiv \langle \bar{\mathcal{T}} \left( \nu_a (x_-) \bar{\nu}_b (y_-) \right) \rangle ~,
\end{align}
\end{subequations}
where $a$ and $b$ denote flavor indices of the neutrinos, while spinor indices are suppressed. The operators $\mathcal{T}$ and $\bar{\mathcal{T}}$ represent the usual time-ordering and reversed time-ordering operations, respectively. It is evident that only two of these four Green functions are independent. For example, by explicitly writing out the time-ordering operators, we obtain
\begin{subequations}
\begin{align}
\left(S_{ab} \right)^{\rm T} (x,y) & = \theta (x^0 - y^0) \left(S_{ab} \right)^> (x,y) - \theta (y^0 - x^0) \left(S_{ab} \right)^< (x,y) ~, \\
\left(S_{ab} \right)^{\rm \bar T} (x,y) & = \theta (y^0 - x^0) \left(S_{ab} \right)^> (x,y) - \theta (x^0 - y^0) \left(S_{ab} \right)^< (x,y) ~.
\end{align}
\end{subequations}
In practice, it is also convenient to introduce the other two dependent propagators to describe the non-equilibrium systems of neutrinos, namely the statistical and spectral functions, defined as
\begin{subequations}
\begin{align}
S^{F}_{ab} (x,y) & \equiv \frac{1}{2} \langle [\nu_a (x), \bar{\nu}_b (y)] \rangle = \frac{1}{2} \left( \left(S_{ab} \right)^> (x,y) - \left(S_{ab} \right)^< (x,y) \right) \,, \\
S^{\mathcal{A}}_{ab} (x,y) & \equiv \frac{i}{2} \langle \{ \nu_a (x), \bar{\nu}_b (y) \} \rangle = \frac{i}{2} \left( \left(S_{ab} \right)^> (x,y) + \left(S_{ab} \right)^< (x,y) \right) \,.
\end{align} \label{eq:def SF}
\end{subequations}
The advantage of this definition is that the statistical and spectral functions have straightforward physical interpretations~\cite{Berges:2004yj,Blaschke:2016xxt,Vlasenko:2013fja}. Specifically, the spectral function provides information about the spectrum of the theory, while the statistical function encodes information about the occupation numbers and quantum coherence of the available states, \ie information about the density matrices. 
Additionally, we note that we do not specify whether the neutrinos are Dirac or Majorana particles here; the distinction between Dirac and Majorana neutrinos will be discussed later.

Therefore, the starting point for the QKEs is the equation of motion for the Green functions $S_{ab} (x,y)$, which can be expressed through the Dyson-Schwinger equation:
\begin{equation}\label{eq: SDeq}
   \left( i \slashed{\partial}^x - m \right) \tilde{S} (x,y) = \mathbb{1} \, i \delta^4(x-y)
+ i \int d^4 z \, \tilde{\Sigma} (x, z) \tilde{S}(z,y) \,.
\end{equation}
Here, $\tilde{S}$ denotes the matrix of Green functions in the Closed-Time-Path (CTP) space, and is defined as
\begin{equation}
     \tilde{S} = \begin{pmatrix}
    \left(S_{ab} \right)^T & \left(S_{ab} \right)^< \\
    \left(S_{ab} \right)^> & - \left(S_{ab} \right)^{\bar{T}}
\end{pmatrix} \,.
\end{equation}
In this context, $m \equiv \mathbb{1}_{2 \times 2} \, m_{ab}$, where $m_{ab}$ is the neutrino mass matrix. The identity matrix preceding the Dirac delta function $\delta^4(x - y)$ is both an identity matrix in the CTP space and in the flavor space, i.e., $\mathbb{1} \, i \delta^4(x-y) = \mathbb{1}_{2 n_f \times 2 n_f} \, i \delta^4(x-y)$, with $n_f$ representing the number of neutrino flavors.
Additionally, $\tilde{\Sigma} (x,y)$ is the neutrino proper self-energy, which is generally a functional of the Green functions and can be decomposed into a local singular term and a matrix in CTP space:
\begin{equation}
\tilde{\Sigma} (x,y)= - \mathbb{1}_{2 \times 2} i \Sigma (x) \delta^{4} (x-y) + \tilde{\Pi} \,, \quad \tilde{\Pi} = \begin{pmatrix}
    \left(\Pi_{ab} \right)^T & \left(\Pi_{ab} \right)^< \\
    \left(\Pi_{ab} \right)^> & - \left(\Pi_{ab} \right)^{\bar{T}}
\end{pmatrix} \,.
\end{equation}
In principle, $\Sigma (x)$ arises from the forward scattering potentials contributed by one-loop diagrams, while $\Pi^\pm (x,y)$ represents contributions from non-forward scatterings and annihilations that occur in at least two loops.

In the following, we suppress the flavor and spinor indices of the Green functions for simplicity. Expanding Eq.~\eqref{eq: SDeq}, the Dyson-Schwinger equation for $S^{\gtrless}$ can be written as
\begin{equation}
    \left( i \slashed{\partial}^x - m - \Sigma(x) \right) S^\gtrless (x,y) = i \int d^4z \left( \Pi^H(x,z) S^\gtrless(z,y) + \Pi^\gtrless(x,z) S^H(z,y) \right) \pm i \, \mathcal{C}_{\rm coll} (x,y) \,,
\end{equation}
which are known as the Kadanoff–Baym (KB) equations. Here, $ S^H \equiv (S^T - S^{\bar T})/2 $ and similarly for $\Pi^H$, while the collision term $\mathcal{C}_{\rm coll}$ is defined as
\begin{equation}
    \mathcal{C}_{\rm coll} (x,y) \equiv \frac{1}{2} \int d^4z \left( \Pi^>(x,z) S^<(z,y) + \Pi^<(x,z) S^>(z,y) \right) \,,
\end{equation}
where the $\pm$ before the collision term corresponds to $\gtrless$, respectively.

Using the definition in Eq.~\eqref{eq:def SF}, we can derive the equations for the statistical function $S^F$ and the spectral function $S^\mathcal{A}$:
\begin{subequations} \label{eq: eom4sfsa}
    \begin{align}
        \left( i \slashed{\partial}^x - m - \Sigma(x) \right) S^F (x,y) &= i \int d^4z \left( \Pi^H(x,z) S^F(z,y) + \Pi^F(x,z) S^H(z,y) \right) - i \, \mathcal{C}_{\rm coll} (x,y) \,, \label{eq: eom4sf}\\
        \left( i \slashed{\partial}^x - m - \Sigma(x) \right) S^\mathcal{A} (x,y) &= i \int d^4z \left( \Pi^H(x,z) S^\mathcal{A}(z,y) + \Pi^\mathcal{A}(x,z) S^H(z,y) \right) \,.
    \end{align}
\end{subequations}
We note that Eq.~\eqref{eq: eom4sf} is equivalent to Eq.~(26) for the statistical function given in Ref.~\cite{Vlasenko:2013fja}. As will be showed, our formulation is more convenient for the Wigner transform.

The purpose of the Wigner transform is to separate the evolution of the Green functions on the microscopic scale with respect to $r \equiv x-y$ from the evolution on the macroscopic scale with respect to $X \equiv (x+y)/2$, which is our primary focus. The (inverse) Wigner transform for any two-point (or self-energy) function $F(x,y)$ is given by
\begin{subequations}
    \begin{align}
        F (k, X) &= \int d^4r \, e^{i k \cdot r} \, F \left(X + \frac{r}{2}, X - \frac{r}{2}\right) \,, \\
        F (x, y) &= \int \frac{d^4k}{(2\pi)^4} \, e^{-i k \cdot (x-y)} \, F \left(k, \frac{x+y}{2}\right) \,.
    \end{align}
\end{subequations}
To transform the equations for statistical and spectral functions, we also need the Wigner transform of the generalized convolution term like $\int d^4z \, F(x,z) \, G(z,y)$. The result is given by
\begin{equation}
    \int d^4z \, F(x,z) \, G(z,y) = \int \frac{d^4k}{(2\pi)^4} \, e^{-i k \cdot (x-y)} \, e^{- i \Diamond} \left\{ F(k,X) \right\} \left\{ G(k,X) \right\} \,,
\end{equation}
where the right-hand side is known as the Moyal product, and the diamond operator $\Diamond$ is defined as
\begin{equation}
\label{diamond}
\Diamond \left\{ F(k,X) \right\} \left\{ G(k,X) \right\} = \frac{1}{2} \left( \frac{\partial F}{\partial X^\mu} \frac{\partial G}{\partial k_\mu} - \frac{\partial F}{\partial k_\mu} \frac{\partial G}{\partial X^\mu} \right) \,.
\end{equation}
The Wigner-transformed equations for the statistical and spectral functions, given by Eq.~\eqref{eq: eom4sfsa}, can then be straightforwardly obtained as
\begin{subequations} \label{eq:sfsa}
    \begin{align}
        \left( \slashed{k} + \frac{i}{2} \slashed{\partial}_X - m \right) S^F(k,X) &= e^{- i \Diamond} \left\{ \Sigma(X) + i \Pi^H (k,X) \right\} \left\{ S^F(k,X) \right\} \\
        &\quad + i \, e^{- i \Diamond} \left\{ \Pi^F (k,X) \right\} \left\{ S^H(k,X) \right\} - i \, \mathcal{C}_{\rm coll} (k,X) \,, \label{eq:sf}\\
        \left( \slashed{k} + \frac{i}{2} \slashed{\partial}_X - m \right) S^\mathcal{A}(k,X) &= e^{- i \Diamond} \left\{ \Sigma(X) + i \Pi^H (k,X) \right\} \left\{ S^\mathcal{A}(k,X) \right\} \\
        &\quad + i \, e^{- i \Diamond} \left\{ \Pi^\mathcal{A} (k,X) \right\} \left\{ S^H(k,X) \right\} \,, \nonumber
    \end{align}
\end{subequations}
with the collision term
\begin{equation}
    \mathcal{C}_{\rm coll} (k,X) = \frac{1}{2} \left( e^{- i \Diamond} \left\{ \Pi^> (k,X) \right\} \left\{ S^< (k,X) \right\} - e^{- i \Diamond} \left\{ \Pi^< (k,X) \right\} \left\{ S^> (k,X) \right\} \right) \,.
\end{equation}

\subsection{Gradient expansion of the kinetic equation}
So far, the equations for statistical and spectral functions \eqref{eq:sfsa} involve infinite gradient expansions and are thus valid to all orders. In practice, we follow the procedure outlined in Ref. \cite{Blaschke:2016xxt} and truncate these infinite series at the leading non-trivial order in a small parameter $\epsilon$. 
To achieve this, we first need to specify the power counting in the regime of interest, as follows:
\begin{itemize}
    \item The variation of physical quantities with respect to the macroscopic coordinate $X$ is small compared to its intrinsic de Broglie frequency, so each derivative $\partial_X$ carries one power of $\epsilon$.
    \item Compared to the neutrino energy, the masses and interaction potentials are small, so terms involving $m$ or $\Sigma$ carry one power of $\epsilon$.
    \item Contributions to the self-energy $\Pi$ appear at least at the 2-loop order in the Feynman diagram expansion, so terms involving $\Pi$ carry two power of $\epsilon$.
    \item The variation of the two-point functions from equilibrium is small, so $S^H = (S^T - S^{\bar{T}})/2$ carries one power of $\epsilon$.
\end{itemize}

In conclusion, our power counting reads
\begin{equation}
    \frac{\partial_X, m, \Sigma, S^H}{E_\nu} = \mathcal{O} (\epsilon) \,, \quad
\frac{\tilde{\Pi}^\pm}{E_\nu} = \mathcal{O} (\epsilon^2) \,,
\end{equation}
where $E_\nu$ is the neutrino energy. Additionally, we note that the contribution from the $\Pi^H$ term can be absorbed into the $\Sigma$ term at the considered order, so we will omit the relevant terms in the following equations.

Next, we expand the equations for statistical and spectral functions \eqref{eq:sfsa} order by order to derive the QKEs for neutrinos and antineutrinos. To do this, we first note that the kinetic equation \eqref{eq:sf} involves the sixteen spinor components (scalar, pseudoscalar, vector, axial-vector, tensor) of the statistical function and the self-energies, which can be decomposed as follows:
\begin{subequations}
    \begin{align}
        S^F  &= \left[ F_S + \left( F_V^R \right)^\mu \gamma_\mu - \frac{i}{4} \left( F_T^L \right)^{\mu \nu} \sigma_{\mu \nu} \right] P_L + 
     \left[ F_S^\dagger + \left( F_V^L \right)^\mu \gamma_\mu + \frac{i}{4} \left( F_T^R \right)^{\mu \nu} \sigma_{\mu \nu} \right] P_R \,, \label{eq:sfcom}\\
      \Pi^F &= \left[ \Pi_S + \Pi_R^\mu \gamma_\mu - \frac{i}{4} \left( \Pi_T^L \right)^{\mu \nu} \sigma_{\mu \nu} \right] P_L + 
     \left[ \Pi_S^\dagger + \Pi_L^\mu \gamma_\mu + \frac{i}{4} \left( \Pi_T^R \right)^{\mu \nu} \sigma_{\mu \nu} \right] P_R \,,
\end{align}
\end{subequations}
where $P_{L,R} \equiv \frac{1 \mp \gamma_5}{2}$ and $\sigma_{\mu\nu} \equiv \frac{i}{2}[\gamma_\mu, \gamma_\nu]$. Due to the hermiticity conditions, we have $F_V^{L\dagger} = F_V^L$, $F_V^{R\dagger} = F_V^R$, and $F_T^{L\dagger} = F_T^R$, with similar conditions applying to the components of $\Pi^F$. An analogous decomposition for the forward scattering potential $\Sigma(x)$ is also permissible.

Therefore, the kinetic equation \eqref{eq:sf} encompasses both the QKEs and the algebraic constraints for various spinor components. The basic strategy to obtain the QKEs is as follows:
\begin{enumerate}
    \item Solve Eq. \eqref{eq:sfsa} at $\mathcal{O}(\epsilon^0)$ to obtain the leading order solutions for the statistical and spectral functions, and establish the relations between the remaining spinor components of the statistical function and the density matrices for Dirac and Majorana neutrinos.
    
    \item Beyond $\mathcal{O}(\epsilon^0)$, all spinor components of $S^F$ receive additional corrections due to neutrino masses and interactions. Thus, solve the kinetic equation \eqref{eq:sf} at $\mathcal{O}(\epsilon)$ to find the constraint relations among these small components.
    
    \item Expand the kinetic equation \eqref{eq:sf} to $\mathcal{O}(\epsilon^2)$ and reorganize it using the constraint relations obtained at $\mathcal{O}(\epsilon)$.
    
    \item Finally, extract the QKEs for Dirac and Majorana neutrinos, incorporating the relations between the spinor components of the statistical function and the density matrices.
\end{enumerate}

The detailed derivation of these steps is presented in the following sub-subsections.

\subsubsection{Kinetic equation to $\mathcal{O} (\epsilon^0)$}

To $\mathcal{O}(\epsilon^0)$, the kinetic equation \eqref{eq:sf} yields $\slashed{k} S^F(k,x) = \mathcal{O}(\epsilon)$. Substituting the general form for $S^F$ from Eq. \eqref{eq:sfcom}, we find that only the left-handed (L) and right-handed (R) vector components, along with two tensor components, are non-zero. These can be parameterized by the real functions $F_{L,R}(k,x)$ and the complex function $\Phi(k,x)$:
\begin{subequations}
\begin{align}
\left(F_V^{L,R} \right)^\mu(k,x) &= \hat{\kappa}^\mu(k) \, F_{L,R}(k,x) \,, \\
\left(F_T^L \right)_{\mu \nu}(k,x) &= e^{-i\varphi(k)} (\hat{\kappa}(k) \wedge \hat{x}^{-}(k))_{\mu \nu} \, \Phi(k,x) \,, \\
\left(F_T^R \right)_{\mu \nu}(k,x) &= e^{i\varphi(k)} (\hat{\kappa}(k) \wedge \hat{x}^{+}(k))_{\mu \nu} \, \Phi^\dagger(k,x) \,.
\end{align}
\end{subequations}
Here, we have introduced a set of basis vectors used to express the Lorentz tensors and components of the two-point functions. The basis consists of two light-like vectors $\hat{\kappa}^\mu(k) = ({\rm sign}(k^0), \hat{k})$ and $\hat{\kappa}'^\mu(k) = ({\rm sign}(k^0), -\hat{k})$, and two transverse four-vectors $\hat{x}_{1,2}(k)$ which satisfy the relations $\hat{\kappa} \cdot \hat{\kappa} = \hat{\kappa}' \cdot \hat{\kappa}' = 0$, $\hat{\kappa} \cdot \hat{\kappa}' = 2$, $\hat{\kappa} \cdot \hat{x}_i = \hat{\kappa}' \cdot \hat{x}_i = 0$, and $\hat{x}_i \cdot \hat{x}_j = -\delta_{ij}$. Additionally, we define $\hat{x}^\pm \equiv \hat{x}_1 \pm i \hat{x}_2$, such that $\hat{x}^+ \cdot \hat{x}^- = -2$.

In addition, $F_{L,R}(k,x)$ and $\Phi(k,x)$ can be organized into a $2 n_f \times 2 n_f$ matrix:
\begin{equation}
\hat{F} = 
\left(
\begin{array}{cc}
F_L &  \Phi \\
\Phi^\dagger &  F_R
\end{array}
\right)~.
\label{eq:Fhat0}
\end{equation}
The advantage of this organization is that it can be directly related to the conventional density matrices. At $\mathcal{O}(\epsilon^0)$, the density matrices for neutrinos and antineutrinos can be obtained via integrals over positive and negative frequencies. For Dirac neutrinos, we have
\begin{subequations}
\label{eq:f4dir}
\begin{align}
\label{eq:moment}
- 2 \int_0^\infty \frac{dk^0}{2 \pi} \hat{F}(k,x) &= F(\vec{k},x) - \frac{1}{2} \id \,, \\
\label{eq:momentbar}
- 2 \int_{-\infty}^0 \frac{dk^0}{2 \pi} \hat{F}(k,x) &= \bar{F}(-\vec{k},x) - \frac{1}{2} \id \,.
\end{align}
\end{subequations}
The density matrices $F$ and $\bar{F}$ are also $2 n_f \times 2 n_f$ matrices:
\begin{equation}
F (\vec{p}, x) =
\left(
\begin{array}{cc}
\vrho_{LL} & \vrho_{LR} \\
\vrho_{RL} & \vrho_{RR}
\end{array}
\right) \,;  
\qquad \qquad \bar{F} (\vec{p}, x) = 
\left(
\begin{array}{cc}
\bvrho_{RR} & \bvrho_{RL} \\
\bvrho_{LR} & \bvrho_{LL}
\end{array}
\right) \,,
\label{eq:FDirac}
\end{equation}
where $n_f$ is the total number of neutrino flavors, and each block is a square $n_f \times n_f$ matrix with $\vrho_{h h'} = \vrho_{h h'}^{ij}$, where $i,\, j$ denote the flavors and $h,\,h' \in \{L, R\}$ denote the chirality.
Among the various components, $\vrho_{LL}$ and $\bvrho_{RR}$ represent the density matrices for active left-handed neutrinos and right-handed antineutrinos, respectively. Conversely, $\vrho_{RR}$ and $\bvrho_{LL}$ correspond to sterile right-handed neutrinos and left-handed antineutrinos. The remaining density matrices account for the quantum coherence between different chiralities of neutrinos and antineutrinos across all flavors. Note that there is no single-particle density matrix for quantum coherence between Dirac neutrinos and antineutrinos, as they are fundamentally distinct particles.

For Majorana neutrinos, the antiparticle is the neutrino itself, and the density matrices for neutrinos and antineutrinos are related by $\vrho_{hh'}^{ij} = \bvrho_{hh'}^{ji}$ through a phase change. Therefore, only the density matrices for active neutrinos are necessary, which can be consolidated into a single $2 n_f \times 2 n_f$ matrix using the definitions $\vrho \equiv \vrho_{LL}$, $\bvrho \equiv \bvrho_{RR} = \vrho_{RR}^T$, and $\phi \equiv \vrho_{LR}$:
\begin{equation}
{\cal F} = 
\left(
\begin{array}{cc}
\vrho &  \phi \\
\phi^\dagger &  \bar{\vrho}^T
\end{array}
\right)~,
\label{eq:FMajorana}
\end{equation}
where $\vrho$ and $\bvrho$ denote the density matrices for active neutrinos and antineutrinos, respectively, and $\phi$ describes the quantum spin coherence between neutrinos and antineutrinos. Similar to the Dirac case, the density matrix for Majorana neutrinos, ${\cal F}$, can be extracted from the two-point functions via
\begin{equation}
    - 2 \int_0^\infty \frac{dk^0}{2 \pi} \hat{F}(k,x) = \mathcal{F}(\vec{k},x) - \frac{1}{2} \id \,.
\label{eq:f4maj}
\end{equation}

Beyond $\mathcal{O}(\epsilon^0)$, the equations \eqref{eq:f4dir} and \eqref{eq:f4maj} serve as the definitions of density matrices for Dirac and Majorana neutrinos, respectively, and are used to extract the QKEs from the evolution equation of $\hat{F}$. 
Additionally, to obtain the QKEs at $\mathcal{O}(\epsilon^2)$, only the $\mathcal{O}(\epsilon^0)$ expression for the vector component of the spectral function is required, which is given by $\hat{\rho}(k) = 2 i \pi |\vec{k}| \delta(k^2) \text{sgn}(k^0)$.

\subsubsection{Kinetic equation to   $\mathcal{O} (\epsilon)$ }

Beyond $\mathcal{O}(\epsilon^0)$, we still consider $F_{L,R}(k,x)$ and $\Phi(k,x)$ as the four independent spinor components of $S^F(k,x)$, which can be isolated using the projections:
\begin{subequations}
\begin{align}
F_{L,R}(k,x) &\equiv \frac{1}{4} \mathrm{Tr} \left( \gamma_\mu P_{L,R} F^{(\nu)}(k,x) \right) \hat{\kappa}'^{\mu}(k)\,, \\
\Phi^{(\dagger)}(k,x) &\equiv \mp \frac{i}{16} \mathrm{Tr} \left( \sigma_{\mu \nu} P_{L/R} F^{(\nu)}(k,x) \right) (\hat{\kappa}'(k) \wedge \hat{x}^\pm(k))^{\mu\nu} e^{\pm i\varphi(k)}\,,
\label{eq:phi}
\end{align}
\end{subequations}
where the upper (lower) signs and indices refer to $\Phi$ ($\Phi^\dagger$). The dispersion relations for the functions $F_{L,R}(k,x)$ and $\Phi(k,x)$, as well as for the other spinor components of $S^F(k,x)$, are modified by neutrino masses and interactions at $\mathcal{O}(\epsilon)$. Therefore, the general statistical function $S^F(k,x)$ can be expressed as
\begin{align}\label{eq:sf4o1}
    S^F &= \left[ F_R \hat{\kappa}^\mu \gamma_\mu - \frac{i}{4} e^{-i\varphi(k)} (\hat{\kappa}(k) \wedge \hat{x}^{-}(k))_{\mu \nu} \Phi \sigma_{\mu \nu} \right] P_L \\
    &+ \left[ F_L \hat{\kappa}^\mu \gamma_\mu + \frac{i}{4} e^{i\varphi(k)} (\hat{\kappa}(k) \wedge \hat{x}^{+}(k))_{\mu \nu} \Phi^\dagger \sigma_{\mu \nu} \right] P_R \nonumber \\
    &+ \left[ \Delta_S + (\Delta_R)^\mu \gamma_\mu - \frac{i}{4} (\Delta_T^L)^{\mu \nu} \sigma_{\mu \nu} \right] P_L \nonumber \\
    &+ \left[ \Delta_S^\dagger + (\Delta_L)^\mu \gamma_\mu + \frac{i}{4} (\Delta_T^R)^{\mu \nu} \sigma_{\mu \nu} \right] P_R\,, \nonumber
\end{align}
where $\Delta$ represents the $\mathcal{O}(\epsilon)$ small components.

Expand the kinetic equation for $S^F(k, x)$ to $\mathcal{O}(\epsilon)$, we obtain
\begin{equation}
    \left( \slashed{k} + \frac{i}{2} \slashed{\partial}_X - m - \Sigma(x) \right) S^F(k, X) = \mathcal{O} (\epsilon^2).
\end{equation}
Substituting the general statistical function $S^F(k, x)$, including the $\mathcal{O} (\epsilon)$ corrections, into the kinetic equation and decomposing into scalar, vector, and tensor components, we derive the dispersion relations for $F_{L,R}(k, x)$ and $\Phi(k, x)$ at $\mathcal{O} (\epsilon)$:
\begin{align}
    \left(k \cdot \hat{\kappa}\right)F_R - \frac{1}{2}\left\{\Sigma_L^\kappa, F_R\right\} &= \mathcal{O} \left(\epsilon^2\right), \\
\left(k \cdot \hat{\kappa}\right)F_L - \frac{1}{2}\left\{\Sigma_R^\kappa, F_L\right\} &= \mathcal{O} \left(\epsilon^2\right), \\
\left(k \cdot \hat{\kappa}\right)P_+^{ij}F_T^j - \frac{1}{2}\left(\Sigma_R^\kappa P_+^{ij}F_T^j + P_+^{ij}F_T^j \Sigma_L^\kappa\right) &= \mathcal{O} \left(\epsilon^2\right), \\
\left(k \cdot \hat{\kappa}\right)P_-^{ij}F_T^j - \frac{1}{2}\left(\Sigma_L^\kappa P_-^{ij}F_T^j + P_-^{ij}F_T^j \Sigma_R^\kappa\right) &= \mathcal{O} \left(\epsilon^2\right),
\end{align}
and the expressions for the small components
\begin{align}
    \Delta_{L/R}^\kappa &= \mathcal{O} \left(\epsilon^2\right)\ \ \ \ \, , \ \ \ \ \ \Delta_T^i = \mathcal{O} \left(\epsilon^2\right) ,
\\
\Delta_S&=\frac{1}{2\left|\vec{k}\right|}\left(m^\dagger F_R + F_L m^\dagger\right)
+\frac{P_+^{ij}}{2\left|\vec{k}\right|}\left(i\partial^i F_T^j-\left(\Sigma_R^i F_T^j-F_T^j\Sigma_L^i\right)\right)
\\
\Delta_T&=-\frac{1}{2\left|\vec{k}\right|}\left(m^\dagger F_R-F_Lm^\dagger\right)
+\frac{P_+^{ij}}{2\left|\vec{k}\right|}\left(\Sigma_R^i F_T^j + F_T^j \Sigma_L^i\right) ,
\\
\Delta_L^i &= \frac{1}{2\left|\vec{k}\right|}\left(m^\dagger P_-^{ij}F_T^j +P_+^{ij}F_T^j m\right)
+\frac{1}{\left|\vec{k}\right|}\left(\frac{1}{2}\epsilon^{ij}\partial^j F_L-\left(P_-^{ij}\Sigma_R^j F_L+F_LP_+^{ij}\Sigma_R^j\right)\right) ,
\\
\Delta_R^i &=-\frac{1}{2\left|\vec{k}\right|}\left(mP_+^{ij}F_T^j + P_-^{ij}F_T^j m^\dagger\right)
-\frac{1}{\left|\vec{k}\right|}\left(\frac{1}{2}\epsilon^{ij}\partial^j F_R+\left(P_+^{ij}\Sigma_L^j F_R+F_RP_-^{ij}\Sigma_L^j\right)\right) .
\end{align}

\subsubsection{Kinetic equation to   $\mathcal{O} (\epsilon^2)$ }
At $\mathcal{O}(\epsilon^2)$, the kinetic equation $S^F (k, x)$ becomes
\begin{equation}
    \left( \slashed{k} + \frac{i}{2}  \slashed{\partial}_x - m - \Sigma(x) \right) S^F(k,X)  =  -\frac{i}{2}   \left(  \Pi^> (k,x) S^< (k,x)  - \Pi^<(k,x) S^> (k,x) \right) .
\end{equation}
Inserting the expressions for the small components obtained at $\mathcal{O}(\epsilon)$ and reorganizing the kinetic equations, we can derive the kinetic equation and the dispersion relation for $\hat{F}$ as
\begin{subequations}
\begin{align}
\partial^\kappa \hat F + \frac{1}{2 |\vec{k}|}  \left\{ \Sigma^i , \partial^i  \hat F \right\}  +  \frac{1}{2}  \left\{  \frac{\partial \Sigma^\kappa}{\partial x^\mu} , 
 \frac{\partial \hat  F}{\partial k_\mu }       \right\}    
&= - i \left[ H, \hat{F} \right]   +   \hat C  
~,\label{eq:qkec0} \\
\left\{  \hat {\kappa} (k) \cdot k - \Sigma^\kappa \ ,  \ \hat{F} \right\} &= 0~, \label{eq:shell}
\end{align}
\end{subequations}
where we have defined $\partial^\kappa \equiv  \hat\kappa (k) \cdot  \partial$ and  $\partial^i \equiv\hat x^i  (k) \cdot \partial$, and $H$ is the Hamiltonian-like operator given by 
\beq
H  =
\left(
\begin{array}{cc}
H_R &  H_{LR}  \\
H_{LR}^\dagger   & H_L
\end{array}
\right)~, \qquad 
\eeq
with 
\begin{subequations}
\label{eq:hs}
\begin{align}
H_R  &=  \Sigma_R^\kappa   + \frac{1}{2 |\vec{k}|}   \left( m^\dagger m  - \epsilon^{ij} \partial^i \Sigma_R^j   + 4 \Sigma_R^+ \Sigma_R^-\right)
\,,\label{eq:HR}
 \\
H_L  &=  \Sigma_L^\kappa   + \frac{1}{2 |\vec{k}|}   \left( m  m^\dagger  + \epsilon^{ij} \partial^i \Sigma_L^j   + 4 \Sigma_L^- \Sigma_L^+\right)
\,,\label{eq:HL}
\\
H_{LR}  &=   - \frac{1}{|\vec{k}|}  \left( \Sigma_R^+  \, m^\dagger - m^\dagger \, \Sigma_L^- \right) ~, 
\label{eq:Hm}
\end{align}
\end{subequations}
where $\Sigma_{L,R}^\pm \equiv (1/2) \,  e^{\pm i \varphi} \,   ( x_1 \pm i x_2)_\mu  \,   \Sigma^\mu_{L,R}$ and 
$\epsilon^{ij}$ is the two-dimensional  Levi-Civita symbol.
Finally, the collision term  $\hat C $ is given by
\beq
\label{eq:Chat}
\hat{C}  = -  \frac{1}{2}  \left\{  \hat \Pi^> , {\hat S}^< \right\}  + \frac{1}{2} \left\{ \hat \Pi^<,  {\hat S}^> \right\}
\,, \\
\eeq 
with 
\beq
\hat{S}^\pm  =  - i  \hat \rho \, \id   \pm \hat F ~,
\eeq
and the  $2 n_f  \times 2 n_f$  self energy matrices $ \hat \Pi^\pm (k)$   
are given by
\begin{align}
    \hat \Pi^\pm (k,x)  &=
\left(\begin{array}{cc}
\Pi_R^{\kappa \pm} & 2P_T^\pm \\
2P_T^{\pm \dagger}  &  \Pi_L^{\kappa \pm} 
\end{array}
\right)~.
\label{eq:def-Pi-matrix}
\end{align}
where the components are obtained by the  projections
\begin{subequations}\label{eq:projections-general}
\begin{align}
\Pi_{L,R}^\kappa (k,x) &= \frac{1}{2}  \hat{\kappa}_\mu  \, \Tr  \left[ \tilde \Pi  (k,x)\, \gamma^\mu P_{L,R} \right]
\,,\\
P_T (k,x)  &=  \frac{ie^{i\vp}}{16} \, (\hat \kappa \wedge\hat x^+)^{\mu \nu}  \,  \, \Tr  \left[ \tilde \Pi (k,x) \, \sigma_{\mu \nu}  P_{R} \right] ~. 
\end{align}
\end{subequations}

\subsection{QKEs for Dirac and Majorana neutrinos}

Finally, integrating Eq.~(\ref{eq:qkec0}) over positive and negative frequencies using Eq.~\eqref{eq:f4dir}, we obtain the quantum kinetic equations (QKEs) for Dirac neutrinos:
\begin{subequations}
\label{eq:qke4dirac}
\begin{align}
 \partial^\kappa F  + \frac{1}{2 |\vec{k}|} \left\{ \Sigma^i , \partial^i F \right\} - \frac{1}{2} \left\{ \frac{\partial \Sigma^\kappa}{\partial \vec{x}}, \frac{\partial F}{\partial \vec{k}} \right\} &= -i [H, F] + {\cal C} \,, \\
 \partial^\kappa \bar{F} - \frac{1}{2 |\vec{k}|} \left\{ \Sigma^i , \partial^i \bar{F} \right\} + \frac{1}{2} \left\{ \frac{\partial \Sigma^\kappa}{\partial \vec{x}}, \frac{\partial \bar{F}}{\partial \vec{k}} \right\} &= -i [\bar{H}, \bar{F}] + \bar{\cal C} \,.
\end{align}
\end{subequations}
The terms on the left-hand side (LHS) of the QKEs generalize the classical space-time derivative, drift, and force terms. The first term on the right-hand side (RHS) of the QKEs describes neutrino flavor oscillations, which are induced by neutrino masses and forward scattering between neutrinos and background particles. This corresponds to the generalization of the well-known Mikheyev-Smirnov-Wolfenstein (MSW) effect. The second term on the RHS represents the collision term with a non-trivial matrix structure:
\begin{subequations}
\label{eq:C-Cbar}
\begin{align}
{\cal C} &= \frac{1}{2} \left\{ \Pi^> , F \right\} - \frac{1}{2} \left\{ \Pi^< , \id - F \right\} \,, \\
\bar{\cal C} &= \frac{1}{2} \left\{ \bar{\Pi}^> , \bar{F} \right\} - \frac{1}{2} \left\{ \bar{\Pi}^< , \id - \bar{F} \right\} \,,
\end{align}
\end{subequations}
where
\begin{align}
\Pi^\pm (\vec{k}) &= \int_0^\infty dk^0 \, \hat{\Pi}^\pm (k^0, \vec{k}) \, \delta (k^0 - \abs{\vec{k}}) \,, \\
\bar{\Pi}^\pm (\vec{k}) &= - \int_{-\infty}^0 dk^0 \, \hat{\Pi}^\mp (k^0, -\vec{k}) \, \delta (k^0 + \abs{\vec{k}}) \,.
\label{eq:Pi-Pibar}
\end{align}

On the other hand, the QKEs for Majorana neutrinos can be derived by integrating Eq.~(\ref{eq:qkec0}) over positive frequencies using Eq.~\eqref{eq:f4maj}, yielding:
\begin{align}
 \partial^\kappa {\cal F} + \frac{1}{2 |\vec{k}|} \left\{ \Sigma^i , \partial^i {\cal F} \right\} - \frac{1}{2} \left\{ \frac{\partial \Sigma^\kappa}{\partial \vec{x}}, \frac{\partial {\cal F}}{\partial \vec{k}} \right\} &= -i [H, {\cal F}] + {\cal C}_M \,,
\label{eq:qkem1}
\end{align}
with
\begin{align}
{\cal C}_M &= \frac{1}{2} \left\{ \Pi^> , {\cal F} \right\} - \frac{1}{2} \left\{ \Pi^< , \id - {\cal F} \right\} \,.
\end{align}

Explicitly, the potentials induced by forward scattering $\Sigma_{R,L}$ and the collision terms for Dirac and Majorana neutrinos should be calculated from the 1-loop and 2-loop self-energy diagrams, depending on the interaction terms for the neutrinos. For active neutrinos in the early universe, we will present the resulting forward potentials and collision terms in the next section.

\section{Neutrino Decoupling in the Presence of Primordial Neutrino Asymmetry}\label{sec: decoupling}

\subsection{QKEs in the early universe}

In the early universe, both the SM thermal bath and the neutrino system can be approximated as homogeneous and isotropic. Under these conditions, the QKEs for neutrinos can be significantly simplified. First, the drift and force terms on the left-hand side are eliminated, and a Hubble term is introduced to account for the expansion of the Universe. Second, the neutrino density matrix is assumed to depend solely on the absolute value of the momentum. Finally, the coherence matrices between left-handed and right-handed neutrinos, denoted as $\vrho_{LR,RL}$ and $\bar{\vrho}_{LR,RL}$ for Dirac neutrinos, and $\phi$ for Majorana neutrinos, vanish, leading to zero collision terms for these components. Consequently, the non-zero density matrices are $\vrho_{LL,RR}$ and $\bar{\vrho}_{LL,RR}$ for Dirac neutrinos, and $\vrho$ and $\bar{\vrho}$ for Majorana neutrinos.

For active neutrinos, the QKEs for $\vrho_{LL}$ and $\bar{\vrho}_{RR}$ in Dirac neutrinos are identical to those for $\vrho$ and $\bar{\vrho}$ in Majorana neutrinos. Therefore, in the subsequent analysis, we do not distinguish between Dirac and Majorana neutrinos and use $\vrho$ and $\bar{\vrho}$ to represent the density matrices for neutrinos and antineutrinos, respectively. In the flavor basis, the QKEs are written as:
\begin{align}
    i \left[ \frac{\partial}{\partial t} - H p \frac{\partial}{\partial p} \right] \vrho &= \Big[ H_\nu , \vrho \Big] + i \mathcal{I} \,, \\
    i \left[ \frac{\partial}{\partial t} - H p \frac{\partial}{\partial p} \right] \bar{\vrho} &= - \Big[ \bar{H}_\nu, \bar{\vrho} \Big] + i \bar{\mathcal{I}} \,,
\end{align}
where $H \equiv \dot{a}/a$ is the Hubble rate, and $H_\nu$ and $\bar{H}_\nu$ are the Hamiltonian terms for neutrinos and antineutrinos, defined as:
\begin{align}\label{eq:QKEs}
    H_\nu &= U \frac{\mathbb{M}^2}{2p}U^\dagger + \sqrt{2} G_F \left( \mathbb{N}_l + \mathbb{N}_\nu \right) - 2 \sqrt{2} G_F p \left( \frac{\mathbb{E}_l + \mathbb{P}_l}{m_W^2} + \frac{4}{3} \frac{\mathbb{E}_\nu}{m_Z^2} \right) \,, \\
    \bar{H}_\nu &= U \frac{\mathbb{M}^2}{2p}U^\dagger - \sqrt{2} G_F \left( \mathbb{N}_l + \mathbb{N}_\nu \right) - 2 \sqrt{2} G_F p \left( \frac{\mathbb{E}_l + \mathbb{P}_l}{m_W^2} + \frac{4}{3} \frac{\mathbb{E}_\nu}{m_Z^2} \right) \,.
\end{align}
Here, $\mathbb{M}$ represents the neutrino mass matrix in the mass basis, and $U$ is the PMNS matrix, which relates flavor eigenstates to mass eigenstates. Using the standard parameterization of the PMNS matrix~\cite{Giunti:2007ry, Gariazzo:2019gyi}, we have:
\begin{equation}
\label{eq:PMNS}
\mathbb{M} = \begin{pmatrix} 
m_{1}^2 & 0 & 0 \\
0 & m_{2}^2 & 0 \\
0 & 0 & m_{3}^2
\end{pmatrix} \, , \quad
U = \begin{pmatrix} 
c_{12} c_{13} & s_{12} c_{13} & s_{13} \\
- s_{12}c_{23} - c_{12}s_{23}s_{13} & c_{12} c_{23} - s_{12}s_{23}s_{13} & s_{23} c_{13} \\
s_{12}s_{23} - c_{12}c_{23}s_{13} & -c_{12}s_{23} - s_{12}c_{23}s_{13} & c_{23} c_{13}
\end{pmatrix} \, ,
\end{equation}
where $c_{ij} = \cos{\theta_{ij}}$, $s_{ij} = \sin{\theta_{ij}}$, and $\theta_{ij}$ are the mixing angles. We neglect the CP-violating phase in this work.

The remaining terms in the Hamiltonian represent matter potentials induced by coherent scattering between neutrinos and other plasma particles. Here, $m_W$ and $m_Z$ are the masses of the $W$ and $Z$ bosons, respectively, and $G_F$ is the Fermi constant. 
The second term in the Hamiltonian reflects the net number densities of charged leptons ($\mathbb{N}_l \equiv \mathrm{diag}(n_e - n_{\bar{e}}, n_\mu - n_{\bar{\mu}}, 0)$) and neutrinos ($\mathbb{N}_\nu \equiv n_\nu - n_{\bar{\nu}}$). 
The third term accounts for the energy densities of charged leptons and neutrinos ($\mathbb{E}_l \equiv \mathrm{diag}(\rho_e + \rho_{\bar{e}}, \rho_\mu + \rho_{\bar{\mu}}, 0)$, $\mathbb{P}_l \equiv \mathrm{diag}(P_e + P_{\bar{e}}, P_\mu + P_{\bar{\mu}}, 0)$, $\mathbb{E}_\nu \equiv \rho_\nu + \rho_{\bar{\nu}}$).
In practice, the net number densities of charged leptons are often negligible due to the small baryon asymmetry. However, the net number densities of neutrinos can significantly affect neutrino flavor oscillations, especially in the presence of primordial neutrino asymmetries. This can result in enhanced neutrino flavor transitions, including the well-known "synchronized oscillations" when this term dominates the Hamiltonian~\cite{Bell:1998ds, Pastor:2001iu, Abazajian:2002qx, Wong:2002fa, Johns:2016enc, Froustey:2021azz, Froustey:2024mgf}.

Finally, the collision terms $\mathcal{I}$ and $\bar{\mathcal{I}}$ account for contributions from neutrino-electron/positron scattering ($\nu e \leftrightarrow \nu e$), annihilation ($\nu \bar{\nu} \leftrightarrow e^+ e^-$), and neutrino-neutrino self-interactions ($\nu \nu \leftrightarrow \nu \nu$, $\nu \bar{\nu} \leftrightarrow \nu \bar{\nu}$). These terms are decomposed as
\begin{equation}
    \mathcal{I} \equiv \mathcal{I}_{\rm sc} + \mathcal{I}_{\rm ann} + \mathcal{I}_{\nu\nu} 
\end{equation}
with a similar decomposition for the antineutrino collision term $\bar{\mathcal{I}}$. The collision terms depend on the neutrino density matrices, and their explicit forms are detailed in Appendix~\ref{sec: collision}. Contributions from other processes, such as $\mu^\pm$ scattering and annihilation, are neglected due to their minimal impact at the temperatures relevant to neutrino decoupling.

In practice, comoving coordinates and their corresponding variables are employed to eliminate the Hubble expansion term in the QKEs. The comoving temperature is defined as $\Tcm \propto a^{-1}$ \cite{Grohs:2015tfy}, with $\Tcm = T_\nu = T_\gamma$ at the initial time before neutrino decoupling, when all species are tightly coupled. Using the comoving temperature, the reduced scale factor, comoving momentum, and dimensionless photon temperature are expressed as \cite{Mangano:2005cc,Froustey:2020mcq,Bennett:2020zkv}:
\begin{equation}
    x = m_e / \Tcm \, , \qquad y = p / \Tcm \, , \qquad z = T_\gamma / \Tcm \, .
\end{equation}
By definition, the dimensionless photon temperature $z$ equals 1 at high temperatures and increases beyond 1 due to the annihilation of electrons and positrons into photons. For convenience in the QKEs, we also define the dimensionless energy density and pressure as $\tilde{\rho} \equiv (x / m_e)^4 \rho$ and $\tilde{P} \equiv (x / m_e)^4 P$, respectively.

The comoving QKEs for neutrinos and antineutrinos are then written as:
\begin{eqnarray}
	\label{eq:drho_dx}
	\frac{{\rm d}\varrho(x, y)}{{\rm d}x}
	&=&
	\sqrt{\frac{3 m^2_{\rm Pl}}{8\pi \tilde{\rho}_{\rm tot}}}
	\left\{
	-i \frac{x^2}{m_e^3}
	\left[ \tilde{H}_\nu
	,
	\varrho
	\right]
	+\frac{m_e^3}{x^4} \tilde{\mathcal{I}}(\varrho,\bvrho) \right\}\,,\nonumber\\
        \frac{{\rm d}\bvrho(x, y)}{{\rm d}x}
	&=&
	\sqrt{\frac{3 m^2_{\rm Pl}}{8\pi\tilde{\rho}_{\rm tot}}}
	\left\{
	-i \frac{x^2}{m_e^3}
	\left[ \tilde{H}_{\bar\nu}
	,
	\bvrho
	\right]
	+\frac{m_e^3}{x^4} \tilde{\bar{\mathcal{I}}}(\bvrho,\varrho) \right\}\,,
	\nonumber
\end{eqnarray}
with the comoving Hamiltonian terms:
\begin{equation}
    \begin{aligned}
       \tilde{H}_\nu &= U \frac{ \mathbb{M}^2}{2y} U^\dagger 
       + \frac{\sqrt{2}G_{\rm F} m_e^4}{x^4} \, \mathbb{N}_\nu
	-\frac{2\sqrt{2}G_{\rm F} y m_e^6}{x^6}\left(
	\frac{\mathbb{E}_\ell+\mathbb{P}_\ell}{m_W^2}
	+\frac{4}{3}\,\frac{\mathbb{E}_\nu}{m_Z^2}
	\right) \, , \\
        \tilde{H}_{\bar\nu} &= U \frac{ \mathbb{M}^2}{2y} U^\dagger
        - \frac{\sqrt{2}G_{\rm F} m_e^4}{x^4} \, \mathbb{N}_\nu
	-\frac{2\sqrt{2}G_{\rm F} y m_e^6}{x^6}\left(
	\frac{\mathbb{E}_\ell+\mathbb{P}_\ell}{m_W^2}
	+\frac{4}{3}\,\frac{\mathbb{E}_\nu}{m_Z^2}
	\right) \, .
    \end{aligned}
\end{equation}
Additionally, $\tilde{\mathcal{I}}$ and $\tilde{\bar{\mathcal{I}}}$ represent the comoving collision terms for neutrinos and antineutrinos, respectively. Calculating these collision terms is the most complex and time-consuming aspect of solving the QKEs. Fortunately, the nine-dimensional collision integrals can be reduced to two-dimensional integrals by exploiting the homogeneous and isotropic conditions and the specific forms of the scattering amplitudes~\cite{Hannestad:1995rs,Dolgov:1997mb,Grohs:2015tfy,Blaschke:2016xxt}. The explicit forms of the reduced collision terms are provided in Appendix \ref{sec: collision}.

Furthermore, as noted in the discussion following the QKEs (Eq.~\eqref{eq:QKEs}), the chemical potential of electrons and the $\mathbb{N}_l$ term in the Hamiltonian are negligible due to the very low baryon-to-photon ratio $\eta$ and the efficient electromagnetic interactions. Consequently, the Hamiltonian is predominantly influenced by the net number densities of neutrinos when primordial lepton asymmetries are sufficiently large and vary among different flavors. This results in synchronized flavor oscillations of neutrinos and antineutrinos~\cite{Bell:1998ds,Pastor:2001iu,Abazajian:2002qx,Wong:2002fa}. As a result, neutrino asymmetries tend to equilibrate among different flavors before the epoch of BBN~\cite{Froustey:2021azz,Froustey:2024mgf}. 
Since this work focuses on the implications of neutrino asymmetries for BBN, CMB, and LSS, we concentrate on primordial neutrino asymmetries with complete flavor equilibration before neutrino decoupling. Specifically, the initial primordial asymmetries are set to $\xi_{\nu_e} = \xi_{\nu_\mu} = \xi_{\nu_\tau} = \xin$, where $\xi_{\nu_\alpha} \equiv \mu_{\nu_\alpha}/\Tcm$ represents the degeneracy parameter for neutrinos and antineutrinos of flavor $\alpha$. By adopting this initial condition, we bypass the intricate mapping between arbitrary primordial neutrino asymmetries and those after flavor equilibration, leaving the investigation of this mapping for future work. Recent studies on this mapping and the implications of distinct neutrino asymmetries among flavors for BBN can be found in~\cite{Froustey:2024mgf}.

On the other hand, it is well known that the annihilation of electrons and positrons shortly after neutrino decoupling partially heats neutrinos, leading to a slightly increased $\Neff$. To accurately determine $\Neff$, finite-temperature QED (FTQED) corrections must be taken into account~\cite{Fornengo:1997wa,Mangano:2001iu,Bennett:2019ewm}. In this work, we address three aspects of FTQED corrections: (1) corrections to the total energy and pressure densities, which primarily affect the Hubble rate; (2) corrections to the electron mass, which mainly influence the collision terms in the QKEs; and (3) corrections to the continuity equation of the Universe, typically expressed as a differential equation for the dimensionless photon temperature $z$,
\begin{equation}
    \frac{\mathrm{d}z}{\mathrm{d}x}=
    \frac{
    r J_2(r)
    + G_1(r)
    - \frac{1}{4\pi^2 z^3}
        \int_0^\infty \mathrm{d}y\, y^3 \sum_{\alpha} \left(\frac{\mathrm{d}\varrho_{\alpha \alpha}}{\mathrm{d}x} + \frac{\mathrm{d}\bvrho_{\alpha \alpha}}{\mathrm{d}x}\right)
    }{
    \left[
    r^2 J_2(r)
    + J_4(r)
    \right]
    + G_2(r)
    + \frac{2\pi^2}{15}
    }\,,
\end{equation}
where $r = x/z$, and $\frac{\mathrm{d}\varrho_{\alpha \alpha}}{\mathrm{d}x}$ and $\frac{\mathrm{d}\bvrho_{\alpha \alpha}}{\mathrm{d}x}$ are the diagonal components of the QKEs for neutrinos and antineutrinos. The explicit forms of the remaining $J_i(r)$ and $G_i(r)$ functions in the equation, as well as other FTQED corrections, can be found in Appendix \ref{sec: FTQED}. In practice, this equation should be solved in conjunction with the QKEs, and we consider corrections up to order $\mathcal{O}(e^3)$ while neglecting logarithmic terms in the FTQED corrections, as their effects are minimal~\cite{Bennett:2019ewm,Bennett:2020zkv}. Further details are provided in Appendix \ref{sec: FTQED}. Other FTQED corrections, such as those to the interaction vertices, are omitted in this work.

Furthermore, we restrict our analysis to the normal mass hierarchy of neutrinos and neglect any possible CP-violating phase in the transformation matrix between the neutrino flavor and mass bases. Based on the most recent values from the Particle Data Group~\cite{pdg}, we adopt the following neutrino mass differences and mixing angles: $\Delta m_{21}^2 = 7.53 \times 10^{-5} \, \text{eV}^2$, $\Delta m_{31}^2 = 2.53 \times 10^{-3} \, \text{eV}^2$, $\sin^2 \theta_{12} = 0.307$, $\sin^2 \theta_{23} = 0.545$, and $\sin^2 \theta_{13} = 0.0218$. The results for neutrino decoupling and BBN are independent of the minimal neutrino mass, provided $m_1 \ll 0.1 \,\text{MeV}$. However, the resulting CMB anisotropies and LSS of the Universe are sensitive to the choice of $m_1$, which will be discussed in more detail in Sections \ref{sec:cmb} and \ref{sec:mcmc}.



\subsection{Numerical Methods}

To numerically solve the QKEs for neutrinos and antineutrinos, we have extended the public code {\tt FortEPiaNO}~\cite{Gariazzo:2019gyi,Bennett:2020zkv}, originally developed for solving QKEs in the Standard Model, to account for primordial neutrino asymmetries. This subsection outlines the primary numerical methods employed in the code.

For the momentum grid, the comoving momentum $y$ is discretized on a linear grid with $y_{\rm min} = 0.01$, $y_{\rm max} = 20$, and a total of $N_y = 40$ points. The integrals are evaluated using the first-order Newton-Cotes (NC) formula. The initial conditions are set at an initial reduced scale factor of $x = 0.005$, corresponding to a comoving temperature of $T_{\rm CM, in} \simeq 102.2~\text{MeV}$. The initial common temperature for all species is calculated by conserving total entropy from a higher temperature, resulting in $z_{\rm in} = 2.81 \times 10^{-6}$. The initial density matrices are specified as:
\begin{equation}
\label{eq:initial_condition}
\vrho(x_{\rm in},y) = \diag \{f_\nu^{({\rm in})}(y), f_\nu^{({\rm in})}(y), f_\nu^{({\rm in})}(y)\}, \quad \bvrho(x_{\rm in},y) = \diag \{f_{\bar\nu}^{({\rm in})}(y), f_{\bar\nu}^{({\rm in})}(y), f_{\bar\nu}^{({\rm in})}(y)\},
\end{equation}
where
\begin{equation}
f_\nu^{({\rm in})}(y) \equiv \frac{1}{e^{(y - \xi_\nu)/z_{\rm in}} + 1}, \quad f_{\bar\nu}^{({\rm in})}(y) \equiv \frac{1}{e^{(y + \xi_\nu)/z_{\rm in}} + 1}.
\end{equation}

The QKEs are solved in conjunction with the energy conservation equation for $z$ using the {\tt DLSODA} solver from the {\tt ODEPACK} \cite{hindmarsh1982odepack} {\tt Fortran} library. Absolute and relative error tolerances are set to $10^{-7}$, ensuring numerical errors remain below $10^{-5}$.

\subsection{Numerical Results of Neutrino Decoupling}

In this subsection, we present the numerical results for neutrino decoupling. First, we focus on the final energy density of neutrinos, typically parameterized by the effective number of relativistic species, $\Neff$, defined as 
\begin{equation}
    \Neff \equiv \frac{7}{8} \left(\frac{11}{4}\right)^{4/3} \frac{\rho_{\nu}}{\rho_\gamma} \,, \quad \rho_\nu = \sum_\alpha \int \frac{d^3 p}{(2 \pi)^3} \, p \, (\vrho_{\nu_\alpha} + \bvrho_{\nu_\alpha}),
\end{equation}
where $\rho_\nu$ is the total neutrino energy density, obtained by solving the QKEs, and $\rho_\gamma$ is the photon energy density. For the SM in the instantaneous decoupling limit, $\Neff$ equals the number of neutrino flavors. However, since the temperature of neutrino decoupling is near that of electron-positron annihilation, neutrinos are also heated through weak interactions with electrons and positrons, leading to a slightly increased $\Neff$. The most precise calculation for $\Neff$ in the SM, including neutrino oscillations, matter effects, FTQED corrections up to $\mathcal{O}(e^3)$, and neutrino-electron and neutrino-neutrino collision terms, gives $N_{\rm eff}^{\rm SM} = 3.0440 \pm 0.0002$~\cite{Froustey:2020mcq,Bennett:2020zkv}.

\begin{figure}\centering
\includegraphics[width=0.49\textwidth]{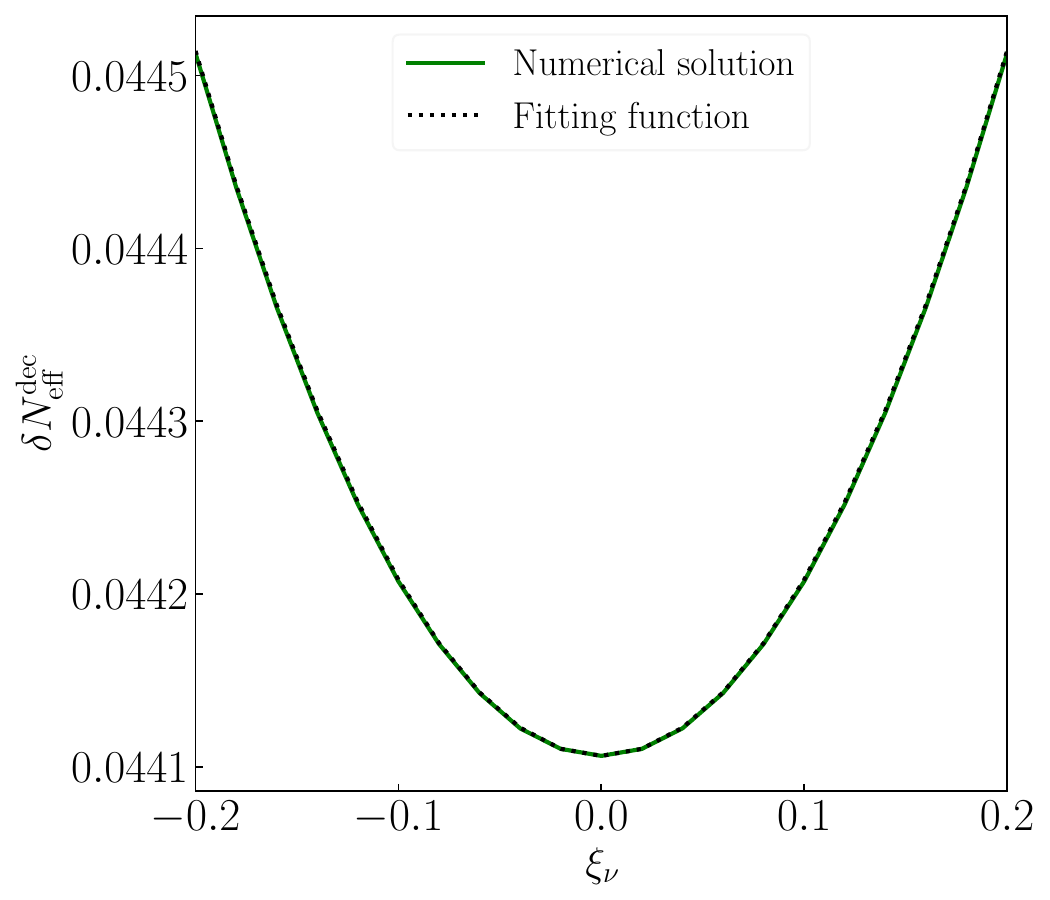} 
\includegraphics[width=0.49\textwidth]{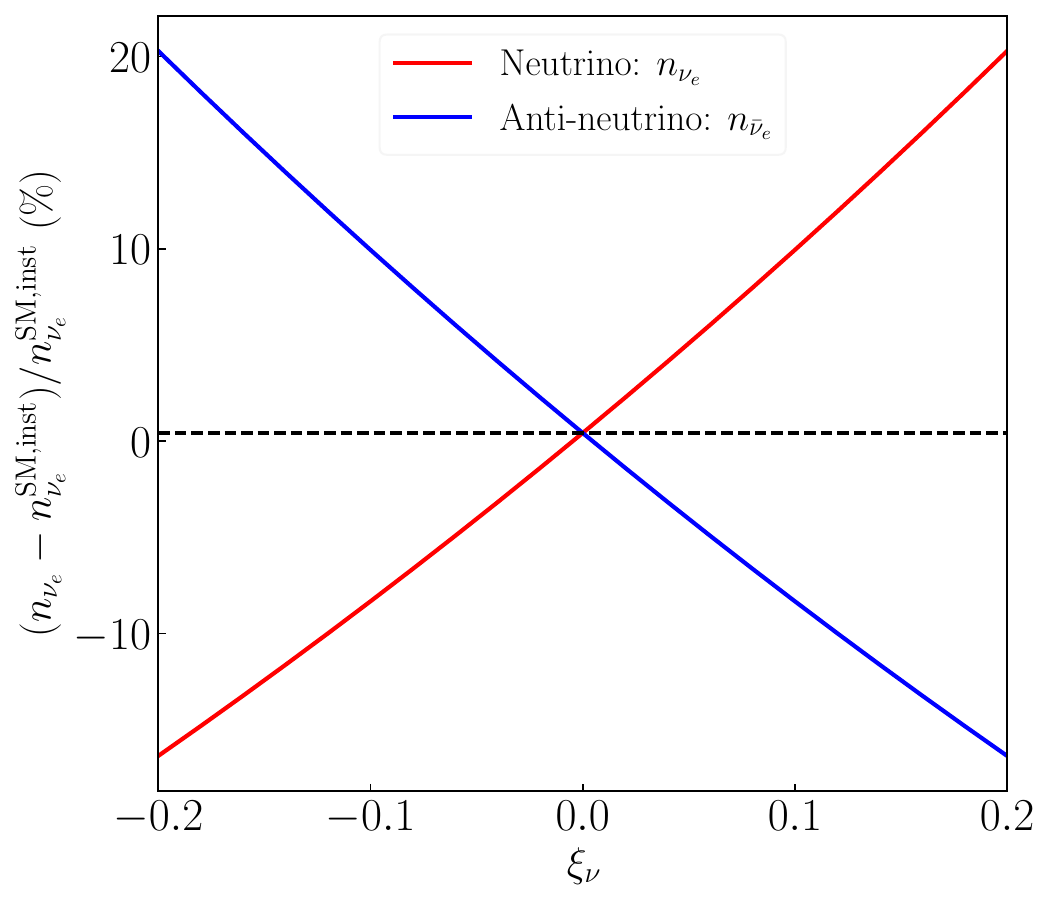} 
\caption{
\textbf{Left panel:} Neutrino non-instantaneous decoupling correction $\delta N_{\rm eff}^{\rm dec} = N_{\rm eff} - N_{\rm eff}^{\rm inst}$ as a function of the neutrino degeneracy parameter $\xi_\nu$. Solid and dashed curves represent the numerical solution and the fitting function of Eq.~\eqref{eq:neff}, respectively.
\textbf{Right panel:} Relative differences in the number densities of electron neutrinos and antineutrinos compared to SM predictions in the instantaneous decoupling limit, $n_{\nu_e, \bar{\nu}_e} \simeq 56 \, \text{cm}^{-3}$.
}
\label{fg:neff}
\end{figure}

In the presence of non-zero primordial neutrino asymmetries, $\Neff$ receives additional contributions from the neutrino chemical potentials. In the instantaneous decoupling limit, the comoving distribution of neutrinos and antineutrinos of flavor $\alpha$ is approximated as $f_{\nu_\alpha, \bar\nu_\alpha} = \frac{1}{e^{y \, \pm \, \xi_{\nu_\alpha}} + 1}$. Therefore, the total $\Neff$ can be estimated by
\begin{equation}\label{eq:neff}
    \Neff^{\rm inst} \simeq 3 + \sum_\alpha \left( \frac{30}{7 \pi^2} \xi_{\nu_\alpha}^2 + \frac{15}{7 \pi^4} \xi_{\nu_\alpha}^4 \right) = 3 + 3 \left( \frac{30}{7 \pi^2} \xi_\nu^2 + \frac{15}{7 \pi^4} \xi_\nu^4 \right),
\end{equation}
where the second equality uses the condition $\xi_{\nu_e} = \xi_{\nu_\mu} = \xi_{\nu_\tau} = \xi_\nu$.
Beyond the instantaneous decoupling limit, we explore the non-instantaneous decoupling corrections, $\delta N_{\rm eff}^{\rm dec} = N_{\rm eff} - N_{\rm eff}^{\rm inst}$, in the presence of the primordial neutrino asymmetry $\xi_\nu$. 
In the left panel of Fig.~\ref{fg:neff}, we show $\delta N_{\rm eff}^{\rm dec}$ as a function of $\xi_\nu$. 
For the SM case with $\xi_\nu = 0$, our calculation yields $\delta N_{\rm eff}^{\rm dec, SM} = 0.0441$, consistent with the literature result, $\delta N_{\rm eff}^{\rm dec, SM} = 0.0440 \pm 0.0002$.\footnote{The $\sim 0.0001$ difference between our result and the central value in previous literature arises mainly from different values of $N_y$ used for discretizing the neutrino momentum grid. Our result agrees with previous results using $N_y = 40$; see Table 2 in \cite{Bennett:2020zkv}.}
For non-zero $\abs{\xi_\nu}$, $\delta N_{\rm eff}^{\rm dec}$ increases beyond $\delta N_{\rm eff}^{\rm dec, SM}$, as expected since a higher $\abs{\xi_\nu}$ results in more energy transfer to neutrinos from electrons and positrons.
We find that $\delta N_{\rm eff}^{\rm dec}$ as a function of $\xi_\nu$ can be well fitted by $\delta N_{\rm eff}^{\rm dec} = \delta N_{\rm eff}^{\rm dec, SM}+0.0102 \, \xi_\nu^2$, as shown in the left panel of Fig.~\ref{fg:neff}.
In conclusion, the final $\Neff$ after neutrino decoupling, in the presence of primordial neutrino asymmetry $\xi_\nu$, is given by
\begin{equation}
    \Neff (\xi_\nu) = N_{\rm eff}^{\rm SM} + 3 \left( \frac{30}{7 \pi^2} \xi_\nu^2 + \frac{15}{7 \pi^4} \xi_\nu^4 \right) + 0.0102 \, \xi_\nu^2,
\end{equation}
where $ N_{\rm eff}^{\rm dec, SM} = 3.0440 \pm 0.0002$.

It is also interesting to discuss the resulting neutrino number densities in the presence of primordial neutrino asymmetries, as these are critical physical quantities for the Cosmic Neutrino Background (C$\nu$B). In particular, experiments like PTOLEMY aim to detect the C$\nu$B via neutrino absorption on tritium: $\nu_e + \rm ^3H \rightarrow ^3He + e^-$~\cite{PTOLEMY:2018jst,Betti:2018bjv,PTOLEMY:2019hkd}, while the key signature of C$\nu$B capture is a peak in the electron spectrum at an energy of $2 m_\nu$ above the beta decay endpoint~\cite{Long:2014zva,Akita:2020jbo}. {  Clearly, a potential measurement of C$\nu$B depends critically on the present electron neutrino number densities. Therefore, the C$\nu$B capture rates would also change in the presence of neutrino asymmetries due to the changes in neutrino number densities. 
In addition, the energy splitting of the electron spin states would also be altered in the presence of neutrino asymmetries, known as the Stodolsky effect\cite{Bauer:2022lri,Duda:2001hd,Domcke:2017aqj}. 
Here we focus on the effect of neutrino asymmetries on neutrino number densities, and note that this is different from the the Stodolsky effect, which vanishes in the absence of neutrino asymmetries.}
In the right panel of Fig.~\ref{fg:neff}, we show the relative difference between the electron neutrino and antineutrino number densities, computed from the QKEs, and the Standard Model predictions under the instantaneous decoupling limit, $n_{\nu_e,\bar{\nu}_e} \simeq 56 \ \rm cm^{-3}$. For the SM case with $\xi_\nu = 0$, the relative difference is approximately $0.43\%$, arising from non-instantaneous decoupling effects. However, for non-zero $\xi_\nu$, the effects of primordial neutrino asymmetries become dominant. As $\xi_\nu$ increases, the electron neutrino number density increases, while the antineutrino number density decreases.
These effects have two significant implications for estimating the PTOLEMY capture rate in the presence of neutrino chemical potentials. First, an enhanced (or suppressed) electron neutrino number density leads to an increased (or decreased) capture rate for positive (or negative) $\xi_\nu$, for both Dirac and Majorana neutrinos. Second, since Majorana antineutrinos are identical to neutrinos, the capture rate for Majorana neutrinos is also affected by variations in the electron antineutrino number density as $\xi_\nu$ changes. This implies that the relative change in the capture rate due to primordial neutrino asymmetries is smaller for Majorana neutrinos than for Dirac neutrinos.

\begin{figure}
\centering
\includegraphics[width=0.49\textwidth]{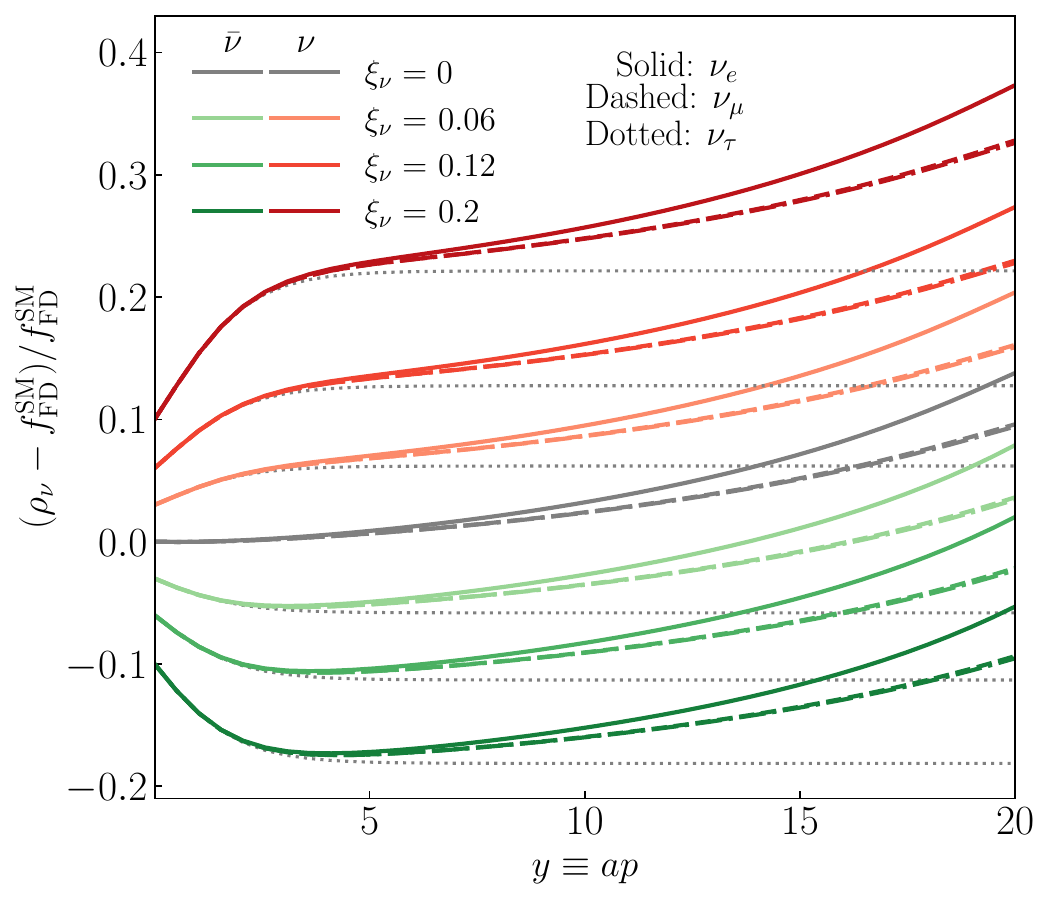}
\includegraphics[width=0.49\textwidth]{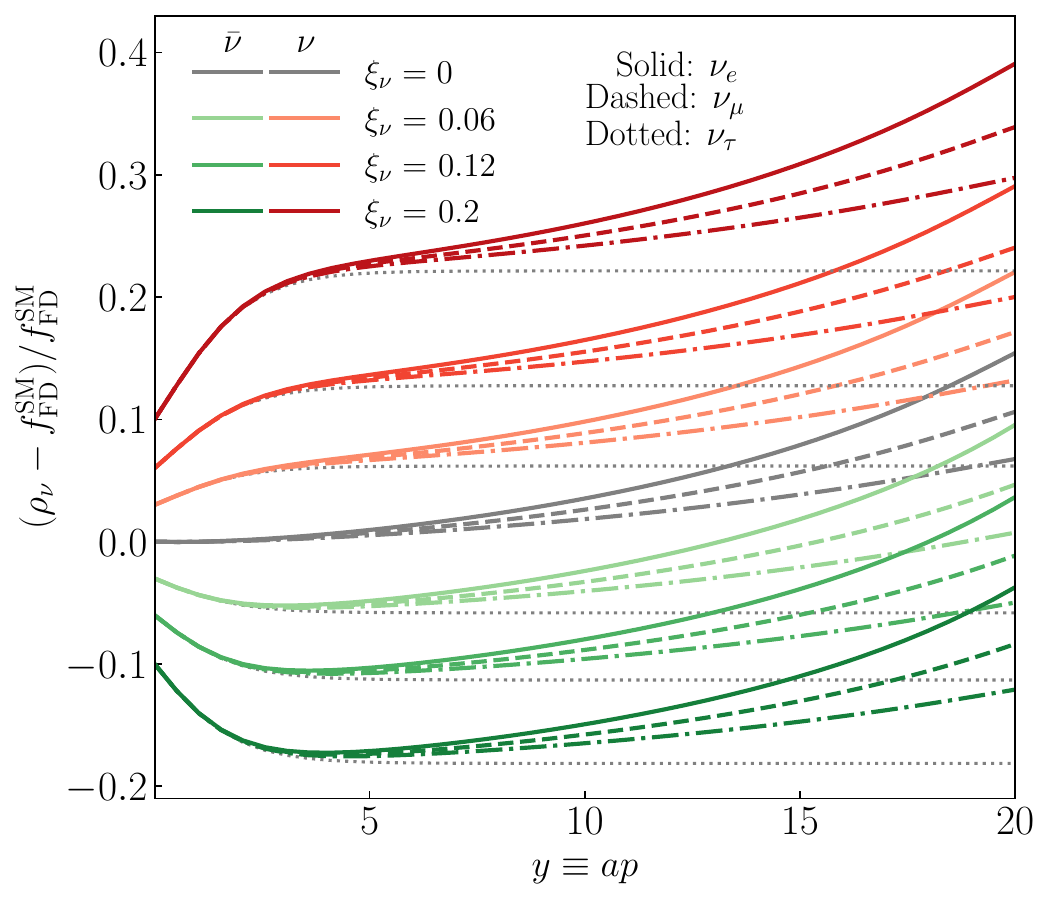} 
\caption{
Density matrices of neutrinos and antineutrinos in the flavor basis (left panel) and mass basis (right panel), normalized to the Fermi-Dirac distribution with $\xi_\nu = 0$. The gray curves represent the density matrices for neutrinos and antineutrinos with $\xi_\nu = 0$, while the red and green curves represent those for neutrinos and antineutrinos with non-zero $\xi_\nu$, respectively. For each case, the different flavor eigenstates ($\nu_e$, $\nu_\mu$, and $\nu_\tau$) and mass eigenstates ($\nu_1$, $\nu_2$, and $\nu_3$) are indicated by solid, dashed, and dash-dotted curves, respectively. Note that the density matrices for $\nu_\mu$ and $\nu_\tau$ nearly overlap. The Fermi-Dirac distributions with corresponding $\xi_\nu$ are also indicated by the gray dotted curves.
}
\label{fg:drho_th}
\end{figure}

We now present the resulting density matrices for neutrinos and antineutrinos after decoupling. The density matrices as a function of the comoving momentum $y \equiv p/\Tcm$ in both the flavor and mass bases for several values of $\xi_\nu$ are shown in the left and right panels of Fig.~\ref{fg:drho_th}, respectively, and are normalized to the Fermi-Dirac distribution for $\xi_\nu=0$. The density matrices for neutrinos and antineutrinos with $\xi_\nu = 0$ coincide and are represented by the gray curves, reproducing results from previous studies.
For non-zero $\xi_\nu$, the density matrices of neutrinos and antineutrinos no longer coincide and are represented by red and green curves, respectively. In each case, the solid, dashed, and dash-dotted curves represent the three flavor eigenstates ($\nu_e$, $\nu_\mu$, and $\nu_\tau$) and the three mass eigenstates ($\nu_1$, $\nu_2$, and $\nu_3$), respectively. The corresponding thermal equilibrium distributions for neutrinos and antineutrinos, $f_{\rm FD}^{\nu, \bar\nu} = 1/(e^{y \pm \xi_\nu} + 1)$, are also shown by the dotted curves.
The density distributions of neutrinos and antineutrinos for a given $\xi_\nu$ coincide with the Fermi-Dirac distributions at small $y$, but deviate at larger $y$. This occurs because the weak interaction rate between neutrinos and electrons/positrons is proportional to the neutrino momentum, causing neutrinos with smaller momenta to decouple earlier and maintain the thermal Fermi-Dirac distribution, while neutrinos with larger momenta decouple later, leading to greater spectral distortions relative to the Fermi-Dirac distribution.
Regarding different flavors, electron neutrinos exhibit larger spectral distortions due to additional charged-current weak interactions, while the spectral distortions of muon and tau neutrinos nearly coincide, as they experience the same weak interactions. In contrast, the spectral distortions of the three neutrino mass eigenstates are uniformly separated, as shown in the right panel.

\begin{figure}
\centering
\includegraphics[width=0.49\textwidth]{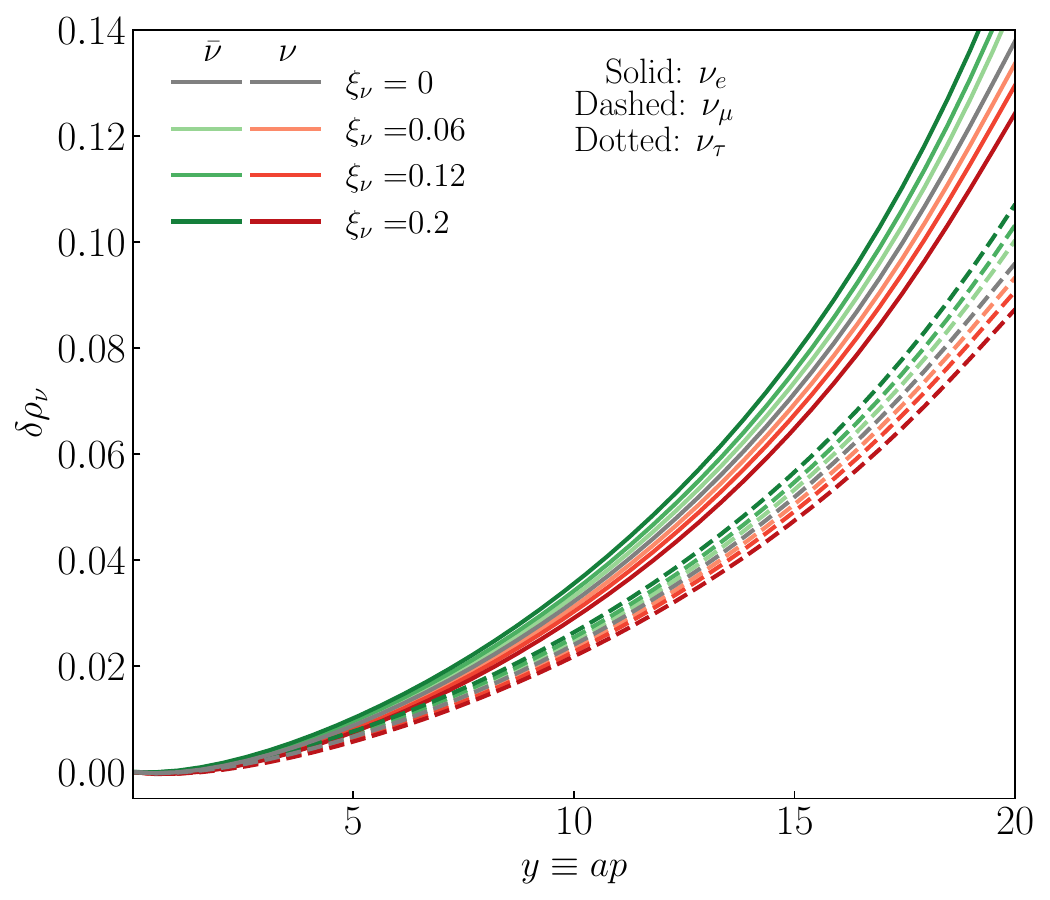} 
\includegraphics[width=0.49\textwidth]{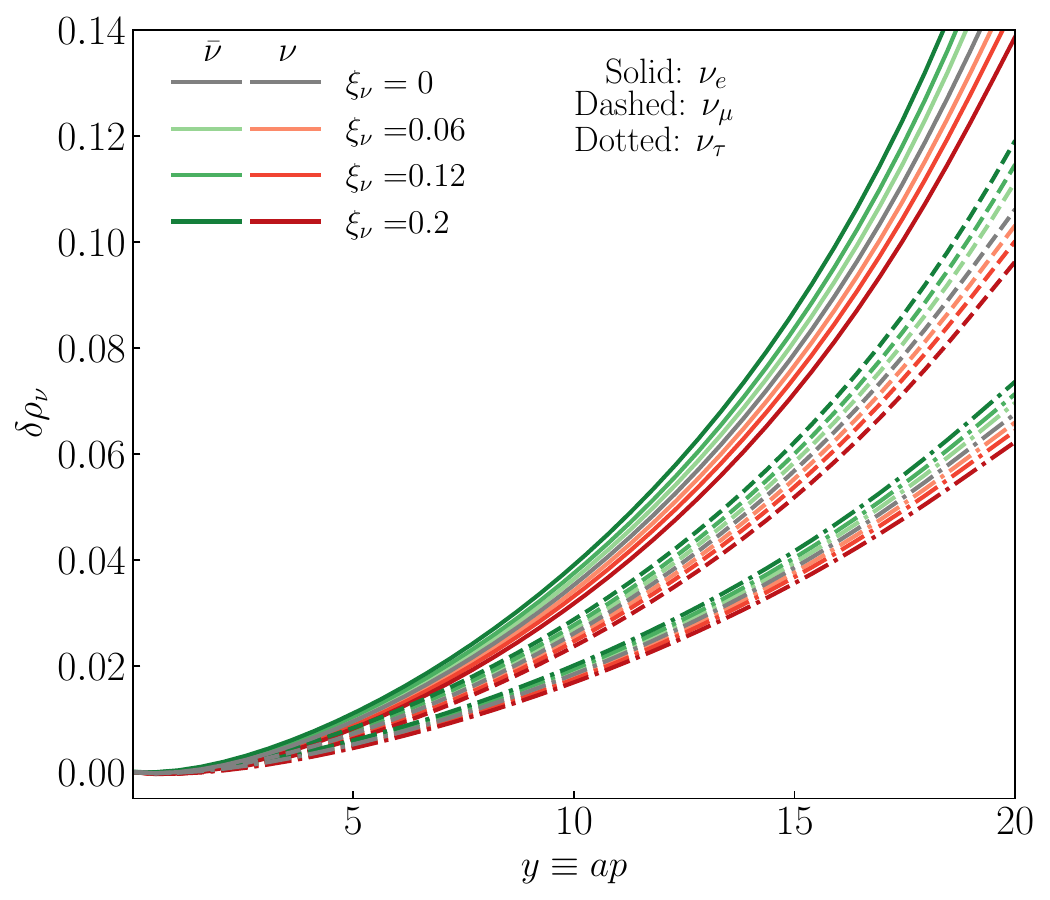}
\caption{
Spectral distortions of neutrinos and antineutrinos in the flavor basis (left panel) and mass basis (right panel), relative to the Fermi-Dirac distribution with the corresponding $\xi_\nu$. The curves follow the same convention as in Fig.~\ref{fg:drho_th}, but the spectral distortions of tau neutrinos in the left panel are omitted for brevity. 
}
\label{fg:drho_norm}
\end{figure}

Since the density distributions of neutrinos and antineutrinos coincide with the thermal Fermi-Dirac distributions at small momenta, it is natural to define the spectral distortions as
\begin{equation}
    \delta\vrho_{\nu,\bar\nu} \equiv \frac{\delta\vrho_{\nu,\bar\nu}}{f_{FD}(\pm \xi)} - 1 \, ,
\end{equation}
where the plus sign corresponds to neutrinos and the minus sign to antineutrinos.
The resulting spectral distortions as a function of momentum $y$ in the flavor and mass bases are shown in the left and right panels of Fig.~\ref{fg:drho_norm}, respectively. The curves follow the same convention as in Fig.~\ref{fg:drho_th}, with tau neutrino spectral distortions omitted in the left panels, as they nearly coincide with those of muon neutrinos.
Notably, the spectral distortions for neutrinos and antineutrinos with non-zero $\xi_\nu$ differ and lie on opposite sides of the spectral distortions for $\xi_\nu = 0$. Antineutrinos with non-zero $\xi_\nu$ exhibit larger spectral distortions compared to those for $\xi_\nu = 0$, while neutrinos display smaller distortions. Moreover, the larger the $\xi_\nu$, the greater the discrepancy between the spectral distortions of neutrinos and antineutrinos. These properties hold in both the flavor and mass bases and arise from the differing Pauli-blocking effects for neutrinos and antineutrinos in the presence of primordial neutrino asymmetries.
Finally, the above conclusions regarding neutrino density matrices also apply to negative $\xi_\nu$, with the density matrices for neutrinos and antineutrinos swapped compared to those for positive $\xi_\nu$. In other words, the density distributions and spectral distortions of antineutrinos in Figs.~\ref{fg:drho_th} and \ref{fg:drho_norm} also apply to neutrinos with opposite $\xi_\nu$.

In summary, we have presented the most accurate results of neutrino decoupling in the presence of primordial neutrino asymmetries and discussed the corresponding $\Neff$ and spectral distortions of neutrinos and antineutrinos. In the following sections, we will explore the effects of primordial neutrino asymmetries on BBN, CMB, and structure formation, focusing on the corrections induced by the spectral distortions of neutrinos and antineutrinos.

\section{Impacts of the Primordial Neutrino Asymmetry on BBN}\label{sec:bbn}

Soon after neutrino decoupling, primordial nucleosynthesis, or Big Bang Nucleosynthesis (BBN), occurs, leading to the production of light elements as protons and neutrons decouple from the Standard Model (SM) thermal bath. The decoupling of protons is influenced by electron neutrinos, causing the resulting abundances of light elements after BBN to depend on the properties of both electron neutrinos and antineutrinos. BBN is impacted by primordial neutrino asymmetries in three ways (for reviews, see, \eg \cite{Pitrou:2018cgg}):
\begin{enumerate}
    \item The primary effect is the alteration of the weak interaction rate at the onset of BBN, which is influenced by the chemical potential of electron neutrinos. The conversion between protons and neutrons is mediated by the weak interaction, leading to a difference in the chemical potentials of protons and neutrons determined by the chemical potential of electron neutrinos, i.e., $\mu_n - \mu_p = \mu_e - \mu_{\nu_e} \simeq - \mu_{\nu_e}$. Consequently, the resulting neutron-to-proton ratio and subsequent BBN processes are modified according to the absolute value and sign of the electron neutrino chemical potential.
    \item Additionally, the expansion rate of the universe during BBN is affected by the increased neutrino energy density contributed by the chemical potentials. As discussed in the previous section, the neutrino energy density following neutrino decoupling is typically represented by $N_{\rm eff}$, which can be expressed as $N_{\rm eff} \simeq 3.044 + 3 \left( \frac{30}{7 \pi^2} \xi_{\nu}^2 + \frac{15}{7 \pi^4} \xi_{\nu}^4 \right)$. Thus, the corresponding corrections in BBN depend on the absolute asymmetries of all three neutrino flavors.
    \item Beyond the approximation of thermal Fermi-Dirac distributions for neutrinos and antineutrinos, spectral distortions arising from the non-instantaneous decoupling of neutrinos will also induce corrections to the weak interaction rate, thereby affecting BBN. Such corrections have often been overlooked in previous studies, particularly in the context of primordial neutrino asymmetries.
\end{enumerate}

In the following sections, we will first review the general treatment of the weak interaction rate and introduce the new corrections induced by neutrino spectral distortions, after which we will discuss the resulting light element abundances from BBN.

\subsection{The Weak Correction from Neutrino Spectral Distortions}

At the onset of BBN, the conversion between neutron ($n$) and proton ($p$) is dominated by the following interaction reactions:
\begin{align}
n+\nu_e &\leftrightarrow p + e^- ,\label{eq:Weak1}\\
n &\leftrightarrow p + e^-+\bar \nu_e ,\label{eq:Weak2}\\
n+e^+ &\leftrightarrow p + \bar \nu_e .\label{eq:Weak3}
\end{align}
In the remainder of this section, we will omit the subscript for the electron neutrino and the superscript for the electron/positron whenever there is no ambiguity. 
Due to the strong coupling between photons, electrons, and positrons, it is appropriate to assume that all particles except for neutrinos are in local thermal equilibrium. Consequently, the number densities of neutrons and protons obey the Boltzmann equations:
\begin{align}
\dot n_n + 3 H n_n &= -n_n \Gamma_{n\to p} + n_p\Gamma_{p\to n},\\
\dot n_p + 3 H n_p &= -n_p\Gamma_{p\to n} + n_n \Gamma_{n\to p}.
\end{align}
Here, the reaction rates $n_n \Gamma_{n\to p}$ and $n_p \Gamma_{p\to n}$ incorporate all relevant collision terms:
\beqn
\Gamma_{n \to p} &=& \Gamma_{n+\nu \to p + e} + \Gamma_{n\to p + e+\bar \nu}+\Gamma_{n+e \to p + \bar \nu}, \\
\Gamma_{p \to n} &=& \Gamma_{p + e\to n+\nu} + \Gamma_{p + e+\bar \nu \to n}+\Gamma_{p + \bar \nu \to n+e}.
\eeqn
The reaction rate for a specific interaction involving nucleon $a$ in the initial state and nucleon $b$ in the final state is given by:
\beqn
n_a \Gamma_{a \to b} = & \int \dd \pi_a \dd \pi_b \dd \pi_e \dd \pi_\nu \delta^{(4)} \left({p}_a-p_b + \coeffa_e p_e + \coeffa_\nu p_{\nu}\right) \left| M\right|^2_{a \to b} \nn  \\
& \times f_a(E_a) f_b(-E_b) f_e(\coeffa_e {E}_e) \tilde f_\nu (E_\nu) ,
\eeqn
where $\dd \pi_i = \frac{1}{(2\pi)^3}\frac{\dd^3 \mathbf{p}_{i}}{2E_{i}}$, $\alpha_e (\alpha_\nu) = 1$ if the electron (neutrino) is in the initial state and $\alpha_e (\alpha_\nu) = -1$ if it is in the final state. Additionally, $f_i(E) \equiv \frac{1}{e^{E/T_{\gamma}}+1}$ for $i=a, b, e$ is the Fermi-Dirac distribution with a vanishing chemical potential, and we have used the relation $1-f_i(E) = f_i(-E)$ to account for the corresponding Pauli blocking factors. Lastly, $\tilde f_\nu (E_\nu)$ equals $\varrho_\nu (E_\nu)$ (or $\varrho_{\bar\nu} (E_\nu)$) for the neutrino (or antineutrino) in the initial state, and equals $1-\varrho_\nu (E_\nu)$ (or $1-\varrho_{\bar\nu} (E_\nu)$) for those in the final state.

To leading order, two additional approximations are typically made. First, we neglect neutrino spectral distortions and use the Fermi-Dirac distribution with the corresponding chemical potentials $f_\nu^{\pm} (E) \equiv \frac{1}{e^{E/T_{\nu} \mp \xi_\nu}+1}$ for neutrinos and antineutrinos, where $T_{\nu}$ is the effective temperature of the neutrino, defined as the temperature that satisfies $3 \frac{7}{8} \frac{\pi^2}{15} T_\nu^4 = \rho_\nu$. Furthermore, since the temperature during the BBN epoch (approximately $0.1 \, {\rm MeV}$) is much smaller than the masses of the neutron and proton (approximately $1 \, {\rm GeV}$), we can work within the infinite nucleon mass limit, also known as the Born approximation. Under these two approximations and following the notation of \cite{Pitrou:2018cgg}, the reaction rates are significantly simplified as follows:
\begin{align}
\bar {\Gamma}_{n \to p} &= K \int_0^\infty p^2 \dd p [\chi_+(E) + \chi_+(-E)]\,, \label{eq:gamma_n2p}\\
\bar {\Gamma}_{p \to n} &= K \int_0^\infty p^2 \dd p [\chi_-(E) + \chi_-(-E)]\,\label{eq:gamma_p2n},
\end{align}
where $E = \sqrt{p^2+m_e^2}$ represents the energy of the electron and $K = 1/(\tau_n \lambda_0 m_e^5)$, with $\tau_n$ being the lifetime of the neutron and $\lambda_0$ the Born coefficient. The $\chi^\pm$ functions are defined as 
\beq\label{eq:chipm}
\chi_\pm(E) \equiv (E_\nu^\mp)^2 f_e(-E) f_\nu^\pm (E^\mp_\nu) \,, \quad E^\mp_\nu \equiv E \mp \Delta,
\eeq
where $\Delta \equiv m_n - m_p$ is the energy gap between the neutron and proton. The first term in \eqref{eq:gamma_n2p} corresponds to the $n \to p + e$ processes \eqref{eq:Weak1} or \eqref{eq:Weak2}, depending on whether $E - \Delta$ is positive or negative, while the second term corresponds to the $n + e \to p$ process \eqref{eq:Weak3}. Similarly, the terms in \eqref{eq:gamma_p2n} can be interpreted. Here, $E^\mp_\nu$ is equal to the energy of the neutrino $E_\nu$ when the neutrino is in the initial state, and $-E_\nu$ when it is in the final state. The term $f_\nu^\pm (E^\mp_\nu)$ in the function $\chi^\pm$ provides the corresponding distribution or Pauli-blocking factor, following the relation $1 - f_\nu^+(E_\nu) = f_\nu^-(-E_\nu)$.

Now, we introduce the corrections induced by the spectral distortions of neutrinos and antineutrinos. In the infinite nucleon mass limit, the corrections to the Born rates can be expressed as follows:
\begin{align}
    \delta \bar {\Gamma}_{n \to p}^{\rm SD} &= K \int_0^\infty p^2 \dd p \left[\delta \chi_+(E) + \delta \chi_+(-E)\right]\,, \label{eq:delta_gamma_n2p}\\
    \delta \bar {\Gamma}_{p \to n}^{\rm SD} &= K \int_0^\infty p^2 \dd p \left[\delta \chi_-(E) + \delta \chi_-(-E)\right]\,\label{eq:delta_gamma_p2n},
\end{align}
where
\begin{align}\label{eq:delta_chipm}
    \delta \chi_\pm(E) \equiv & (E_\nu^\mp)^2 f_e(-E) \, {\rm sgn}(E^\mp_\nu) \nonumber \\
&\times \left( \frac{1+\Theta(\pm E^\mp_\nu) \delta\varrho_\nu + \Theta(\mp E^\mp_\nu) \delta\varrho_{\bar\nu}}{e^{\abs{E^\mp_\nu}/T_{\rm CM} \mp {\rm sgn}(E^\mp_\nu) \xi_\nu}+1} - \frac{1}{e^{\abs{E^\mp_\nu}/T_{\nu} \mp {\rm sgn}(E^\mp_\nu) \xi_\nu}+1} \right)\,, 
\end{align}
where $\Theta(x)$ is the Heaviside step function, and $\delta \varrho_\nu$ and $\delta\varrho_{\bar\nu}$ represent the spectral distortions of electron neutrinos and antineutrinos, respectively. For a given process, the various sgn and $\Theta$ functions together characterize the nature and state of the neutrino particle, indicating whether it is a neutrino or an antineutrino and whether it is in the initial or final state. Additionally, we note that $T_{\rm CM}$ and $T_\nu$ exhibit slight differences after neutrino decoupling. If we ignore this difference and set $\xi_\nu = 0$, resulting in $\delta \varrho_\nu = \delta \varrho_{\bar\nu}$, then \eq\eqref{eq:delta_chipm} coincides with \eq(29) in \cite{Froustey:2019owm}, where corrections to the weak rates from neutrino spectral distortions in the SM were considered.

Finally, we provide a brief overview of other corrections to the weak interaction rates considered in this work. First, we address the zero-temperature radiative corrections, which include pure radiative corrections corresponding to the emission and absorption of a virtual photon during the interaction, as well as bremsstrahlung corrections arising from the emission of a real photon. Second, we consider finite-temperature radiative corrections related to the finite temperature effects on proton-electron interactions and electron energy. Third and fourth, we include corrections due to the finite nucleon mass and weak magnetism. The total weak reaction rates, combining all these corrections, are expressed as
\beq\label{eq: n2ptot}
\Gamma_{n \to p} = \Gamma^{\rm RC0}_{n \to p}+ \delta\Gamma^{\rm SD, RC0}_{n \to p} + \Gamma^{T}_{n \to p}+\delta\Gamma^{\rm BS}_{n \to p}+\delta\Gamma^{\rm FM}_{n \to p}\,,
\eeq
where the superscripts 'RC0', 'SD', 'T', 'BS', and 'FM' denote the zero-temperature radiative correction, spectral distortion, finite-temperature radiative correction, bremsstrahlung, and finite nucleon mass correction, respectively. The form for the $p\to n$ rates is analogous. 

It is important to note that the effects of different corrections may be coupled. For instance, the impact of zero-temperature radiative corrections on those induced by spectral distortions is also addressed in this work. Specifically, the Born rates with zero-temperature radiative correction are given by
\begin{align}\label{eq:CCRn}
    {\Gamma}^{{\rm RC} 0}_{n \to p} &= K \int_0^\infty p^2 \dd p\left[\mathcal{F}_+(E)R(E,|\Delta-E|)\chi_+(E) + \mathcal{F}_+(-E)R(E,\Delta+E)\chi_+(-E)\right]\,, \\
    {\Gamma}^{{\rm RC} 0}_{p \to n} &= K \int_0^\infty p^2 \dd p\left[\mathcal{F}_-(E)R(E, \Delta+E)\chi_-(E) + \mathcal{F}_-(-E)R(E,|E-\Delta|)\chi_-(-E)\right]\,.
\end{align}
The explicit expressions for the Fermi function $F_\pm$ and the radiative correction factor $R(E,k_{\rm max})$ can be found in, e.g., \cite{Pitrou:2018cgg}, and are not repeated here for brevity. It is evident that the zero-temperature radiative correction modifies only the coefficients of the $\chi_\pm$ functions. The terms $\delta\Gamma^{\rm RC0, SD}_{n \to p}$ and $\delta\Gamma^{\rm RC0, SD}_{p \to n}$ can be derived simply by replacing $\chi_\pm$ with $\delta \chi_\pm$. 
In principle, other corrections are also influenced by the spectral distortions of neutrinos and antineutrinos. However, these effects are expected to be negligible, as the corrections themselves are already sub-leading. Thus, we disregard these interferences in this work.

\subsection{Results of BBN Abundances}

In this subsection, we present the resulting light element abundances, incorporating all corrections discussed previously. Specifically, we have implemented the corrections induced by neutrino spectral distortions into the public BBN code {\tt PRIMAT} \cite{Pitrou:2018cgg} and computed the abundances of the light elements. For this computation, we utilized the latest values of relevant physical and cosmological parameters; for instance, the neutron lifetime is taken to be $\tau_n = 879.4 \, \text{s}$, the axial coupling of the nucleons is $g_A = 1.2756$, and the baryon density is $\omega_b = 0.0224$.
For convenience, we define the mass fraction of isotope $i$ as $X_i \equiv A_i \frac{n_i}{n_b}$, where $A_i$ and $n_i$ are the baryon number and number density of isotope $i$, respectively, and $n_b$ is the total baryon number density. We also customarily denote $Y_P \equiv X_{\hef}$ and $i/\rm H \equiv n_i/n_{^1 \rm H}$.

\begin{figure}\centering
\includegraphics[width=0.8\textwidth]{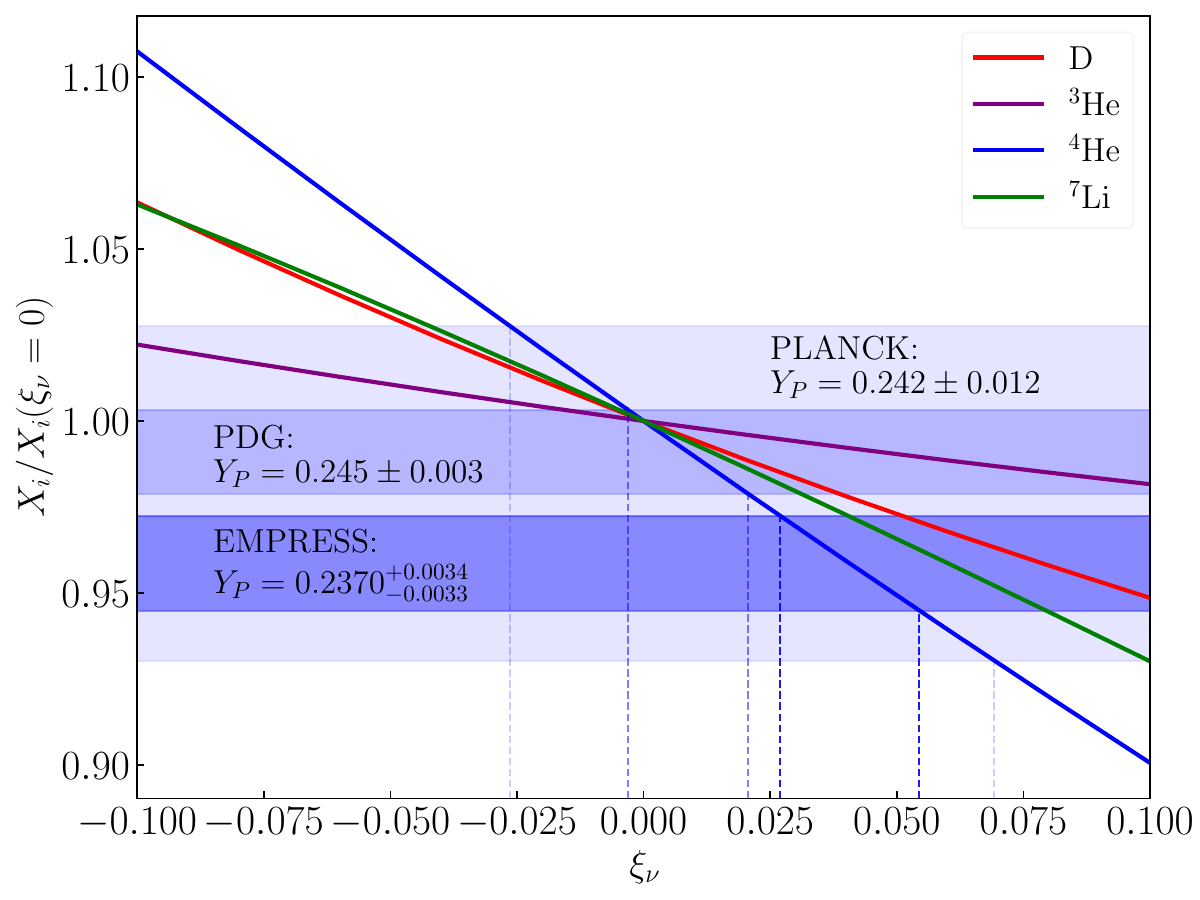}
\caption{
Light element abundances as a function of the neutrino degeneracy parameter $\xi_\nu$, normalized with the results for $\xi_\nu = 0$. The baryon density is fixed at $\omega_b = 0.0224$, and the nuclear rates of the BBN are driven by the PRIMAT approach. The $\het$ notation refers to $\het + \rm T$, and $\lis$ denotes $\lis + \rm^7 Be$, accounting for slow radioactive decays. All optional weak rate corrections in {\tt PRIMAT} (see \cite{Pitrou:2018cgg}) and the new corrections from neutrino spectral distortions are included in the calculations, see text for more details. The $68\%$ constraints on the helium abundance $Y_P$ from the CMB ($Y_P = 0.242\pm0.012$), the PDG recommended value ($Y_P = 0.245 \pm 0.003$) and the recent EMPRESS experiment ($Y_P = 0.2370^{+0.0034}_{-0.0033}$) are indicated by the light and dark blue shaded regions, respectively.}
\label{fg:bbn_xi}
\end{figure}

Fig. \ref{fg:bbn_xi} illustrates the resulting light element abundances, incorporating all weak rate corrections from \eqref{eq: n2ptot} and the QED corrections up to $\mathcal{O}(e^3)$ for plasma thermodynamics and certain nuclear rates, as a function of the neutrino degeneracy parameter $\xi_\nu$. The abundance of $\het$ refers to $\het + \rm T$, while $\lis$ corresponds to $\lis + \rm ^7 Be$, accounting for slow radioactive decays. Since the abundances of different elements can vary by several orders of magnitude, we normalize the element abundances to their respective values at $\xi_\nu = 0$. Consequently, all curves in \fig\ref{fg:bbn_xi} intersect at $\xi_\nu = 0$.
We observe that the abundances of $\rm D, \, \het, \, \hef$, and $\lis$ decrease with increasing $\xi_\nu$ due to the corresponding reduction in the neutron-to-proton number density ratio. Furthermore, the variation in $\hef$ abundance is most sensitive to changes in $\xi_\nu$, while the $\het$ abundance is the least sensitive.
Additionally, \fig\ref{fg:bbn_xi} shows the constraints on helium abundance derived from the CMB observation, the EMPRESS observation {  and the PDG recommended value:
\begin{align}
    & Y_P |_{\rm CMB} = 0.242\pm0.012, \\
    & Y_P |_{\rm EMPRESS} = 0.2370^{+0.0034}_{-0.0033}, \\
    & Y_P |_{\rm PDG} = 0.245 \pm 0.003
\end{align}
as represented by the blue shaded regions. Through our numerical results with the fixed baryon density $\omega_b = 0.0224$, the constraints on $Y_P$ within the $68\%$ confidence interval can be easily translated into constraints on the neutrino degeneracy parameter $\xi_\nu$:
\begin{align}
    {\rm CMB}: \quad -0.026 \leq & \,\xi_\nu \leq 0.069 \, ,\\
    {\rm EMPRESS}: \quad 0.027 \leq & \, \xi_\nu \leq 0.054 \, , \\
    {\rm PDG}: \quad -0.003 \leq & \,\xi_\nu \leq 0.021 \, .
\end{align}
Clearly, while the CMB observation yields a significantly weaker constraint on the helium abundance $Y_P$, the constraint from the EMPRESS observation has similar uncertainties but a much smaller central value with respect to the PDG recommended observation. Consequently, the EMPRESS observation favours a non-zero positive $\xi_\nu$, while the CMB and PDG recommended observations remain compatible with the Standard Model case of $\xi_\nu = 0$. }

\renewcommand{\arraystretch}{1.4}
\begin{table}[t]
\centering
\begin{tabular}{|cccc|}
\hline
\multicolumn{4}{|c|}{\textbf{Key Nuclear Reactions}}     \\ 
\hline
\hline
n+p $\to$ D+$\gamma$ & D+p $\to$ $^{3}$He+$\gamma$ &
D+D $\to$ $^{3}$He+n & D+D $\to$ $^{3}$H+p \\
$^{3}$H+p $\to$ $^{4}$He+$\gamma$ & $^{3}$H+D $\to$ $^{4}$He+n &
$^{3}$H+$^{4}$He $\to$ $^{7}$Li+$\gamma$ & $^{3}$He+n $\to$ $^{3}$H+p \\
$^{3}$He+D $\to$ $^{4}$He+p & $^{3}$He+$^{4}$He $\to$ $^{7}$Be+$\gamma$ &
$^{7}$Be+n $\to$ $^{7}$Li+p & $^{7}$Li+p $\to$ $^{4}$He+$^{4}$He \\
\hline
\end{tabular}
\caption{ The key nuclear reactions that possessing different reaction rates in the PRIMAT and NACRE II approaches. }
\label{tab:nuclear}
\end{table}

\begin{figure}\centering
\includegraphics[width=0.8\textwidth]{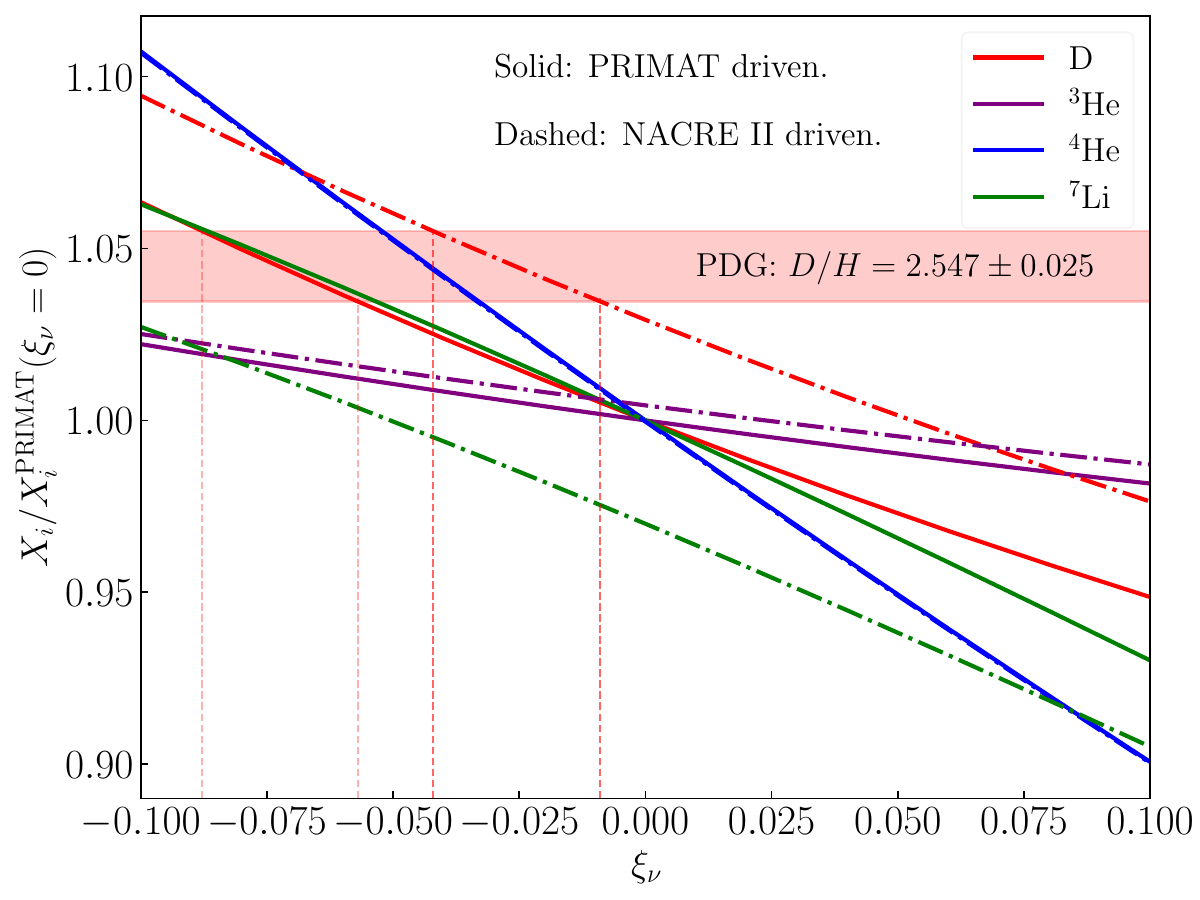}
\caption{
Light element abundances as a function of the neutrino degeneracy parameter $\xi_\nu$ that driven by the PRIMAT approach (solid curves) and the NACRE II approach (dash-dotted curves), normalized with the results for $\xi_\nu = 0$ in the PRIMAT approach. The $\het$ notation refers to $\het + \rm T$, and $\lis$ denotes $\lis + \rm^7 Be$, accounting for slow radioactive decays. All optional weak rate corrections in {\tt PRIMAT} (see \cite{Pitrou:2018cgg}) and the new corrections from neutrino spectral distortions are included in the calculations, see text for more details. { The $68\%$ constraints on the deuterium abundance $D/H $ of the PDG recommended value ($ D/H \times 10^5 = 2.547 \pm 0.025$) are indicated by the light red shaded region.}}
\label{fg:bbn_doh}
\end{figure}

{  It is also interesting to consider the deuterium abundance constraint. For this purpose, we first briefly discuss the two most widely used databases for thermonuclear rates relevant to BBN, as these are very relevant to the deuterium abundance predicted by BBN.
The first database is tabulated by the public BBN code {\tt PRIMAT} with an extensive catalogue of more than 400 reactions\cite{Pitrou:2018cgg}, implemented according to the refined statistical analyses of Refs. \cite{Descouvemont:2004cw,Iliadis:2016vkw,InestaGomez:2017eya}. We will refer to the predicted BBN abundances obtained with this database as the PRIMAT approach, and use it to obtain most of the numerical results in this paper.
On the other hand, the NACRE~II database~\cite{Xu:2013fha} gives an evaluation of 12 key nuclear reaction rates with mass number $A<16$, based on the nuclear cross sections described by the so-called potential model~\cite{Angulo:1999zz}, as shown in Tab. \ref{tab:nuclear}. With the same nuclear rates of the other reactions with the PRIMAT approach, we will denote it as the NACRE II approach.

We present the resulting BBN abundances as a function of the neutrino degeneracy parameter $\xi_\nu$ in these two approaches in Fig. \ref{fg:bbn_doh}, where the solid curves refer to the PRIMAT approach, which is identical to the results in Fig. \ref{fg:bbn_xi}, and the dashed curves refer to the NACRE II approach. All curves are normalized using the resulting for $\xi_\nu = 0$ in the PRIMAT approach, and the $68\%$ constraints on the deuterium abundance $D/H$ of the PDG recommended value ($ D/H \times10^5 = 2.547 \pm 0.025$) are indicated by the light red shaded region.
Comparing the light element abundances obtained by the two approaches, we find that they give exactly the same predictions for $Y_P$ and similar predictions for $\het/H$.
However, the NACRE II approach predicts a a significantly larger $D/H$ and smaller $\lis/H$ than the PRIMAT approach.
Consequently, the constraints on the neutrino degeneracy parameter $\xi_\nu$ from the PDG constraint $ D/H \times 10^5 = 2.547 \pm 0.025$ with the fixed baryon density $\omega_b = 0.0224$ are also significantly different in the two approaches:
\begin{align}
    {\rm PRIMAT}: \quad -0.088 \leq & \,\xi_\nu \leq -0.057 \, ,\\
    {\rm NACRE \, II}: \quad -0.042 \leq & \, \xi_\nu \leq -0.009 \, .
\end{align}
We see that, while the PRIMAT approach points to a much smaller $\xi_\nu$, both approaches favor a negative negative neutrino asymmetry $\xi_\nu$.
Such results indicate a mild tension between current observations of helium and deuterium abundances.
In addition, we note that despite the mild tension between $Y_P$ and $D/H$ observations, the preference for positive $\xi_\nu$ from the EMPRESS observation on $Y_P$ is more reliable, since $Y_P$ is more sensitive to the neutrino asymmetry $\xi_\nu$, while $D/H$ is more sensitive to the baryon density $\omega_b$, which is fixed in our upper analysis.
Therefore, the preference for a negative $\xi_\nu$ from the $D/H$ observation can be greatly mitigated with a slightly different $\omega_b$.
Consequently, an overall MCMC analysis combining helium and deuterium observations still indicates a preference for positive $\xi_\nu$, as we will see in Sec. \ref{sec:mcmc}.
For the sake of simplicity, we will also only consider the PRIMAT approach in the following, and leave a detailed comparison between different nuclear reaction rates to our subsequent companion paper \cite{Li:2025rjr}.
}

\begin{figure}\centering
\includegraphics[width=0.8\textwidth]{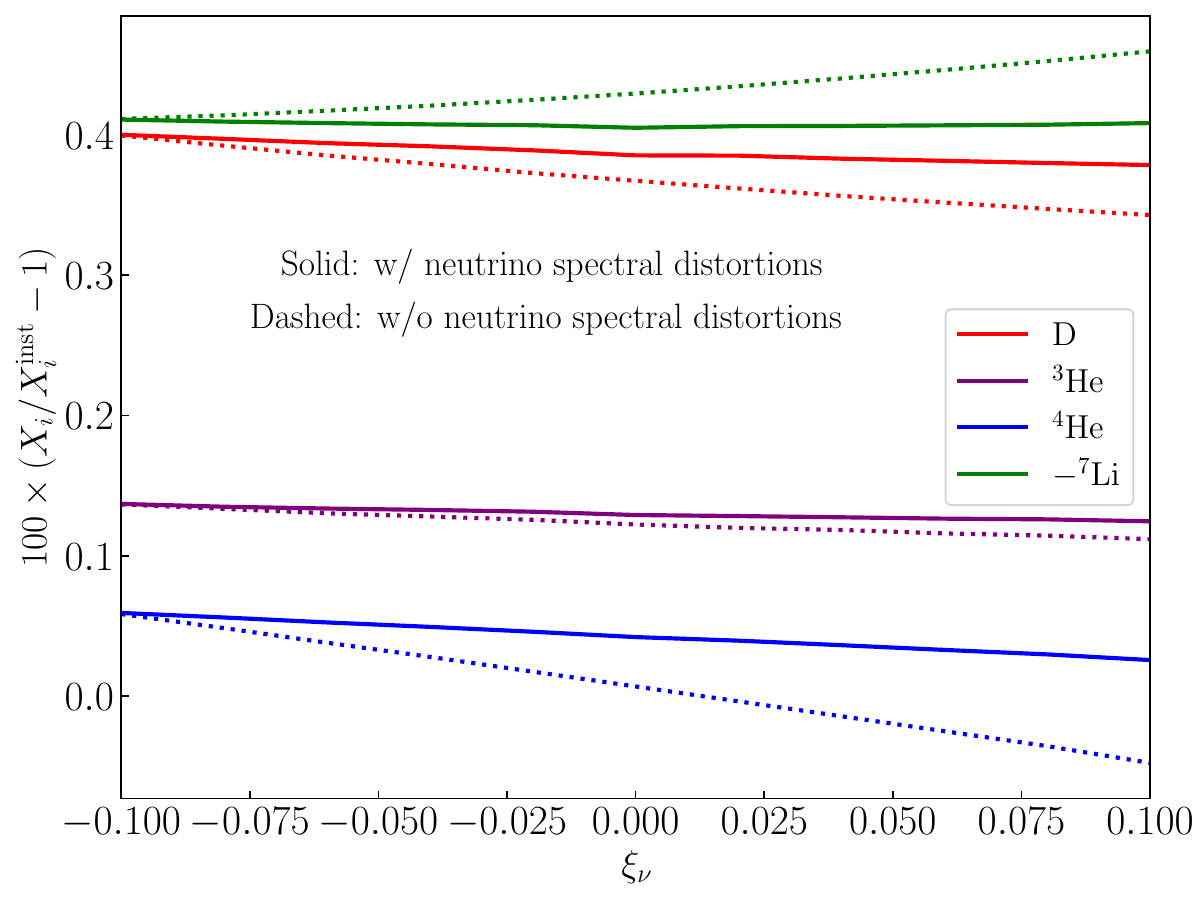}
\caption{
The relative difference in light element abundances compared to the results in the neutrino instantaneous decoupling limit as a function of the neutrino degeneracy parameter $\xi_\nu$. The solid and dotted curves correspond to results with and without weak corrections induced by neutrino spectral distortions, respectively. The color scheme for the curves is consistent with that in \fig \ref{fg:bbn_xi}, except that negative results for $\lis$ are presented here for simplicity.}
\label{fg:bbn_sd}
\end{figure}

\renewcommand{\arraystretch}{1.4}

\begin{table}[!ht]
	\centering
	\begin{tabular}{|l|cccc|}
  	\hline 
 BBN Framework & $\yp$  & ${\rm D}/{\rm H} \times 10^5$ &  ${\het}/{\rm H} \times 10^5$ &  ${\lis}/{\rm H} \times 10^{10}$ \\
  \hline \hline
  Inst. dec, $\xi_\nu=-0.1$ & 0.27359  &  2.58214  &  1.06094  &  5.87974   \\
  Inst. dec, $\xi_\nu=0$ & 0.24711  &  2.42833  &  1.03799  &  5.53146 \\ 
  Inst. dec, $\xi_\nu=0.1$ & 0.22260  &  2.30372  &  1.01897  &  5.14551 \\ \hline
  No-SD, $\xi_\nu=-0.1$& 0.27375  &  2.59246  &  1.06239  &  5.85555 \\
  No-SD, $\xi_\nu=0$& 0.24712  &  2.43725  &  1.03926  &  5.50770 \\
  No-SD, $\xi_\nu=0.1$& 0.22250  &  2.31162  &  1.02011  &  5.12186 \\ \hline
  Actual-SD, $\xi_\nu=-0.1$ & 0.27375  &  2.59247  &  1.06239  &  5.85558 \\
  Actual-SD, $\xi_\nu=0$& 0.24721  &  2.43769  &  1.03933  &  5.50905 \\
  Actual-SD, $\xi_\nu=0.1$& 0.22266  &  2.31244  &  1.02024  &  5.12449 \\ \hline
\end{tabular}
	\caption{Light element abundances, including all weak rate corrections in \eqref{eq: n2ptot}, for various implementations of neutrino decoupling and degeneracy parameters. $\het$ denotes $(\het + \mathrm{T})$ and $\lis$ denotes $(\lis + \rm Be)$ to account for slow radioactive decays.}\label{tab:bbn}
\end{table}

\renewcommand{\arraystretch}{1.4}

\begin{table}[!ht]
	\centering
	\begin{tabular}{|l|cccc|}
  	\hline 
 BBN Framework & $\delta \yp \,(\%)$  & $\delta {\rm D}/{\rm H} \,(\%) $ &  $ \delta {\het}/{\rm H} \,(\%)$ &  $\delta {\lis}/{\rm H} \,(\%)$ \\
  \hline \hline
    No-SD, $\xi_\nu=-0.1$& 0.058  &  0.4  &  0.137  &  -0.411 \\
    No-SD, $\xi_\nu=0$& 0.007  &  0.367  &  0.122  &  -0.430 \\
    No-SD, $\xi_\nu=0.1$& -0.048  &  0.343  &  0.112  &  -0.460  \\
    \hline
    Actual-SD, $\xi_\nu=-0.1$& 0.059  &  0.4  &  0.137  &  -0.411 \\
    Actual-SD, $\xi_\nu=0$& 0.042  &  0.385  &  0.129  &  -0.405 \\
    Actual-SD, $\xi_\nu=0.1$& 0.026  &  0.379  &  0.125  &  -0.408 \\ \hline
\end{tabular}
	\caption{Relative differences in light element abundances, comparing cases with and without corrections induced by neutrino spectral distortions to results in the neutrino instantaneous decoupling limit. $\het$ denotes $(\het + \mathrm{T})$ and $\lis$ denotes $(\lis + \rm Be)$ to account for slow radioactive decays.}\label{tab:delta_bbn}
\end{table}

It is also interesting to discuss the effects of corrections induced by neutrino spectral distortions on BBN. We examine the impact of different treatments of the neutrino decoupling process, considering three implementations: (1) the neutrino instantaneous decoupling limit, \ie $\Neff^{\rm inst}$, as discussed in the previous section, where thermal Fermi-Dirac distributions for neutrinos and antineutrinos are used in the BBN calculations; (2) the no-SD case, which employs the actual $\Neff$ and neutrino temperatures, incorporating the non-instantaneous decoupling effect while only using thermal Fermi-Dirac distributions for neutrinos and antineutrinos. This treatment is the most common in current BBN studies; (3) the actual SD case, which utilizes the actual $\Neff$, accounting for the non-instantaneous decoupling effect and incorporating actual density distributions with spectral distortions for neutrinos and antineutrinos. This case corresponds to the results shown in \fig \ref{fg:bbn_xi}.

Using the BBN results in the instantaneous decoupling limit as a benchmark, we present the relative differences for the no-SD and actual-SD cases, indicated by dotted and solid curves in \fig \ref{fg:bbn_sd}, respectively. Note that we have plotted the negative of the relative differences for $\lis$ in \fig \ref{fg:bbn_sd} due to the specific nuclear reaction rates utilized in {\tt PRIMAT}. We report the explicit abundances of helium-4, deuterium, helium-3, and lithium-7 for the degeneracy parameters $\xi_\nu = 0, \pm 0.1$ in table \ref{tab:bbn}, along with the associated relative variations compared to the baseline abundance in table \ref{tab:delta_bbn}. Our results with $\xin = 0$ are consistent with those reported in \cite{Froustey:2020mcq}, which also considered the effects of neutrino spectral distortions on BBN within the standard model (SM).
As shown in \fig \ref{fg:bbn_sd} and the accompanying tables, the actual-SD and no-SD cases yield nearly identical light element abundances for $\xin=-0.1$. The BBN abundances in the actual-SD case show minimal dependence on $\xi_\nu$, while the traditional no-SD treatment predicts a stronger dependence for both elements. Notably, the no-SD case predicts negative variations for the helium abundance with positive $\xin$ compared to the instantaneous decoupling limit. In contrast, the more precise results with spectral distortions indicate positive variations in helium abundance, demonstrating that incorporating neutrino spectral distortions is crucial for accurate predictions of BBN abundances.

Finally, we provide fitting functions for the light element abundances as follows:
\begin{align}\label{eq:bbnfit}
 Y_{\mathrm{P}}\left(\xi_{\nu}\right) &\simeq \left. Y_{\mathrm{P}}\right|_{\xin=0} \times e^{-1.033\xi - 0.131\xi^2}, \, \,\\
 \mathrm{D/H}\left(\xi_{\nu}\right)  &\simeq \left. \mathrm{D/H}\right|_{\xin = 0} \times e^{-0.57 \, \xi_{\nu} (1-\xi_{\nu})} \,, \\
 ^3\mathrm{He/H}\left(\xi_{\nu}\right) &\simeq \left. ^3\mathrm{He/H}\right|_{\xin=0} \times e^{-0.19 \, \xi_{\nu} (1-\xi_{\nu})}, \, \, \\
 ^7\mathrm{Li/H}\left(\xi_{\nu}\right) &\simeq \left. ^7\mathrm{Li/H}\right|_{\xin=0} \times e^{-0.63 \, \xi_{\nu} (1+\xi_{\nu})},
\end{align}
where the subscript $\xin=0$ indicates results obtained in the SM case using actual density matrices, as detailed in Table \ref{tab:bbn}. These fitting functions facilitate further studies, yielding fractional errors smaller than $0.01\%$ within the range $-0.1 \leq \xin \leq 0.1$. \footnote{The coefficients of our fitting functions differ slightly from those in~\cite{Pitrou:2018cgg,Froustey:2024mgf}, which is mainly due to the different neutron lifetimes we have chosen. In addition, our fitting function is more precise.}


\section{Impacts of Primordial Neutrino Asymmetry on the CMB and LSS}\label{sec:cmb}

In this section, we examine the effects of primordial neutrino asymmetries on the evolution of the late universe, particularly focusing on the cosmic microwave background (CMB) and large-scale structure (LSS). Primordial neutrino asymmetry affects the CMB and LSS in three primary ways:
\begin{enumerate}
    \item The expansion rate of the universe is altered by the increased $\Neff$ resulting from primordial neutrino asymmetries. Specifically, the additional radiation density delays the time of matter-radiation equality, enhancing the early integrated Sachs-Wolfe effect in the CMB angular power spectrum.
    \item As shown in \fig\ref{fg:bbn_xi}, the electron neutrino asymmetry significantly modifies the helium abundance, which subsequently affects the late-time evolution of the universe. This impacts features such as the tail of the CMB angular power spectrum due to diffusion damping and the baryon acoustic oscillations (BAO).
    \item Additionally, to account for neutrino oscillations during neutrino decoupling, neutrino masses must be considered. These masses influence the late-time evolution of the universe, particularly affecting matter-radiation equality, even in the standard model (SM) scenario with thermal Fermi-Dirac distributions. For the SM case, CMB and BAO observations can place stringent constraints on the sum of neutrino masses, yielding $\sum m_\nu < 0.07 \, \ev$~\cite{DESI:2024mwx}. We will revisit these constraints in the next section. Furthermore, the neutrino Boltzmann hierarchy for massive neutrinos is modified by the neutrino degeneracy parameter and spectral distortions.
\end{enumerate}

In the following, we first review the impact of massive neutrinos on the universe and present the Boltzmann hierarchies for electrons, photons, and neutrinos. We then discuss the effects of primordial neutrino asymmetries on CMB temperature anisotropies and the linear matter power spectrum, computed using the Einstein-Boltzmann code {\texttt{CLASS}}~\cite{Lesgourgues:2011re,Lesgourgues:2011rh}.

\subsection{Effects of Primordial Neutrino Asymmetries in Structure Formation}

To accurately describe flavor oscillations during neutrino decoupling, neutrino mass must be considered, which has significant implications for structure formation. Massive neutrinos are the only species in the SM that behave as radiation in the early universe and as matter in the late universe. 

In the early universe, neutrinos and antineutrinos remain relativistic after decoupling, as long as their effective temperatures are much higher than their masses. In this regime, their energy densities contribute to the total radiation energy density, parameterized by $\Neff$. As the universe expands and the effective temperatures decrease as $a^{-1}$, neutrinos and antineutrinos become non-relativistic when their effective temperatures fall below their masses, at which point they contribute to the matter content. The transition to the non-relativistic phase and the free-streaming effects of massive neutrinos leave noticeable imprints on CMB anisotropies and LSS. 

As discussed, primordial neutrino asymmetries modify this scenario. In addition to indirect changes from the altered Hubble rate and helium abundance, we emphasize several direct effects of primordial neutrino asymmetries.

\subsubsection{Effects on Neutrino Abundance}

First, the total relic abundance of neutrinos and antineutrinos increases directly due to primordial neutrino asymmetries. At present, the total neutrino relic abundance is given by
\begin{equation}
\begin{aligned}
    \Omega_\nu h^2 &= \frac{\sum \limits_i T_{\rm CM, 0}^4 \int_0^{\infty} \dd y\, y^2 \sqrt{y^2 + m_{i}} \left( \varrho_i(y) + \bvrho_i(y) \right)/(2\pi^2)}{\rho_{\rm crit}}  \\
    &\overset{\delta\vrho=0}{\simeq}  \frac{\sum \limits_i m_i }{93.12 \,\ev} \left( - \frac{{\rm Li}_3(-\exp(-\xin)) + {\rm Li}_3(-\exp(\xin)) }{3 \zeta(3)/2 } \right),
\end{aligned}
\label{eq: omega_nu}
\end{equation}
where $y = p/ T_{\rm CM}$ is the comoving momentum defined in section \ref{sec: decoupling}, $T_{\rm CM,0}$ is the current comoving temperature, $\rho_{\rm crit}$ is the critical energy density, and $\varrho_i(y)$ and $\bvrho_i(y)$ represent the comoving density distributions of neutrinos and antineutrinos, respectively, for $i=1,2,3$, corresponding to the three mass eigenstates with mass $m_i$. 

The present neutrino abundance is further modified by spectral distortions of neutrinos and antineutrinos, though the integrals over their distributions cannot be carried out analytically in this case. The effect of primordial neutrino asymmetries becomes clearer when spectral distortions are neglected, allowing an analytical evaluation of the integrals, as shown in the second row of \eqref{eq: omega_nu}. The first factor, $\frac{\sum \limits_i m_i }{93.12 \,\ev}$, represents the usual neutrino relic abundance in the SM, while the second factor accounts for the effects of neutrino asymmetries, with ${\rm Li}_j (x)$ being the polylogarithms of order $j$ and $\zeta(x)$ the Riemann zeta function. 
In the SM case, ${\rm Li}_3 (-1) = -3 \zeta(3)/4$, so the second factor equals 1. For non-zero $\xin$, the second factor depends on the absolute value of $\xin$, with larger $\xin$ leading to a larger factor and, consequently, a larger neutrino abundance. However, the corrections from primordial neutrino asymmetries are small. For instance, the relative difference in $\Omega_\nu h^2$ for $\xin=0.2$ is only $1.54\%$. Similarly, the corrections to the neutrino abundance from spectral distortions of neutrinos and antineutrinos are negligible.

\subsubsection{Effects on the Neutrino Boltzmann Hierarchies}

To obtain accurate CMB spectral power and matter power spectra, we must solve the coupled Boltzmann-Einstein equations for perturbations in the metric, photons, baryons, cold dark matter, and massive neutrinos. Here, we briefly review how neutrino asymmetries and spectral distortions affect the Boltzmann-Einstein equations.
Following Ref.~\cite{Ma:1995ey}, the perturbed distribution functions for the three mass eigenstate neutrinos can be written as
\begin{equation}\label{eq:f_1st}
\vrho_i \left(\vec{x}, \vec{p}, \tau\right) = \vrho_i^0(y) \left[1 + \Psi_i\left(\vec{x}, y, \hat{n}, \tau\right)\right],
\end{equation}
where $\tau \equiv \int dt/a(t)$ is the conformal time, $y \equiv p/\Tcm$ is the comoving momentum, $\hat{n} \equiv \vec{p} / p$ is the direction of the momentum, and $\vrho_i^0(y)$ represents the unperturbed zeroth-order neutrino density distributions obtained in Section \ref{sec: decoupling}.

In this work, we adopt the synchronous gauge, where the line element is written as
\begin{equation}
ds^2 = a^2(\tau) \left\{ - d\tau^2 + (\delta_{ij} + h_{ij}) dx^i dx^j \right\},
\end{equation}
with the metric perturbation $h_{ij}$ decomposed as $h_{ij} = h \delta_{ij} + h_{ij}^{\parallel} + h_{ij}^{\perp} + h_{ij}^{T}$, where $h \equiv h_{ii}$ is the trace of $h_{ij}$, and $h_{ij}^{\parallel}$, $h_{ij}^{\perp}$, and $h_{ij}^{T}$ are the traceless components, satisfying
\begin{equation}
\epsilon_{ijk} \partial_j \partial_l h_{lk}^{\parallel} = 0, \quad \partial_i \partial_j h_{ij}^{\perp} = 0, \quad \partial_i h_{ij}^{T} = 0.
\end{equation}
By definition, $h_{ij}^{\parallel}$ and $h_{ij}^{\perp}$ can be written in terms of a scalar field $\mu$ and a divergenceless vector $\vec{A}$ as
\begin{equation}
\begin{aligned}
    h_{ij}^{\parallel} &= \left( \partial_i \partial_j - \frac{1}{3} \delta_{ij} \nabla^2 \right) \mu, \\
    h_{ij}^{\perp} &= \partial_i A_j + \partial_j A_i, \quad \partial_i A_i = 0.
\end{aligned}
\end{equation}
Thus, the scalar modes of the metric perturbations are characterized by the two scalar fields $h$ and $\mu$, while the vector and tensor modes are characterized by $\vec{A}$ and $h_{ij}^{T}$, respectively.
In this work, we focus solely on the scalar modes of $h_{ij}$, which can be Fourier transformed as
\begin{equation}
 h_{ij} (\vec{x}, \tau) = \int d^3 k~e^{i \vec{k} \cdot \vec{x}} \left\{ \frac{k_i k_j}{k^2} h(\vec{k}, \tau) + 6 \eta(\vec{k}, \tau) \left( \frac{k_i k_j}{k^2} - \frac{1}{3} \delta_{ij} \right) \right\},
\end{equation}
where $h(\vec{k}, \tau)$ and $\eta(\vec{k}, \tau)$ represent the trace and traceless parts of the scalar modes in Fourier space, related to the scalar fields $h$ and $\mu$ in real space.

Since neutrinos are collisionless after decoupling, the evolution of $\Psi_i$ follows the first-order collisionless Boltzmann equation in Fourier space:
\begin{equation}\label{eq:Boltzmann_1st}
\frac{\partial \Psi_i}{\partial \tau} + i \frac{q}{\epsilon}(\vec{k} \cdot \hat{n}) \Psi_i + \frac{d \ln \vrho_i^0}{d \ln q} \left[\dot{\eta} - \frac{\dot{h} + 6 \dot{\eta}}{2} (\hat{k} \cdot \hat{n})^2\right] = 0,
\end{equation}
where $\epsilon \equiv a E = \sqrt{q^2 + a^2 m^2_{\nu_i}}$ is the comoving energy. Since Eq.~\eqref{eq:Boltzmann_1st} is independent of the azimuthal angle, we can expand $\Psi_i$ in a Legendre series:
\begin{equation}\label{eq:Psi_l}
\Psi_i(\vec{k}, \hat{n}, q, \tau) = \sum_{l=0}^{\infty} (-i)^l (2 l + 1) \Psi_{i,l}(k, q, \tau) P_l(\hat{k} \cdot \hat{n}).
\end{equation}
Substituting Eq.~\eqref{eq:Psi_l} into Eq.~\eqref{eq:Boltzmann_1st} and using the orthonormality of Legendre polynomials, we obtain the Boltzmann hierarchies for $\Psi_{i,l}$:
\begin{equation}
\begin{aligned}
\dot{\Psi}_{i,0} &= -\frac{q k}{\epsilon} \Psi_1 + \frac{\dot{h}}{6} \frac{d \ln \vrho_i^0}{d \ln q}, \\
\dot{\Psi}_{i,1} &= \frac{q k}{3 \epsilon} \left(\Psi_0 - 2 \Psi_2\right), \\
\dot{\Psi}_{i,2} &= \frac{q k}{5 \epsilon} \left(2 \Psi_1 - 3 \Psi_3\right) - \left(\frac{\dot{h}}{15} + \frac{2 \dot{\eta}}{5}\right) \frac{d \ln \vrho_i^0}{d \ln q}, \\
\dot{\Psi}_{i, l \geq 3} &= \frac{q k}{(2 l + 1) \epsilon} \left[l \Psi_{l-1} - (l + 1) \Psi_{l+1}\right].
\end{aligned}
\end{equation}
The effect of neutrino asymmetries and spectral distortions manifests through the neutrino density distribution term $\frac{d \ln \vrho_i^0}{d \ln q}$. Due to primordial neutrino asymmetries, the density matrices for neutrinos and antineutrinos of different flavors differ, leading to distinct Boltzmann hierarchies for the perturbations $\Psi_i$ and $\bar{\Psi}_i$, which must be treated separately.

With the moments $\Psi_{i,l}$, perturbations in physical quantities such as energy density, pressure, energy flux, and shear stress can be computed straightforwardly by integrating different moments. Specifically, the fractional density perturbation for neutrinos of family $i$ is given by
\begin{equation}
\delta_i(\vec{x}, t) \equiv \frac{\rho_i(\vec{x}, t) - \rho_{i,0}(t)}{\rho_{i,0}(t)} = \frac{4 \pi \Tcm^{4} \int dy~y^2 \vrho_i^0(y)\Psi_{i,0} \epsilon}{\rho_{i,0}(t)},
\end{equation}
where $\rho_{i,0}(t)$ represents the zeroth-order homogeneous background energy density for neutrinos. The effects of neutrino masses and asymmetries across different length scales are reflected in the evolution of the Fourier modes $\delta_{i,k}$, which we obtain via the \texttt{CLASS} code.

\subsubsection{Effects on the BAO}
Following Ref.~\cite{Ma:1995ey}, the evolution of photon perturbations is governed by the Boltzmann hierarchy:
\begin{equation}
    \begin{aligned}\label{photon}
        \dot{\delta}_\gamma &= -\frac{4}{3} \theta_\gamma - \frac{2}{3} \dot{h}, \\
     \dot{\theta}_\gamma &= k^2 \left(\frac{1}{4} \delta_\gamma - \sigma_\gamma\right)
	+ a n_e \sigma_T (\theta_b - \theta_\gamma),\\
     \dot{F}_{\gamma\,2} &= 2\dot\sigma_\gamma = \frac{8}{15} \theta_\gamma - \frac{3}{5} k F_{\gamma\,3} + \frac{4}{15} \dot{h} + \frac{8}{5} \dot{\eta} - \frac{9}{5} a n_e \sigma_T \sigma_\gamma + \frac{1}{10} a n_e \sigma_T (G_{\gamma\,0} + G_{\gamma\,2}), \\
     \dot{F}_{\gamma\,l} &= \frac{k}{2l+1} \left[ l F_{\gamma\,(l-1)} - (l+1) F_{\gamma\,(l+1)} \right] - a n_e \sigma_T F_{\gamma\,l}, \quad l \geq 3, \\
     \dot{G}_{\gamma\,l} &= \frac{k}{2l+1} \left[ l G_{\gamma\,(l-1)} - (l+1) G_{\gamma\,(l+1)} \right]
        + a n_e \sigma_T \left[ -G_{\gamma\,l} + \frac{1}{2} \left( F_{\gamma\,2} + G_{\gamma\,0} + G_{\gamma\,2} \right) \left( \delta_{l0} + \frac{\delta_{l2}}{5} \right) \right].
    \end{aligned}
\end{equation}
Here, $\delta_\gamma$ is the photon density fluctuation, $\theta_\gamma$ is the divergence of photon velocity, and $F_{\gamma\,l}$ represents the photon density perturbation. Additionally, $n_e$ is the electron number density, and $\sigma_T$ is the Thomson scattering cross section.

The Boltzmann equations for baryons are
\begin{eqnarray}
\label{baryon}
	\dot{\delta}_b &=& -\theta_b - \frac{1}{2} \dot{h}, \nonumber\\
	\dot{\theta}_b &=& -\frac{\dot{a}}{a} \theta_b + c_s^2 k^2 \delta_b + \frac{4 \bar{\rho}_\gamma}{3 \bar{\rho}_b} a n_e \sigma_T (\theta_\gamma - \theta_b),
\end{eqnarray}
where $\delta_b$ and $\theta_b$ are the baryon density fluctuation and the divergence of baryon fluid velocity, respectively, and $c_s^2$ is the square of the baryon sound speed.
Due to electromagnetic interactions between photons and electrons, the evolution of photons and baryons differs before and after recombination. Prior to recombination, photons and baryons are tightly coupled and can be treated as a single photon-baryon fluid, known as the "tight-coupling approximation." During this "tight-coupling epoch," the perturbations of baryons and photons undergo harmonic motion with slowly decaying amplitude~\citep{Peebles:1970ag,Sunyaev:1970bma}. After recombination, photons decouple from baryons, allowing the baryon acoustic oscillations (BAO) to freeze, and baryon perturbations grow as $\propto a$. The BAO feature is then imprinted on the gravitational potential and matter fluctuations, influencing the galaxy power spectrum, which depends on the sound horizon at decoupling and the angular diameter distance to the last-scattering surface.

Primordial neutrino asymmetries modify the matter-radiation equality via their contribution to $\Neff$, altering the BAO properties. Moreover, neutrino asymmetries affect the evolution of neutrino perturbations, which in turn influence photon and baryon evolution through metric perturbations encoded in $h$ and $\eta$, despite the collisionless nature of neutrinos. As a result, primordial neutrino asymmetries significantly impact both the CMB angular power spectrum and the BAO in the matter power spectrum, as we will explore later.

\subsection{Results of the CMB Anisotropies and the Matter Power Spectrum}

\begin{figure}\centering
\includegraphics[width=0.495\textwidth]{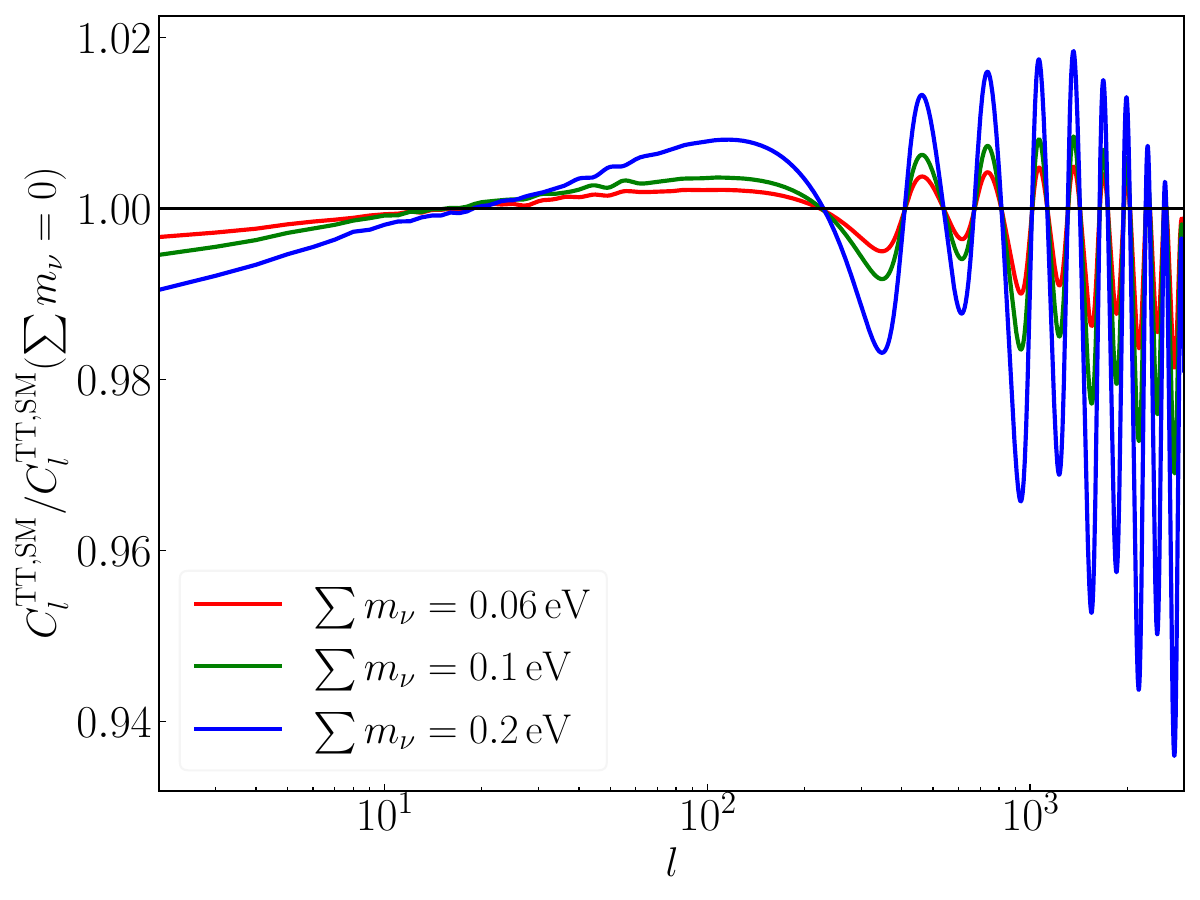} 
\includegraphics[width=0.495\textwidth]{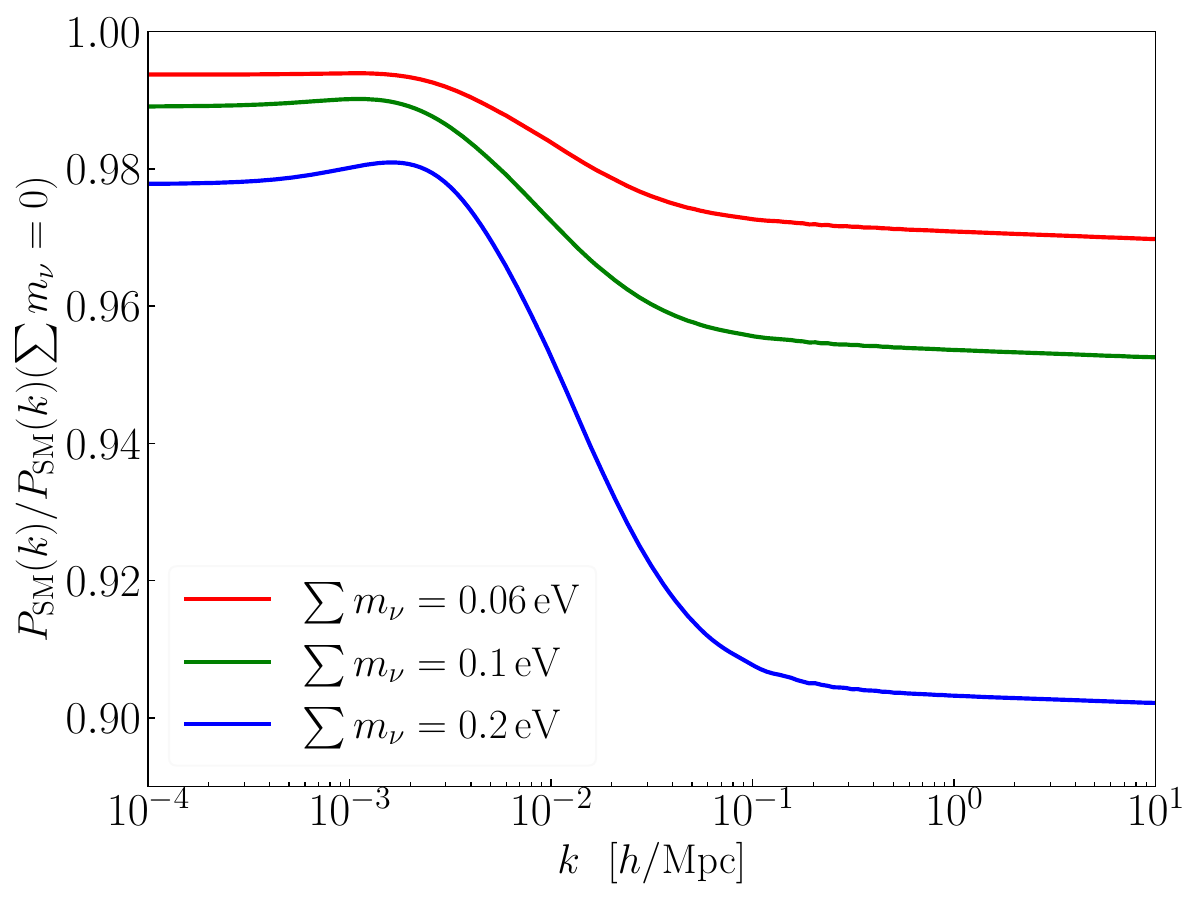} 
\caption{
\textbf{Left panel:} The CMB angular power spectrum $C_l^{\rm TT}$ in the SM for different total neutrino masses, compared to massless neutrinos. \textbf{Right panel:} The linear matter power spectrum $P(k)$ in the SM for different total neutrino masses, compared to massless neutrinos.}
\label{fg:lss_mass}
\end{figure}

This subsection presents the numerical results for the CMB and LSS obtained using the \texttt{CLASS} code. Since both observables are influenced by neutrino masses and asymmetries, we first analyze the effect of neutrino masses on the CMB and LSS within the SM. 

In Fig.~\ref{fg:lss_mass}, we show the ratio of the CMB angular power spectrum, $C_l^{\rm TT}/C_l^{\rm TT}(\sum m_\nu = 0)$, and the matter power spectrum, $P_k/P_k(\sum m_\nu = 0)$, for several choices of total neutrino mass. The left and right panels depict the results for the CMB and matter power spectra, respectively. This analysis assumes the normal hierarchy for neutrino masses, where $\m_{\nu_1}$ is adjusted for each total mass, while $\m_{\nu_2}$ and $\m_{\nu_3}$ are derived from the fixed mass differences between neutrinos.
Focusing on the left panel, the ratio of $C_l^{\rm TT}$ between massive and massless neutrinos can be categorized into three stages. At large angular scales (lower multipoles, $\ell \lesssim 20$), massive neutrinos cause a small suppression due to the late-time integrated Sachs-Wolfe effect. At intermediate scales ($20 \lesssim \ell \lesssim 500$), dominated by acoustic oscillations in the baryon-photon fluid, the first peak of $C_l^{\rm TT}$ is influenced by the modified gravitational potential evolution, resulting from the interaction between massive neutrinos and metric perturbations. Notably, the height of the first peak, due to the early integrated Sachs-Wolfe effect, is enhanced for larger neutrino masses compared to massless neutrinos. At smaller scales ($\ell \gtrsim 500$), the acoustic peaks are similarly affected, with their height increasing for larger neutrino masses.

The matter power spectrum with massive neutrinos can also be divided into three stages. At large scales ($k \lesssim 10^{-3} \, h/{\rm Mpc}$), neutrinos remain outside the horizon, and the suppression of $P(k)$ results from the shift in the matter-radiation equality, caused by the transition of neutrinos to non-relativistic speeds. At intermediate scales ($10^{-3} \, h/{\rm Mpc} \lesssim k \lesssim 1 \, h/{\rm Mpc}$), neutrinos gradually enter the horizon, and their free-streaming produces additional suppression in the matter power spectrum, with $P_k/P_k(\sum m_\nu = 0)$ decreasing as $k$ increases. Finally, at small scales ($k \gtrsim 1 \, h/{\rm Mpc}$), most neutrinos freely stream, leading to an almost constant suppression of the power spectrum. The larger the total neutrino mass, the greater the suppression. 
It is worth noting that the matter power spectrum is more sensitive to neutrino masses than the CMB, as shown in the plots. Thus, LSS observations are more effective in constraining neutrino masses, a point that will be further discussed in the next section.

\begin{figure}\centering
\includegraphics[width=0.495\textwidth]{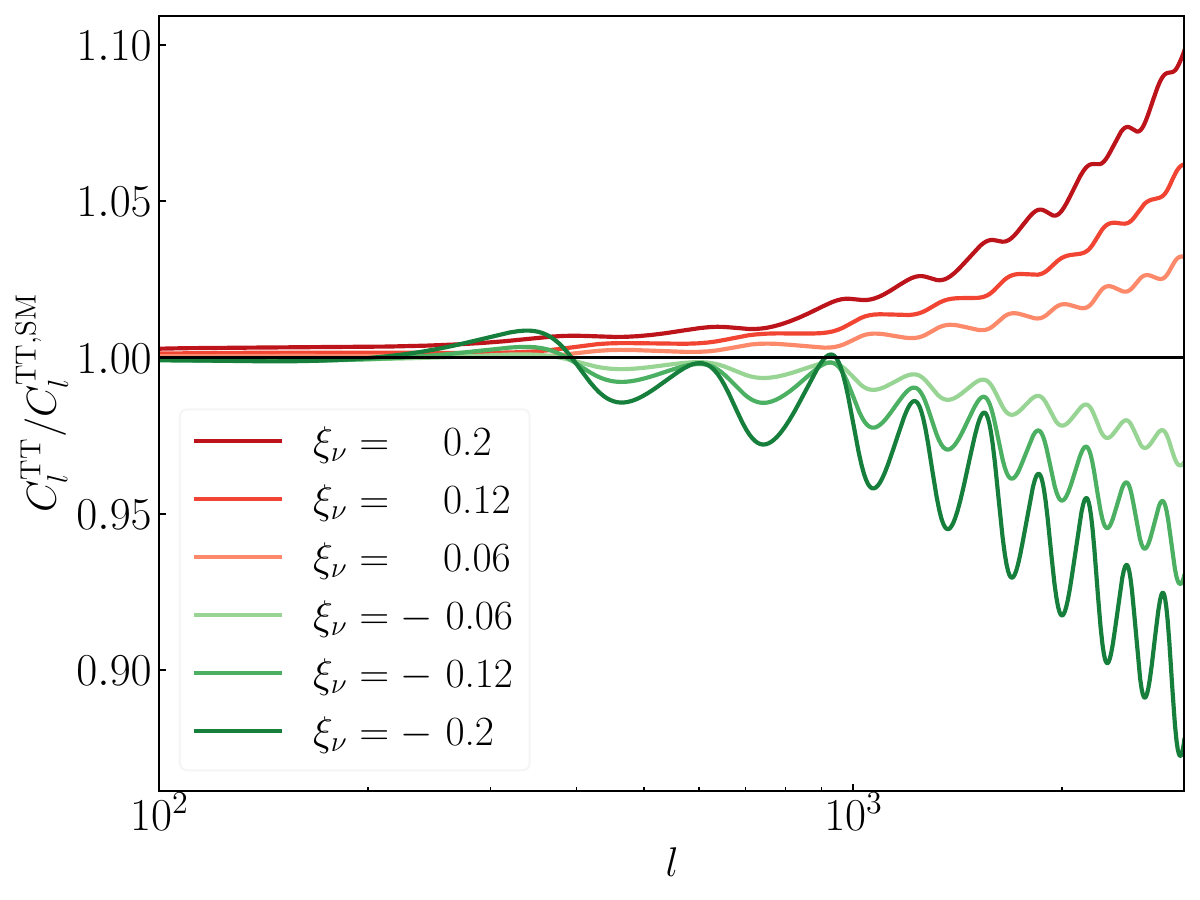} 
\includegraphics[width=0.495\textwidth]{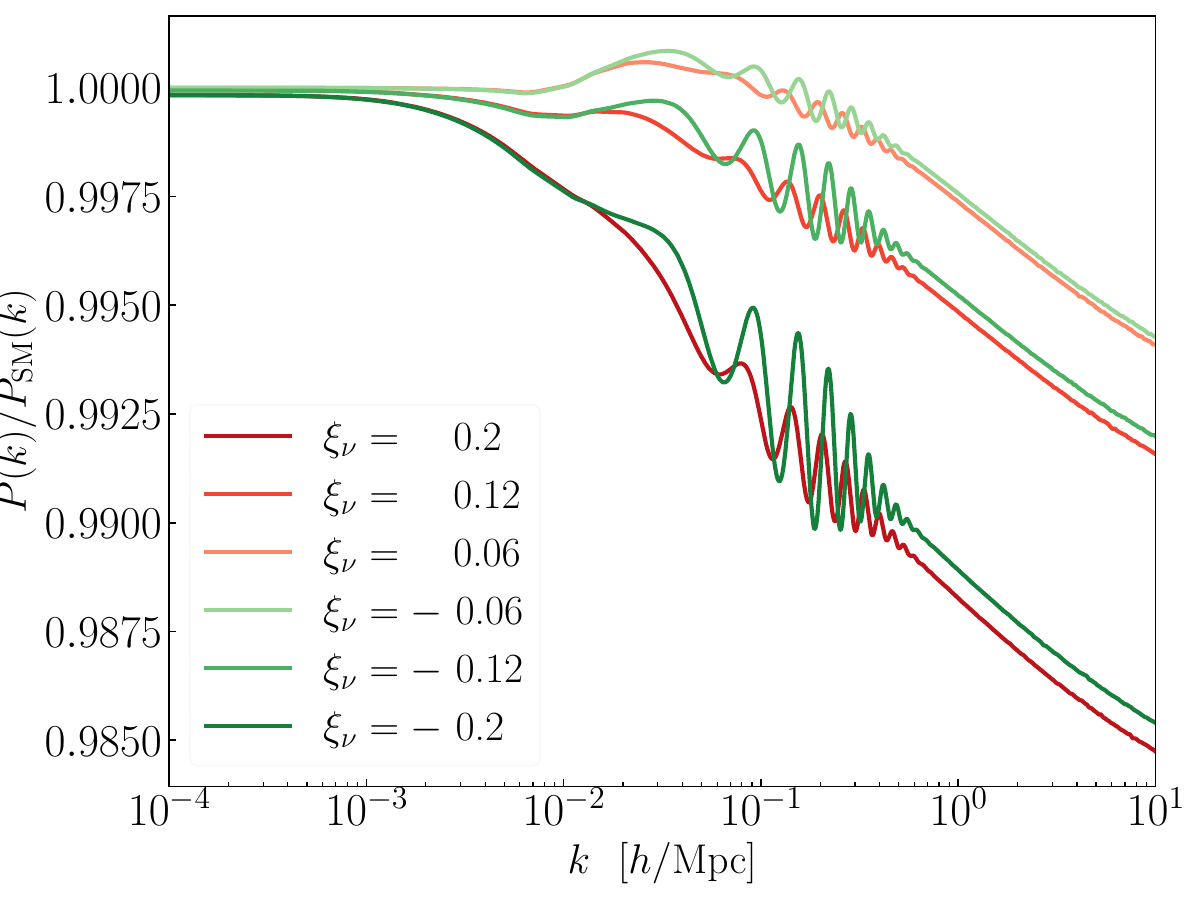} 
\caption{
\textbf{Left panel:} The CMB angular power spectrum $C_l^{\rm TT}$ for different neutrino degeneracy parameters $\xi_\nu$, compared to the result for $\xi_\nu = 0$. \textbf{Right panel:} The linear matter power spectrum $P(k)$ for different neutrino degeneracy parameters $\xi_\nu$, compared to the result for $\xi_\nu = 0$. The total neutrino mass is fixed at $\sum m_\nu = 0.1 \, \rm{eV}$ in both panels.}
\label{fg:lss_xi}
\end{figure}

\begin{figure}\centering
\includegraphics[width=0.5\textwidth]{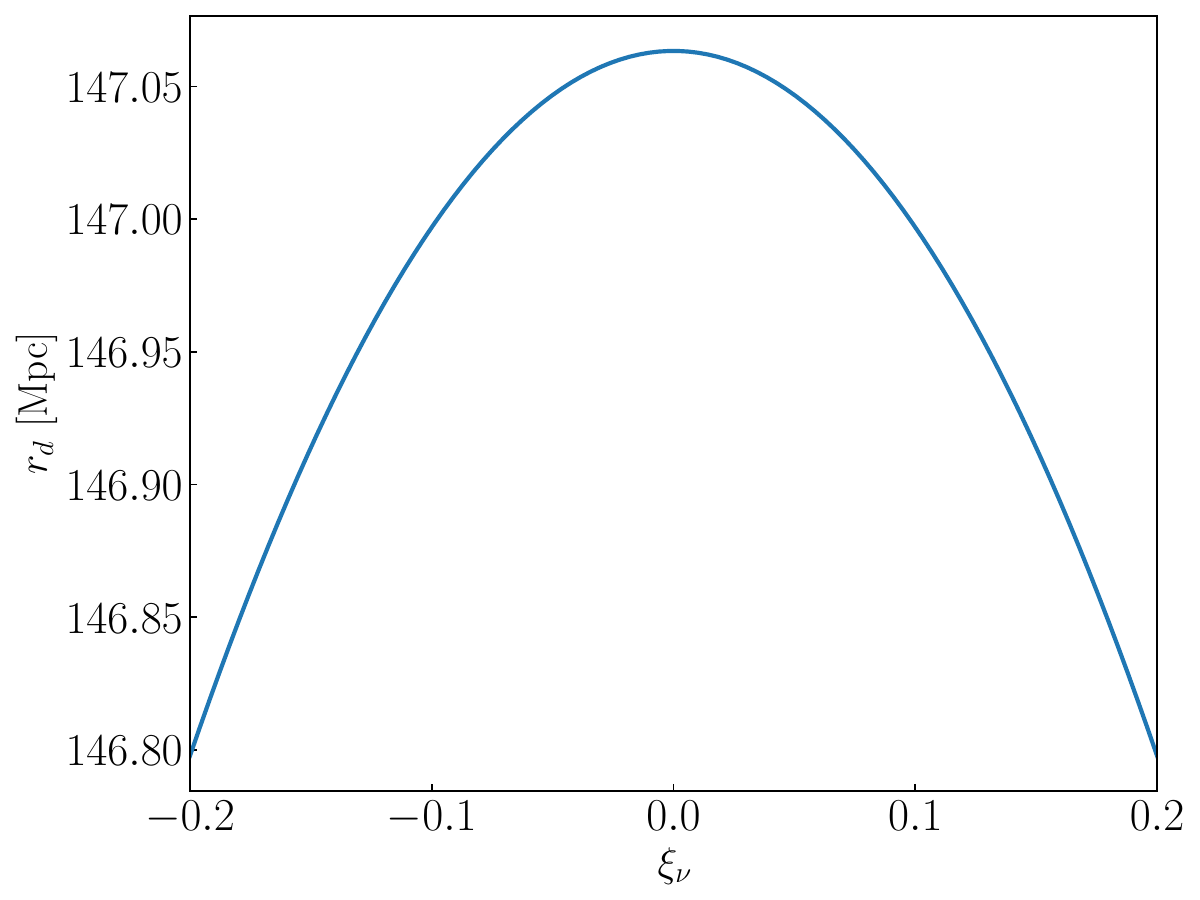}
\caption{
The sound horizon at the baryon drag epoch, $r_d$, as a function of the primordial neutrino asymmetry $\xi_\nu$.}
\label{fg:rd_xi}
\end{figure}

We now present the results for non-zero $\xi_\nu$. To isolate the effects of neutrino asymmetry from the total neutrino mass, the left panel of Fig.~\ref{fg:lss_xi} shows the ratio of the CMB angular power spectrum for $\xi_\nu \neq 0$ to that for $\xi_\nu = 0$, with $\sum m_{\nu_i} = 0.1 \, \text{eV}$. The corresponding ratio for the matter power spectrum is shown in the right panel. As seen in the left panel, the effects of neutrino asymmetry on the CMB angular power spectrum are primarily observed at small scales ($\ell \gtrsim 400$). For $\ell \lesssim 400$, $C_l^{\rm TT}$ remains nearly identical for both $\xi_\nu \neq 0$ and $\xi_\nu = 0$. However, as $\ell$ increases, deviations due to $\xi_\nu$ become apparent. These deviations depend on both the magnitude and sign of $\xi_\nu$: positive $\xi_\nu$ enhances the small-scale power, while negative $\xi_\nu$ leads to suppression. This behavior is primarily driven by changes in helium abundance, which affect the CMB spectrum through diffusion damping. 
In addition to the overall enhancement or suppression, the acoustic peaks are also influenced by $\xi_\nu$, as larger values modify the oscillations at the last scattering surface. The amplitude of these modifications increases with the absolute value of $\xi_\nu$, with negative $\xi_\nu$ producing more pronounced effects.

For the linear matter power spectrum, shown in the right panel of Fig.~\ref{fg:lss_xi}, we observe that at large scales, the results for non-zero $\xi_\nu$ are identical to those for $\xi_\nu = 0$. At smaller scales, however, the free-streaming of neutrinos is enhanced with increasing $\xi_\nu$, resulting in a slightly stronger suppression of the matter power spectrum, while this suppression is largely insensitive to the sign of $\xi_\nu$. A more notable effect of non-zero $\xi_\nu$ is the oscillations in the matter power spectrum on scales $10^{-1} \, h/{\rm Mpc} \gtrsim k \gtrsim 10^{0} \, h/{\rm Mpc}$, which reflect the influence of the BAO. Similar to the acoustic oscillations in the CMB angular power spectrum, the amplitude of BAO in the matter power spectrum increases with the absolute value of $\xi_\nu$, with negative $\xi_\nu$ leading to a larger effect than positive $\xi_\nu$. 

In addition to BAO features in the matter power spectrum, Fig.~\ref{fg:rd_xi} shows the sound horizon at the baryon drag epoch, $r_d$, as a function of the primordial neutrino asymmetry $\xi_\nu$. Due to the additional contribution to $\neff$ for non-zero $\xi_\nu$, as per Eq.~\ref{eq:neff}, larger $\abs{\xi_\nu}$ corresponds to a smaller $r_d$, resulting in significant imprints on the matter power spectrum. These results imply that BAO experiments can provide stringent constraints on the primordial neutrino asymmetry, as will be discussed in Section \ref{sec:mcmc}, a consideration often overlooked in previous studies.
Note that these conclusions are drawn with a fixed total neutrino mass of $\sum m_{\nu_i} = 0.1 \, \text{eV}$; however, they remain valid for different total neutrino masses, with only slight variations in the ratios $C_l^{\rm TT}/C_l^{\rm TT, \, \xi_\nu = 0}$ and $P(k)/P^{\xi_\nu = 0}(k)$ at intermediate scales due to the non-relativistic transition of massive neutrinos.

\begin{figure}\centering
\includegraphics[width=0.495\textwidth]{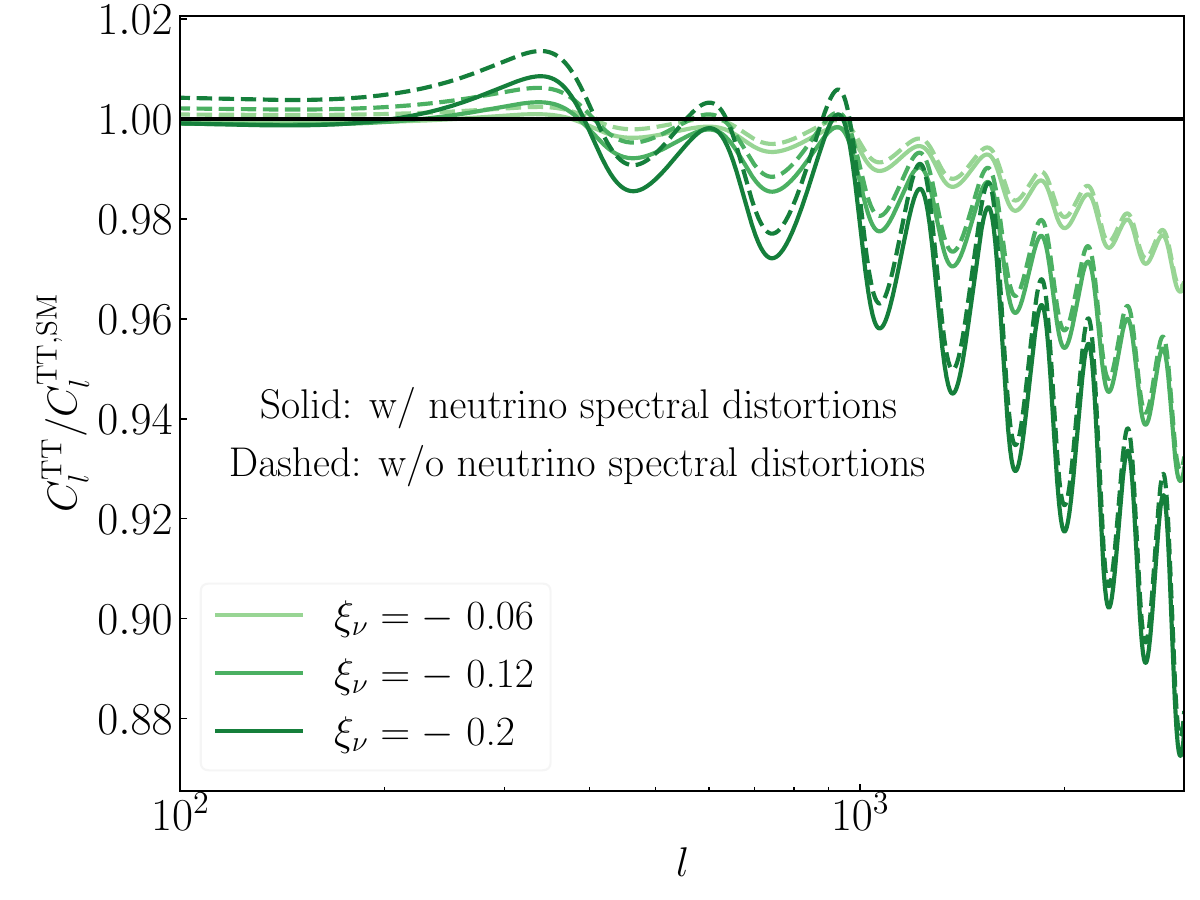} 
\includegraphics[width=0.495\textwidth]{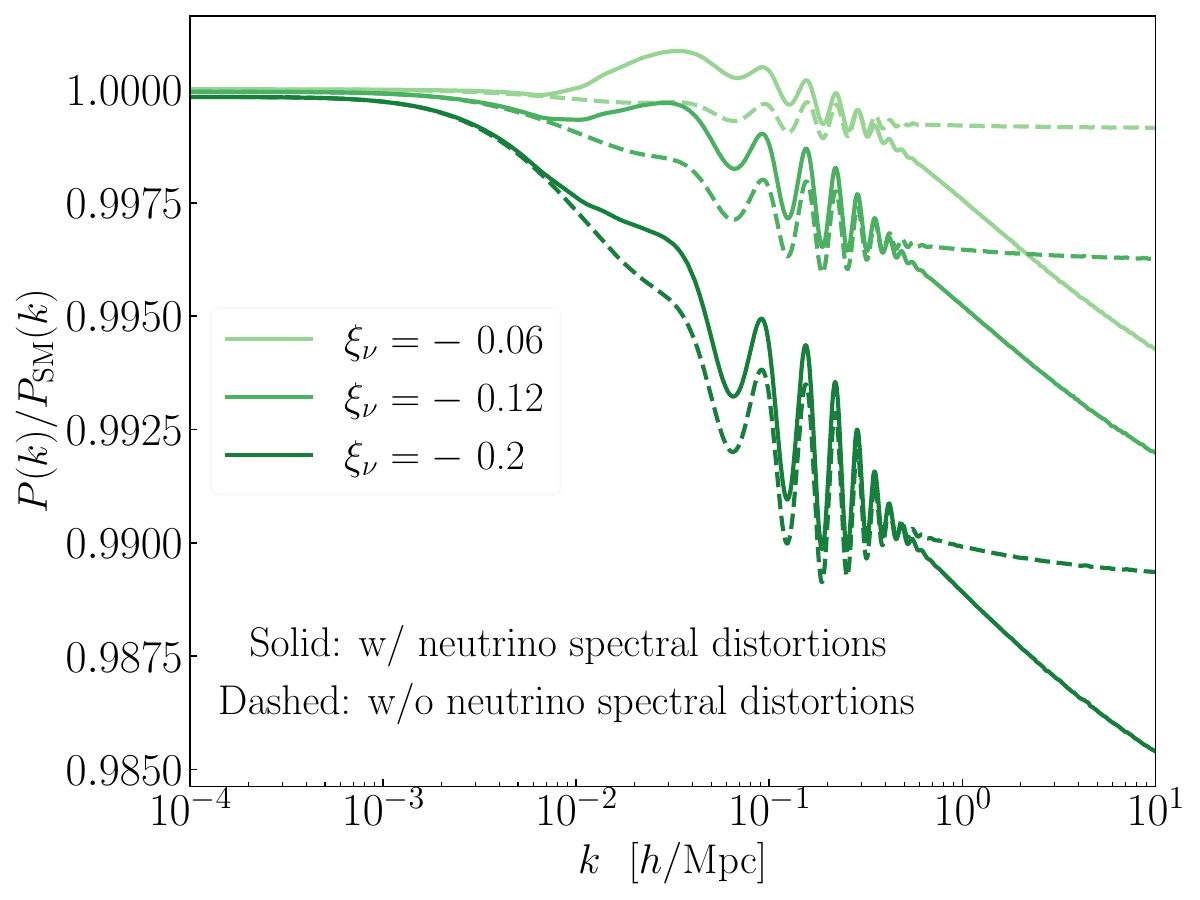} 
\caption{
The CMB angular power spectrum $C_l^{\rm TT}$ (left panel) and the linear matter power spectrum $P(k)$ (right panel) for different neutrino degeneracy parameters $\xi_\nu$, compared to the corresponding results with $\xi_\nu = 0$. The total neutrino mass is fixed to $\sum m_\nu = 0.1 \, \text{eV}$ for both panels. The solid curves are identical to those in Fig.~\ref{fg:bbn_xi}, representing the results including spectral distortions of the neutrinos and antineutrinos, while the dashed curves represent results without spectral distortions.}
\label{fg:lss_dist}
\end{figure}

Finally, we briefly discuss the effects of spectral distortions of neutrinos and antineutrinos on the CMB and LSS. As in Fig.~\ref{fg:lss_xi}, we present in the left and right panels of Fig.~\ref{fg:lss_dist} the CMB angular power spectrum $C_l^{\rm TT}$ and the linear matter power spectrum $P(k)$ for non-zero $\xi_\nu$, compared to the corresponding results with $\xi_\nu = 0$. The solid curves correspond to results that include spectral distortions, as shown in Fig.~\ref{fg:lss_xi}, while the dashed curves correspond to results without the neutrino spectral distortions. Comparing $C_l^{\rm TT}$ with and without spectral distortions, we see that the primary effect of including these distortions is a slight suppression of the first three acoustic peaks, with negligible impact on the smaller-scale peaks. For the linear matter power spectrum, the effect on the BAO is similar to that on the CMB spectrum, with the first two acoustic peaks slightly enhanced while the effects on subsequent peaks remain minimal. 
Furthermore, we find that calculations that exclude spectral distortions fail to accurately capture the behavior of the matter power spectrum at small scales. Without distortions, the results predict a near-constant suppression at $k \gtrsim 1 \, h/{\rm Mpc}$, whereas more precise calculations including distortions show greater suppression at these scales. However, this analysis is limited to the linear matter power spectrum. A comprehensive study of the impact of neutrino spectral distortions on LSS in the nonlinear regime would require future N-body simulations, which is beyond the scope of this paper and left for future work.

\section{Combined Analysis of Primordial Neutrino Asymmetries}\label{sec:mcmc}

In previous sections, we have demonstrated that primordial neutrino asymmetry significantly influences the neutrino decoupling process, leaving observable imprints on the Universe, such as the abundance of light elements produced during BBN, the anisotropies in the CMB, and the BAO features in the LSS. These cosmological observations can be used to constrain primordial neutrino asymmetries. 
In this section, we consolidate the results from previous sections and discuss the constraints on primordial neutrino asymmetries derived from present BBN, CMB, and BAO observations. Furthermore, we briefly explore the associated constraints on UV model parameters by using a specific model that generates primordial neutrino asymmetries in the early universe.

\subsection{Observational Constraints on Primordial Neutrino Asymmetry}

\begin{table}[t]
\begin{center}
{\def\arraystretch{1.4}
\begin{tabular}{|c|cccc|}
\hline
\multicolumn{5}{|c|}{\textbf{Bounds for the model $\mathbf\Lambda$CDM + $\mathbf{\xi_\nu}$ + $\mathbf{m_{\nu_1}}$}}     \\ 
\hline \hline
 \textbf{Data Sets}  & $\qquad \,\,\, {\bf \xi_{\nu}} \,\,\, \qquad$ & $\omega_{b}$ &  $\omega_{cdm}$ & $ {\sum m_{\nu} \,\, (\mathrm{eV}) }$ \\
\hline 
    {\bf BBN}   &  $0.039\pm 0.013$  & $2.155\pm 0.016 $  & -- &  --   \\ 
    {\bf BBN + CMB}   & $0.028 \pm 0.012$   & $2.189\pm 0.012$  & $0.1225\pm 0.0013$ &  0.414   \\ 
    {\bf BBN + CMB + BAO}   & $0.024 \pm 0.012 $   & $2.200\pm 0.011$  & $0.1205\pm 0.0009 $ &  0.131   \\ 
\hline \hline
\multicolumn{5}{|c|}{\textbf{Bounds from the datasets BBN + CMB + BAO}}     \\ 
\hline \hline
\textbf{Models}  & $\qquad \,\,\, {\bf \xi_{\nu}} \,\,\, \qquad$ & $\omega_{b}$ &  $\omega_{cdm}$ & $ {\sum m_{\nu} \,\, (\mathrm{eV}) }$ \\
\hline 
    {\bf $\mathbf\Lambda$CDM + $\mathbf{m_{\nu_1}}$}   &  --  & $2.211\pm 0.010 $  & $0.1203 \pm 0.0009 $&  0.136   \\ 
    {\bf $\mathbf\Lambda$CDM + $\mathbf{\xi_\nu}$}   & $0.023 \pm 0.012$   & $2.200\pm 0.011$  & $0.1207\pm 0.0009$ &  --   \\ 
    {\bf $\mathbf\Lambda$CDM + $\mathbf{\xi_\nu}$ + $\mathbf{m_{\nu_1}}$}   & $0.024 \pm 0.012 $   & $2.200\pm 0.011$  & $0.1205\pm 0.0009 $ &  0.131   \\ 
\hline 
\end{tabular}
}
\end{center}
\vspace{-0.3cm}
\caption{Summary of constraints on the primordial lepton asymmetry $\xi_{\nu}$, the baryon density $\omega_b$, the cold dark matter density $\omega_{cdm}$, and the total neutrino mass $\sum m_\nu$ in normal hierarchy from various datasets and cosmological models.}
\label{tab:CurrentConstraints}
\end{table}

The datasets used for BBN, CMB, and BAO analyses in this work are as follows:
\begin{description}
\item[BBN:] The EMPRESS measurement of helium abundance, $\yp = 0.2370^{+0.0034}_{-0.0033}$~\cite{empress}, and the deuterium abundance from the PDG, $10^5 \times {\rm D/H} = 2.547 \pm 0.025$~\cite{pdg}, based on~\cite{Cooke:2017cwo}.
\item[CMB:] The Planck 2018 low-$l$ and high-$l$ TT, TE, and EE spectra, along with the reconstructed CMB lensing spectrum~\cite{Planck:2018vyg,Planck:2018lbu,Planck:2019nip}.
\item[BAO:] The BOSS DR12~\cite{BOSS:2016wmc} with its full covariance matrix for BAO, as well as the 6dFGS~\cite{Beutler:2011hx} and MGS of SDSS~\cite{Ross:2014qpa}.
\end{description}
It should be noted that the helium abundance from the BBN dataset primarily constrains primordial neutrino asymmetries, while the deuterium abundance is more sensitive to the baryon density $\omega_b$. The BAO dataset primarily constrains the total neutrino mass $\sum m_\nu$, but also provides additional constraints on primordial neutrino asymmetries.

To extract constraints from these datasets, we perform MCMC analyses using the Montepython engine~\cite{Audren:2012wb,Brinckmann:2018cvx}. For the analysis, $\Neff$ is derived from Eq.~\eqref{eq:neff} as discussed in Section~\ref{sec: decoupling}. The predicted helium and deuterium abundances from BBN are calculated using our fitting formulas in Eq.~\eqref{eq:bbnfit}. As shown in \fig\ref{fg:lss_dist}, the impact of neutrino spectral distortions on the CMB angular power spectrum and BAO in the matter power spectrum is minor. Therefore, for simplicity, (anti)neutrino spectral distortions are neglected in the MCMC analysis.
For the CMB and BAO datasets, the parameter set includes the six independent parameters of the $\Lambda$CDM model, along with the neutrino degeneracy parameter $\xin$ and the mass of the lightest neutrino in the normal hierarchy $m_{\nu_1}$, i.e., $\{\omega_b, \omega_{cdm}, \theta_s, A_s, n_s, \tau_\mathrm{reio}, \xi_\nu, m_{\nu_1}\}$. Here, $\theta_s$ is the sound horizon, $A_s$ is the amplitude of primordial curvature perturbations at the pivot scale $k=0.05\,\mathrm{Mpc}^{-1}$, $n_s$ is the scalar spectral index, and $\tau_\mathrm{reio}$ is the optical depth to reionization. For the BBN dataset, we consider only $\{\omega_b, \xi_\nu\}$, as the predicted light element abundances from BBN are largely independent of other cosmological parameters within the relevant range.

\begin{figure}\centering
\includegraphics[width=0.97\textwidth]{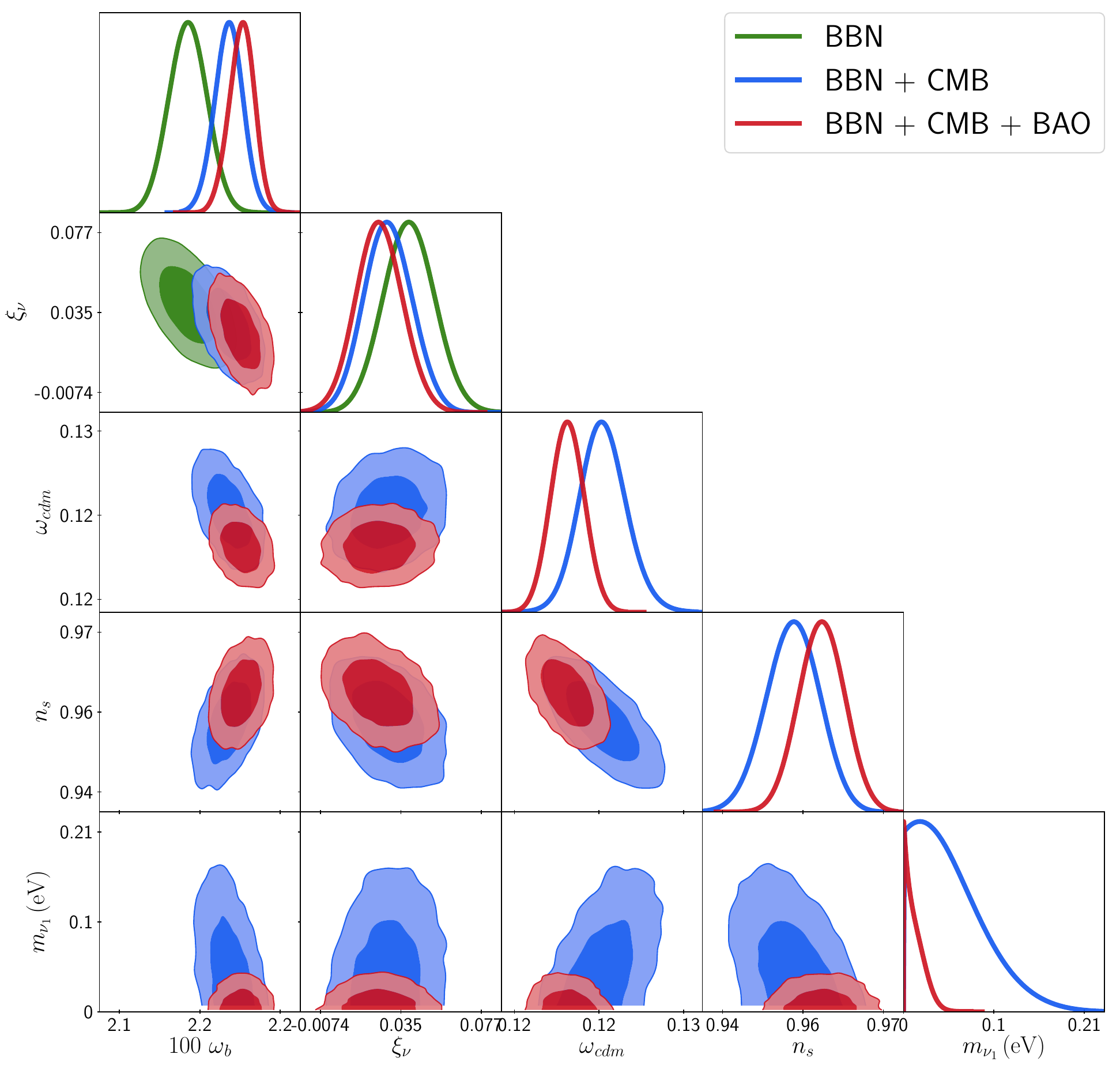} 
\caption{2D and 1D marginalized posterior distributions for the baryon abundance $\omega_b$, the neutrino degeneracy parameter $\xi_\nu$, the cold dark matter abundance $\omega_{cdm}$,  the scalar spectral index $n_s$, and the mass of the lightest neutrino for the normal hierarchy $m_{\nu_1}$. The green, blue and red regions represent constraints from BBN, CMB + BAO, and their combination, respectively.}
\label{fg:mcmc_data}
\end{figure}

\begin{figure}\centering
\includegraphics[width=0.97\textwidth]{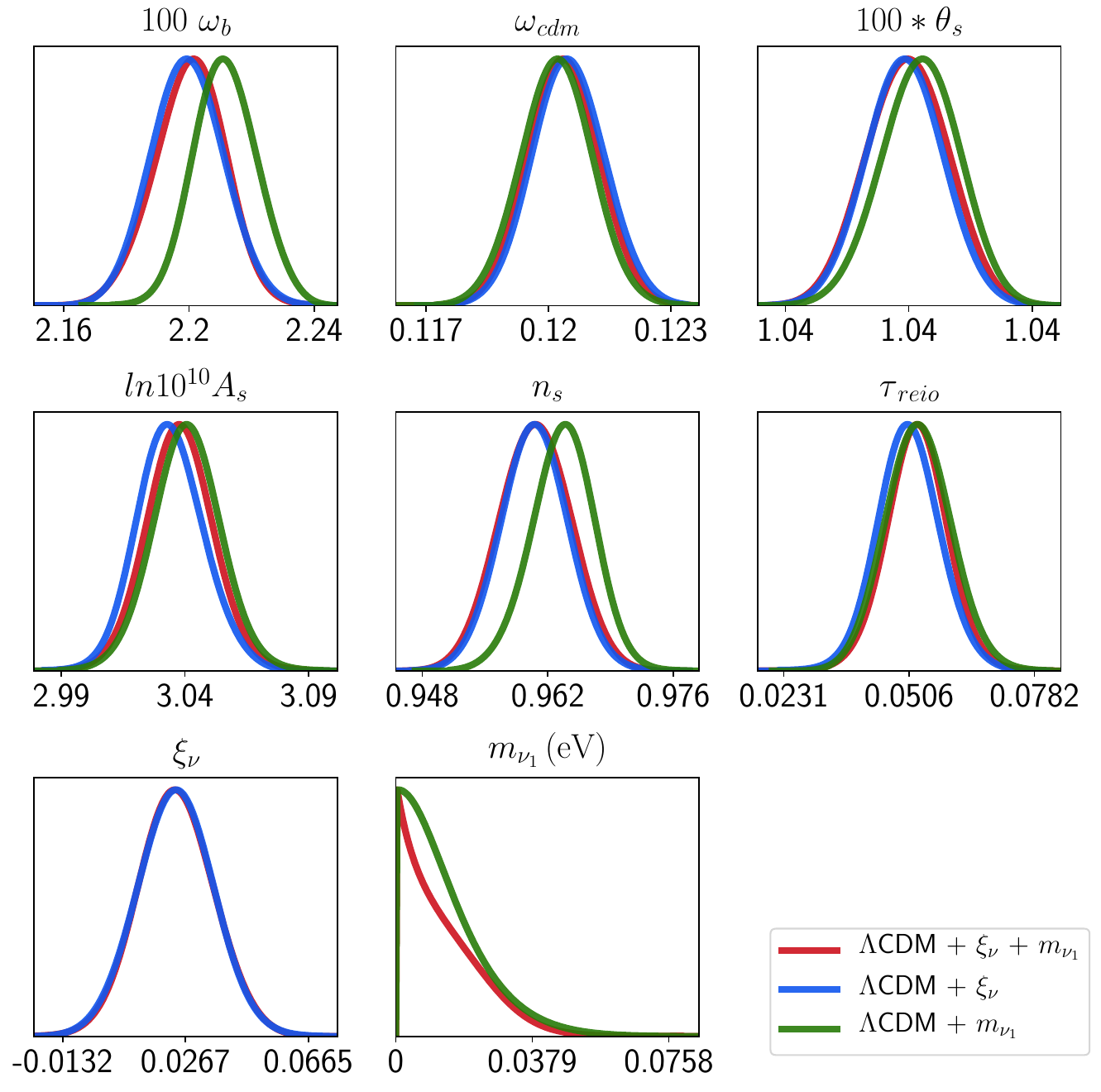} 
\caption{1D marginalized distributions for different parameter sets derived from the BBN + CMB + BAO datasets.}
\label{fg:mcmc_param}
\end{figure}

Using the specified parameters and datasets, we modified the MontePython sampler and CLASS codes to perform the MCMC analysis. All chains converged with a Gelman-Rubin criterion of $R-1 < 0.01$. The resulting $1\,\sigma$ cosmological constraints are summarized in Table~\ref{tab:CurrentConstraints}, where the $95\%$ confidence limits on the total neutrino mass are also presented, derived from constraints on the mass of the lightest neutrino for the normal hierarchy, $m_{\nu_1}$. 
For the model $\Lambda$CDM + ${\xi_\nu}$ + ${m_{\nu_1}}$, we present the 1D and 2D marginalized posterior distributions of key parameters that show significant variations across different datasets in Fig.~\ref{fg:mcmc_data}, including the baryon abundance $\omega_b$, the neutrino degeneracy parameter $\xi_\nu$, the cold dark matter abundance $\omega_{cdm}$, the sound horizon $\theta_s$, and the mass of the lightest neutrino for the normal hierarchy $m_{\nu_1}$. 

Regarding constraints on primordial neutrino asymmetries, the EMPRESS data in the BBN dataset indicates a positive primordial neutrino asymmetry, $\xi_\nu=0.039 \pm 0.013$, which differs from zero at $3.0 \, \sigma$ significance. Combining BBN and CMB data significantly reduces the constraint to $\xi_\nu=0.028 \pm 0.012$, which differs from zero at $2.3 \, \sigma$ significance. Compared to previous results ($\xi_\nu=0.034 \pm 0.014$) using the same BBN and CMB datasets~\cite{Escudero:2022okz}, our results show an 18\% reduction in the central value and a slight decrease in uncertainty. Incorporating the BOSS BAO observations further reduces the constraint on primordial neutrino asymmetry to $\xi_\nu = 0.024 \pm 0.012$, representing a 29\% reduction in the central value relative to previous findings, while the significance of a positive neutrino asymmetry decreases to $2.0 \, \sigma$. Our analysis highlights the additional constraining power of BAO experiments on primordial neutrino asymmetries, which was not fully explored in previous studies.

We find that constraints on other parameters, such as $\omega_b$, $\omega_{cdm}$, $n_s$, and $m_{\nu_1}$, also vary significantly depending on the dataset used, as shown in Fig.~\ref{fg:mcmc_data}. For $\omega_b$, the discrepancy is mainly attributed to the choice of nuclear reaction rates in the {\tt PRIMAT} code during the BBN process, which predicts a deuterium abundance lower than current constraints from PDG, leading to a significantly lower $\omega_b$~\cite{Pitrou:2018cgg,Burns:2022hkq,Escudero:2022okz}. While constraints on the primordial neutrino asymmetry $\xi_\nu$ do not strongly depend on the choice of nuclear reaction rates, as discussed in~\cite{Escudero:2022okz}, a detailed study using different nuclear reaction rates will be presented in our companion paper \cite{Li:2025rjr}. The inclusion of BOSS BAO data significantly modifies the constraints on $\omega_{cdm}$, $n_s$, and $m_{\nu_1}$ due to its sensitivity to the evolution of the late Universe. In particular, the $95\%$ upper limit on the total neutrino mass tightens from $0.414$ eV (BBN + CMB) to $0.131$ eV (BBN + CMB + BAO), as BAO data help break the CMB degeneracy. Recently, DESI BAO observations have placed even stronger constraints on neutrino masses, $\sum m_{\nu_i} \lesssim 0.07 \, \mathrm{eV}$~\cite{DESI:2024mwx}. Relevant studies on primordial neutrino asymmetry and neutrino masses using DESI BAO data will be addressed in our companion paper \cite{Li:2025rjr}.

Finally, we briefly discuss the impact of different parameter sets on the resulting constraints, particularly the interplay between primordial neutrino asymmetry and total neutrino mass. Using the BBN + CMB + BAO dataset, we present in Fig.~\ref{fg:mcmc_param} the 1D marginalized distributions for three parameter sets: $\Lambda$CDM + $m_{\nu_1}$, $\Lambda$CDM + $\xi_\nu$, and $\Lambda$CDM + $m_{\nu_1}$ + $\xi_\nu$. As shown, while different parameter sets lead to significant modifications in constraints on the six $\Lambda$CDM parameters, especially $\omega_b$ and $n_s$, the resulting constraints on additional parameters are consistent across different parameter sets. The corresponding constraints from different parameter sets are also summarized in Table~\ref{tab:CurrentConstraints}.
We find that the constraints on the total neutrino mass are only slightly tightened when $\xi_\nu$ is included as a free parameter, while constraints on $\xi_\nu$ remain nearly unchanged when $\sum m_\nu$ is a free parameter. Therefore, we conclude that constraints on neutrino masses and primordial neutrino asymmetries can be analyzed independently if one is not concerned with constraints on other $\Lambda$CDM parameters. However, if constraints on other $\Lambda$CDM parameters are of interest, a parameter set including both of neutrino mass and primordial neutrino asymmetry must be considered.

\subsection{Constraints on UV Model Parameters}

In this subsection, we discuss the constraints on UV model parameters based on the constraints on primordial neutrino asymmetries. Specifically, we consider a Q-ball decay model~\cite{Kawasaki:2022hvx}, where Q-balls are produced through the Affleck-Dine mechanism~\cite{Kusenko:1997si,Enqvist:1997si,Kasuya:1999wu}, which operates along flat directions in the minimal supersymmetric standard model (MSSM). When a flat direction associated with lepton number reaches a large field value during inflation, it generates a lepton number via field dynamics after inflation. At the same time, Q-balls form, and the generated lepton number is confined within them. These Q-balls, referred to as L-balls, possess a large lepton number. Since the sphaleron process is inactive inside L-balls, the lepton asymmetry generated by the Affleck-Dine mechanism is preserved without conversion to a baryon asymmetry~\cite{Kawasaki:2002hq,Gelmini:2020ekg}. After the electroweak scale, the L-balls decay into neutrinos, resulting in a significant lepton asymmetry in the universe.

Following~\cite{Kawasaki:2022hvx}, the primordial lepton asymmetry $\eta_L$ generated by the decay of Q-balls can be written as\footnote{The difference between the expression here and the expression in ~\cite{Kawasaki:2022hvx} is due to the different definition of the primordial lepton asymmetry $\eta_L$.}
\begin{equation}
    \eta_L = \frac{3 \pi^4 g_\star T_\mathrm{D}}{495 \zeta(3) m_{3/2}},
    \label{eq: lepton asymmetry for delayed Q-ball}
\end{equation}
where $g_\star = 10.75$ is the relativistic degree of freedom at neutrino decoupling, $m_{3/2}$ is the gravitino mass and $T_\mathrm{D}$ is the L-balls' decay temperature, which is a function of other UV model parameters:
\begin{align}
    T_\mathrm{D} 
    &\simeq
    2.69~\mathrm{MeV} \,
\nonumber\\
    & \hspace{15pt} \times 
    \left( \frac{g_*}{10.75} \right)^{-1/4}
    \left( \frac{\beta}{6 \times 10^{-4}} \right)^{-5/8}
    \left( \frac{m_{3/2}}{0.5~\mathrm{GeV}} \right)^{5/2}
    \left( \frac{M_F}{5 \times 10^6~\mathrm{GeV}} \right)^{-2}
    \left( \frac{N_\ell}{3} \right)^{1/2}
    \left( \frac{\zeta}{2.5} \right)^{1/2},
    \label{eq: decay temperature formula for delayed}
\end{align}
where $g_*(T_\mathrm{D})$ is the relativistic degrees of freedom at $T_\mathrm{D}$, $\beta \simeq 6 \times 10^{-4}$ is a dimensionless constant, $M_F$ is the SUSY breaking scale, and $N_\ell$ represents the number of decay channels. Hence, the cosmological constraints on lepton asymmetry can be translated into constraints on the UV model parameters.

Our analysis of the BBN + CMB + BAO datasets yields a constraint on the lepton asymmetry of $ \eta_L = 0.0025 \pm 0.0013$. The corresponding constraint on the L-balls' decay temperature is:
\begin{align}
    T_\mathrm{D}
    &=
    \frac{4}{3} \eta_L m_{3/2} \simeq
    3.3~\mathrm{MeV} \,
    \left( \frac{m_{3/2}}{0.5~\mathrm{GeV}} \right)
    \left( \frac{\eta_L}{5 \times 10^{-3}} \right)
\nonumber\\
    &
    \simeq
    \left( \frac{m_{3/2}}{0.5~\mathrm{GeV}} \right) \, (1.65 \pm 0.86) ~\mathrm{MeV}   \quad \, [\textbf{BBN+CMB+BAO}],
\end{align}
and, conversely,
\begin{align}
    m_{3/2}
    &=
    \frac{3 T_\mathrm{D}}{4 T\eta_L} \simeq
    0.5~\mathrm{GeV} \,
    \left( \frac{T_\mathrm{D}}{3.3~\mathrm{MeV}} \right)
    \left( \frac{5 \times 10^{-3}}{\eta_L} \right)
\nonumber\\
    &
    \simeq
    \left( \frac{T_\mathrm{D}}{3.3~\mathrm{MeV}} \right) (1.00^{+1.08}_{-0.34})~\mathrm{GeV}  \quad \, [\textbf{BBN+CMB+BAO}].
\end{align}
These constraints can be further translated into constraints on other UV model parameters, such as the SUSY breaking scale, as shown in~\cite{Kawasaki:2022hvx}.



\section{Conclusion}\label{sec:conclusion}

The existence of primordial neutrino asymmetries prior to the neutrino decoupling epoch has significant implications for the neutrino decoupling process and subsequent cosmological phenomena. These asymmetries can therefore be constrained through observations of BBN, CMB, and LSS. However, the effects of primordial neutrino asymmetries on neutrino decoupling, particularly with respect to the resulting $\Neff$ and spectral distortions of (anti)neutrinos—including the impacts of flavor oscillations and FTQED corrections-remain uncertain. These uncertainties make it difficult to interpret subsequent cosmological processes accurately. Consequently, studies of primordial neutrino asymmetries often rely on approximations, such as the use of thermal Fermi-Dirac distributions for (anti)neutrinos, without rigorous verification.

In this work, we address these ambiguities by numerically solving the neutrino decoupling process and analyzing the resulting (anti)neutrino density matrices in the context of BBN, CMB, and LSS. We rederive the quantum kinetic equations (QKEs) for neutrinos and antineutrinos using the Closed Time Path (CTP) formalism for two-point functions. 
{  Focusing on the complete flavor equilibration cases with $\xi_{\nu_e} = \xi_{\nu_\mu} = \xi_{\nu_\tau} = \xin$}
, we numerically solve the QKEs in the early universe using an extended version of the public code {\tt FortEPiaNO}~\cite{Gariazzo:2019gyi,Bennett:2020zkv}. Our calculations include the effects of neutrino oscillations, matter interactions, FTQED corrections up to {  $\mathcal{O}(e^3)$}, and comprehensive neutrino-electron and neutrino-neutrino collision terms. The final $\Neff$ is then computed based on the resulting density matrices and can be expressed as $\Neff \simeq \Neff^{\rm SM} + 3 \left( \frac{30}{7 \pi^2} \xi_{\nu}^2 + \frac{15}{7 \pi^4} \xi_{\nu}^4 \right) + 0.0102 \xi_{\nu}^2 $, 
where $\Neff^{\rm SM} = 3.0440 \pm 0.0002$. The second and third terms represent corrections from nonzero $\xin$ in the instantaneous decoupling limit and non-instantaneous effects, respectively. We also confirm that spectral distortions of neutrinos and antineutrinos with nonzero $\xin$ differ due to distinct Pauli-blocking effects, in agreement with preliminary studies~\cite{Grohs:2016cuu}. {  The formalism we develop can also be straightforwardly applied to the study of flavor equilibration processes, see \cite{Pastor:2008ti,Mangano:2011ip,Froustey:2024mgf} for previous relevant discussions. }

Using the neutrino and antineutrino density matrices, we further investigate the effects of primordial neutrino asymmetries on BBN, CMB, and LSS. For BBN, we introduce a novel correction to the weak interaction rates arising from distinct neutrino and antineutrino spectral distortions. We calculate the light element abundances using the public BBN code PRIMAT~\cite{Pitrou:2018cgg}, incorporating additional corrections to weak rates. According to the EMPRESS measurement of the helium abundance $Y_P |_{\rm EMPRESS} = 0.2370^{+0.0034}_{-0.0033}$, our results suggest a positive primordial neutrino asymmetry in the range $0.032 \leq \xin \leq 0.052$, assuming a fixed baryon abundance $\omega_b=0.024$. 
Our analysis also highlights the importance of spectral distortions for precise BBN predictions, particularly for Helium-4 abundance. Compared to results in the instantaneous limit, the traditional treatment using Fermi-Dirac distributions predicts a negative relative difference for positive primordial neutrino asymmetry, whereas our precise treatment with spectral distortions shows a positive relative difference. Additionally, we provide semi-analytic expressions for these abundances as functions of $\xin$, which will be valuable for future studies.
For CMB and LSS, our study reveals that primordial neutrino asymmetries influence the amplitudes of acoustic oscillations in both the CMB and baryon acoustic oscillations (BAO), suggesting that BAO observations can serve as a tool for constraining these asymmetries.

To further constrain primordial asymmetries, we perform a joint MCMC analysis including the neutrino degeneracy parameter $\xin$ and the total neutrino mass $\sum m_\nu$, using data from EMPRESS BBN, Planck CMB, and BOSS BAO observations. Combining the EMPRESS BBN and Planck CMB datasets, we constrain the primordial neutrino asymmetry to $\xin = 0.028 \pm 0.012$, a reduction of $\sim 18\%$ in the central value and a slight reduction in the uncertainty compared to previous results of $\xin = 0.034 \pm 0.014$~\cite{Escudero:2022okz}. Incorporating BOSS BAO observations further refines the constraint to $\xin = 0.024 \pm 0.012$, yielding a $\sim 29\%$ reduction in the central value and a $2.0 \sigma$ significance for a positive neutrino asymmetry. We also discuss the interplay between $\sum m_\nu$ and $\xin$ when both are treated as free parameters, and how these constraints translate into bounds on UV model parameters, using a specific model where large neutrino asymmetries are generated by Q-ball decay~\cite{Kawasaki:2002hq,Gelmini:2020ekg}.

Our work establishes a framework for exploring the implications of physics beyond the Standard Model and $\Lambda$CDM for neutrino decoupling and related cosmological processes, providing corresponding constraints from current and future BBN, CMB, and LSS observations. It paves the way for more precise studies of primordial neutrino asymmetries and other new physics scenarios, such as synchronous neutrino oscillations~\cite{Froustey:2021azz,Froustey:2024mgf} and non-standard neutrino interactions~\cite{Du:2021idh,Du:2023upj}, driven by increasingly precise cosmological observations.

\acknowledgments

We thank J. Froustey and C. Pitrou for helpful communications about their relevant work.
This work is supported by the NSFC under Grants No. 12347105, No. 12375099 and No. 12447101, and the National Key Research and Development Program of China Grant No. 2020YFC2201501, No. 2021YFA0718304. The authors gratefully acknowledge the use of publicly available codes {\tt FortEPiaNO}~\cite{Gariazzo:2019gyi,Bennett:2020zkv}, {\tt PRIMAT}~\cite{Pitrou:2018cgg}, {\tt CLASS}~\cite{Lesgourgues:2011re,Lesgourgues:2011rh} and {\tt MontePython}\cite{Audren:2012wb,Brinckmann:2018cvx}.

\appendix
\section{Collision Terms for Neutrinos and Antineutrinos} \label{sec: collision}
\subsection{Complete expressions for the collision terms}

In this work, we consider the interaction processes of neutrino-electron/positron scattering
($\nu{} e\leftrightarrow \nu{} e$)
and annihilation ($\nu\bar\nu\leftrightarrow e^+e^-$),
plus neutrino-neutrino self-interactions ($\nu\nu\leftrightarrow \nu\nu$, $\nu\bar\nu\leftrightarrow \nu\bar\nu$). Therefore, the total collision term for neutrinos $\mathcal{I}$ can be decomposed as
\begin{equation}
    \mathcal{I}
	\equiv
	\mathcal{I}_{\rm sc}
	+ \mathcal{I}_{\rm ann}
	+ \mathcal{I}_{\nu\nu}
\end{equation}
where $\mathcal{I}_{\rm sc}$ and $\mathcal{I}_{\rm ann}$ represent the collision terms for neutrino-electron/positron scattering and annihilation respectively, while $\mathcal{I}_{\nu\nu}$ represents the collision terms for neutrino self-interactions.
The collision term for antineutrinos $\bar{\mathcal{I}}$ also admits a similar decomposition. 
And the contributions from other interaction processes such as $\mu^\pm$ scattering and annihilation are neglected, as their contributions are highly suppressed at the temperature of neutrino decoupling.

In the following we present the explicit form of the collision terms, which are also presented in various studies (see \eg~\cite{Blaschke:2016xxt}).
The collision terms for the neutrino-electron/positron scattering processes ($\nu{} e\leftrightarrow \nu{} e$) and the annihilation processes ($\nu\bar\nu\leftrightarrow e^+e^-$) take the form as
\begin{align}\label{eq:C_sc_ann_full}
\mathcal{I}_{\rm sc} = & \frac12 \frac{2^5 G_F^2}{2 p_1}\int{\dd \pi_2 \, \dd \pi_3 \, \dd \pi_4 \,  (2 \pi)^4 \delta^{(4)}(p_1 + p_2 - p_3 - p_4)}  \nonumber\\ 
&\Big[ 4 (p_1 \cdot p_2)(p_3 \cdot p_4) \left(F_\mathrm{sc}^{LL}(\nu^{(1)},e^{(2)},\nu^{(3)},e^{(4)}) + F_\mathrm{sc}^{RR}(\nu^{(1)},\bar{e}^{(2)},\nu^{(3)},\bar{e}^{(4)})\right)  \nonumber\\
&+ 4 (p_1 \cdot p_4)(p_2 \cdot p_3) \left(F_\mathrm{sc}^{RR}(\nu^{(1)},e^{(2)},\nu^{(3)},e^{(4)}) + F_\mathrm{sc}^{LL}(\nu^{(1)},\bar{e}^{(2)},\nu^{(3)},\bar{e}^{(4)}) \right) \nonumber\\
&- 2 (p_1 \cdot p_3) m_e^2 \left(F_\mathrm{sc}^{LR}(\nu^{(1)},e^{(2)},\nu^{(3)},e^{(4)}) + F_\mathrm{sc}^{LR}(\nu^{(1)},\bar{e}^{(2)},\nu^{(3)},\bar{e}^{(4)}) + \{L \leftrightarrow R \}\right) \Big] \, , \\
\mathcal{I}_{\rm ann} = & \frac12 \frac{2^5 G_F^2}{2 p_1}\int{\dd \pi_2 \, \dd \pi_3 \, \dd \pi_4 \,  (2 \pi)^4 \delta^{(4)}(p_1 + p_2 - p_3 - p_4)} \nonumber\\
&\Big[ 4 (p_1 \cdot p_4)(p_2 \cdot p_3) F_\mathrm{ann}^{LL}(\nu^{(1)},\bar{\nu}^{(2)},e^{(3)},\bar{e}^{(4)})  \nonumber\\
&+ 4 (p_1 \cdot p_3)(p_2 \cdot p_4) F_\mathrm{ann}^{RR}(\nu^{(1)},\bar{\nu}^{(2)},e^{(3)},\bar{e}^{(4)}) \nonumber\\
&+ 2 (p_1 \cdot p_2) m_e^2 \left(F_\mathrm{ann}^{LR}(\nu^{(1)},\bar{\nu}^{(2)},e^{(3)},\bar{e}^{(4)}) + F_\mathrm{ann}^{RL}(\nu^{(1)},\bar{\nu}^{(2)},e^{(3)},\bar{e}^{(4)}) \right) \Big] \, .
\end{align}
where $\dd \pi_i = \frac{\\dd^3 p_i}{(2 \pi)^3 2 E_i}$ with $E_i = p_i$ if the particle $i$ is an (anti)neutrino or $E_i = \sqrt{p_i^2 + m_e^2}$ if the particle $i$ is an electron/positron.
The statistical factors incorporating the density distributions for all participating particles are 
\begin{align}
    F_\mathrm{sc}^{AB}(\nu^{(1)},e^{(2)},\nu^{(3)},e^{(4)}) &= f_e^{(4)}(1-f_e^{(2)})\left[G^A\varrho^{(3)}G^B(1-\varrho^{(1)})+(1-\varrho^{(1)})G^B\varrho^{(3)}G^A\right]  \nonumber\\
    & - f_e^{(2)}(1-f_e^{(4)})\left[\varrho^{(1)}G^B(1-\varrho^{(3)})G^A+G^A(1-\varrho^{(3)})G^B\varrho^{(1)}\right] \, ,  \label{eq:F_ab_sc}
\end{align}
\begin{align}
    F_\mathrm{ann}^{AB}(\nu^{(1)},\bar{\nu}^{(2)},e^{(3)},\bar{e}^{(4)}) &= f_e^{(3)} \bar{f}_e^{(4)}\left[G^A(1-\bar{\varrho}^{(2)})G^B(1-\varrho^{(1)})+(1-\varrho^{(1)})G^B(1-\bar{\varrho}^{(2)})G^A\right]
    \nonumber\\
    &- (1-f_e^{(3)})(1-\bar{f}_e^{(4)})\left[G^A\bar{\varrho}^{(2)}G^B\varrho^{(1)}+\varrho^{(1)}G^B\bar{\varrho}^{(2)}G^A\right] \, , \label{eq:F_ab_ann}
\end{align}
where $a, b \in \{L, R\}$ denote the charalitics for the electrons/positrons involved in the processes, $f(\bar{f})_e^{(i)} = 1/(\exp{p_i/T_\gamma}+1)$ are the distribution functions for electrons/positrons, and $\varrho^{(i)}, \bar{\varrho}^{(i)}$ are the corresponding density matrices for neutrinos and antineutrinos with momentum $p_i$.
The interaction matrices used in the collision terms are
\begin{equation}
	\label{eq:gLR}
	G^L=\text{diag}(g_L, \tilde g_L, \tilde g_L)\,,
	\qquad
	G^R=\text{diag}(g_R, g_R, g_R)\,,
\end{equation}
where $g_L=\sin^2\theta_W+1/2$, $\tilde g_L=\sin^2\theta_W - 1/2$, $g_R=\sin^2\theta_W$,
and $\theta_W$ is the weak mixing angle. 

The collision term for the neutrino self-interactions, including the scattering between neutrinos and (anti)neutrinos ($\nu\nu\leftrightarrow \nu\nu$, $\nu\bar\nu\leftrightarrow \nu\bar\nu$) and the annihilation between a pair of neutrinos and antineutrinos ($\nu\bar\nu\leftrightarrow \nu\bar\nu$), can be written as
\begin{equation}
\label{eq:C_nn}
\begin{aligned}
\mathcal{I}_{\nu  \nu} = &\frac12 \frac{2^5 G_F^2}{2 p_1} \int{\dd \pi_2 \, \dd \pi_3 \, \dd \pi_4 \, (2 \pi)^4 \delta^{(4)}(p_1 + p_2 - p_3 - p_4)} \\
&\Big[ (p_1 \cdot p_2)(p_3 \cdot p_4) F_\mathrm{sc}(\nu^{(1)},\nu^{(2)},\nu^{(3)},\nu^{(4)})  \\
&+ (p_1 \cdot p_4)(p_2 \cdot p_3) \left( F_\mathrm{sc}(\nu^{(1)},\bar{\nu}^{(2)},\nu^{(3)},\bar{\nu}^{(4)}) + F_\mathrm{ann}(\nu^{(1)},\bar{\nu}^{(2)},\nu^{(3)},\bar{\nu}^{(4)}) \right) \Big] \, ,
\end{aligned} 
\end{equation}
with the statistical factors for scattering and annihilation processes:
\begin{multline}
\label{eq:F_sc_nn}
F_\mathrm{sc}(\nu^{(1)},\nu^{(2)},\nu^{(3)},\nu^{(4)}) =  \\
\left[ \varrho^{(4)} (\Id - \varrho^{(2)}) + \Tr(\cdots) \right] \varrho^{(3)} (\Id -\varrho^{(1)}) + (\Id - \varrho^{(1)}) \varrho^{(3)} \left[ (\Id - \varrho^{(2)}) \varrho^{(4)} + \Tr(\cdots)\right]  \\
- \left[ (\Id - \varrho^{(4)}) \varrho^{(2)}  + \Tr(\cdots)\right] (\Id -\varrho^{(3)})  \varrho^{(1)} - \varrho^{(1)}  (\Id -\varrho^{(3)})  \left[\varrho^{(2)}(\Id -\varrho^{(4)})  + \Tr(\cdots)\right]  \, ,
\end{multline}
\begin{multline}
\label{eq:F_sc_nbn}
F_\mathrm{sc}(\nu^{(1)},\bar{\nu}^{(2)},\nu^{(3)},\bar{\nu}^{(4)}) = \\
\left[ (\Id - \bar{\varrho}^{(2)}) \bar{\varrho}^{(4)} + \Tr(\cdots) \right] \varrho^{(3)} (\Id -\varrho^{(1)}) + (\Id - \varrho^{(1)}) \varrho^{(3)} \left[ \bar{\varrho}^{(4)} (\Id - \bar{\varrho}^{(2)}) + \Tr(\cdots)\right]  \\
- \left[ \bar{\varrho}^{(2)} (\Id -\bar{\varrho}^{(4)}) + \Tr(\cdots) \right] (\Id -\varrho^{(3)}) \varrho^{(1)} - \varrho^{(1)} (\Id -\varrho^{(3)}) \left[ (\Id -\bar{\varrho}^{(4)}) \bar{\varrho}^{(2)} + \Tr(\cdots)\right] \, ,
\end{multline}
\begin{multline}
\label{eq:F_ann_nn}
F_\mathrm{ann}(\nu^{(1)},\bar{\nu}^{(2)},\nu^{(3)},\bar{\nu}^{(4)}) = \\
\left[ \varrho^{(3)} \bar{\varrho}^{(4)} + \Tr(\cdots) \right] (\Id -\bar{\varrho}^{(2)}) (\Id -\varrho^{(1)}) + (\Id - \varrho^{(1)}) (\Id -\bar{\varrho}^{(2)}) \left[ \bar{\varrho}^{(4)} \varrho^{(3)} + \Tr(\cdots)\right]  \\
- \left[ (\Id -\varrho^{(3)}) (\Id -\bar{\varrho}^{(4)}) + \Tr(\cdots) \right] \bar{\varrho}^{(2)} \varrho^{(1)} - \varrho^{(1)} \bar{\varrho}^{(2)} \left[ (\Id -\bar{\varrho}^{(4)}) (\Id -\varrho^{(3)}) + \Tr(\cdots)\right] \, ,
\end{multline}
where, as before, $\varrho^{(i)}, \bar{\varrho}^{(i)}$ are the corresponding density matrices for neutrinos and antineutrinos with momentum $p_i$, and $\Tr(\cdots)$ means the trace of the term in front of it.

Finally, the collision terms for the antineutrinos $\bar{\mathcal{I}}
	\equiv
	\bar{\mathcal{I}}_{\rm sc}
	+ \bar{\mathcal{I}}_{\rm ann}
	+ \bar{\mathcal{I}}_{\nu\nu}$ can be obtained simply by the transformations $\vrho \leftrightarrow \bvrho$ and $L \leftrightarrow R$.

\subsection{Reduced expressions for the comoving collision terms}
The comoving collision terms $\tilde{\mathcal{I}}$ in \eqref{eq:drho_dx} are just the above collision terms expressed with the comoving momentum and comoving energies differing by a total factor.
In addition, for computational convenience, the nine-dimensional integrals in the collision terms are usually simplified to two-dimensional integrals, using the homogeneous and isotropic condition and the special form of the scattering amplitudes~\cite{Hannestad:1995rs,Dolgov:1997mb,Grohs:2015tfy,Blaschke:2016xxt}.
As a result, the collision term can be simplified as
\begin{eqnarray}
	\tilde{\mathcal{I}}
	&=&
	\frac{G_F^2}{(2\pi)^3y^2} \left( \tilde{\mathcal{I}}_{\rm sc} 
	+ \tilde{\mathcal{I}}_{\rm ann} 
	+ \tilde{\mathcal{I}}_{\nu\nu}  \right)  \,,
	\label{eq:collint}
	\\
	\tilde{\mathcal{I}}_{\rm sc}  
	&=&
	\int {\rm d}y_2 {\rm d}y_3 \frac{y_2}{E_2}
	\label{eq:I_sc}
	\\
	&&
	\left\{\left(\Pi_2^s(y, y_4)+\Pi_2^s(y, y_2)\right)
	\left[
	F_\mathrm{sc}^{LL}(\nu^{(1)},e^{(2)},\nu^{(3)},e^{(4)}) + F_\mathrm{sc}^{RR}(\nu^{(1)},\bar{e}^{(2)},\nu^{(3)},\bar{e}^{(4)})\right]\right.
	\nonumber\\
	&&\left.-2(x^2+\delta m_e^2)\Pi_1^s(y,y_3)
	\left[
	F_\mathrm{sc}^{LR}(\nu^{(1)},e^{(2)},\nu^{(3)},e^{(4)}) + F_\mathrm{sc}^{LR}(\nu^{(1)},\bar{e}^{(2)},\nu^{(3)},\bar{e}^{(4)})
	\right]\nonumber
	\right\}\,,
	\\
	\tilde{\mathcal{I}}_{\rm ann} 
	&=&
	\int {\rm d}y_2 {\rm d}y_3 \frac{y_3}{E_3}
	\label{eq:I_ann}\\
	&&\left\{\Pi_2^a(y, y_4)F_{\rm ann}^{LL}(\nu^{(1)},\bar{\nu}^{(2)},e^{(3)},\bar{e}^{(4)})
	+\Pi_2^a(y, y_3)F_{\rm ann}^{RR}(\nu^{(1)},\bar{\nu}^{(2)},e^{(3)},\bar{e}^{(4)})\right.
	\nonumber\\
	&&\left.+ (x^2+\delta m_e^2)\Pi_1^a(y,y_2)
	\left[
	F_{\rm ann}^{RL}(\nu^{(1)},\bar{\nu}^{(2)},e^{(3)},\bar{e}^{(4)})
	+F_{\rm ann}^{LR}(\nu^{(1)},\bar{\nu}^{(2)},e^{(3)},\bar{e}^{(4)})
	\right]\nonumber
	\right\}\,,
	\\
	\tilde{\mathcal{I}}_{\nu\nu}  
	&=&
	\frac{1}{4}
	\int {\rm d}y_2 {\rm d}y_3\, \\
         &&\left\{
	\Pi_2^\nu(y, y_2)
	F_{\nu\nu}(\nu^{(1)},\bar{\nu}^{(2)},\nu^{(3)},\bar{\nu}^{(4)})
         + \Pi_2^\nu(y, y_4)
	F_{\nu\bar\nu}(\nu^{(1)},\bar{\nu}^{(2)},\nu^{(3)},\bar{\nu}^{(4)})
         \right\}   \nonumber
	\,,
	\label{eq:I_nunu}
\end{eqnarray}
where $E^2_i = \sqrt{x^2+y_i^2+\delta m_e^2}$ is the comoving energy for electrons/positrons taking into account the mass contribution induced by FTQED corrections (see Appendix \ref{sec: FTQED} for more details) and the collision kernels in the collision terms are
\begin{eqnarray}
	\Pi_1^s(y,y_3)
	&=&
	y\,y_3\,D_1+D_2(y,y_3,y_2,y_4),
	\\
	\Pi_1^a(y,y_2)
	&=&
	y\,y_2\,D_1-D_2(y,y_2,y_3,y_4),
	\\
	\Pi_2^s(y,y_2)/2
	&=&
	y\,E_2\,y_3\,E_4\,D_1 + D_3 - y\,E_2 D_2(y_3,y_4,y,y_2) \\
 && \nonumber - y_3\,E_4 D_2(y,y_2,y_3,y_4),
	\\
	\Pi_2^s(y,y_4)/2
	&=&
	y\,E_2\,y_3\,E_4\,D_1 + D_3 + E_2\,y_3 D_2(y,y_4,y_2,y_3) \\
 && \nonumber + y\,E_4 D_2(y_2,y_3,y,y_4),
	\\
	\Pi_2^a(y,y_3)/2
	&=&
	y\,y_2\,E_3\,E_4\,D_1 + D_3 + y\,E_3 D_2(y_2,y_4,y,y_3) \\
 && \nonumber  + y_2\,E_4 D_2(y,y_3,y_2,y_4),
	\\
	\Pi_2^a(y,y_4)/2
	&=&
	y\,y_2\,E_3\,E_4\,D_1 + D_3 + y_2\,E_3 D_2(y,y_4,y_2,y_3) \\
 && \nonumber+ y\,E_4 D_2(y_2,y_3,y,y_4),
	\\
	\Pi_2^\nu(y,y_2)/2
	&=&
	y\,y_2\,y_3\,y_4\,D_1 + D_3 - y\,y_2 D_2(y_3,y_4,y,y_2)  - y_3\,y_4 D_2(y,y_2,y_3,y_4),
	\\
	\Pi_2^\nu(y,y_4)/2
	&=&
	y\,y_2\,y_3\,y_4\,D_1 + D_3 + y_2\,y_3 D_2(y,y_4,y_2,y_3)  + y\,y_4 D_2(y_2,y_3,y,y_4),
\end{eqnarray}
where the functions $D_i$ are defined as \cite{Dolgov:1997mb}:
\begin{eqnarray}
	D_1(a,b,c,d)
	&=&
	\frac{16}{\pi}
	\int_0^\infty
	\frac{{\rm d}\lambda}{\lambda^2}
	\prod_{i=a,b,c,d}\sin(\lambda i)
	\,,\\
	D_2(a,b,c,d)
	&=&
	-\frac{16}{\pi}
	\int_0^\infty
	\frac{{\rm d}\lambda}{\lambda^4}
	\prod_{i=a,b}\Big[\lambda i \cos(\lambda i)-\sin(\lambda i)\Big]
	\prod_{j=c,d}\sin(\lambda j)
	\,,\\
	D_3(a,b,c,d)
	&=&
	\frac{16}{\pi}
	\int_0^\infty
	\frac{\mathrm{d}\lambda}{\lambda^6}
	\prod_{i=a,b,c,d}\Big[\lambda i \cos(\lambda i)-\sin(\lambda i)\Big]
	\,.
\end{eqnarray} 

Again, the comoving collision terms for the antineutrinos $\tilde{\bar{\mathcal{I}}}
	\equiv
	\tilde{\bar{\mathcal{I}}}_{\rm sc}
	+ \tilde{\bar{\mathcal{I}}}_{\rm ann}
	+ \tilde{\bar{\mathcal{I}}}_{\nu\nu}$ can be obtained simply by the transformations $\vrho \leftrightarrow \bvrho$ and $L \leftrightarrow R$.

\section{Finite Temperature QED Corrections} \label{sec: FTQED}

The finite temperature QED (FTQED) corrections on neutrino decoupling must be taking into account to obtain accurate density matrices and the resulting $\Neff$
\cite{Fornengo:1997wa,Mangano:2001iu,Bennett:2019ewm}.
In this work we consider the following three aspects of the FTQED corrections.

\subsection{FTQED correction to the continuity equation}

The continuity equation of the Universe with the FTQED corrections is usually expressed as a differential equation for the dimensionless photon temperature $z$ and are solved combing with the QKEs in practice,
\begin{equation}
    \frac{\mathrm{d}z}{\mathrm{d}x}=
    \cfrac{
    r J_2(r)
    + G_1(r)
    - \cfrac{1}{4\pi^2z^3}
        {\displaystyle \int_0^\infty \mathrm{d}y\,y^3\sum_{\alpha} \left(\cfrac{\mathrm{d}\varrho_{\alpha \alpha}}{\mathrm{d}x} +\cfrac{\mathrm{d}\bvrho_{\alpha \alpha}}{\mathrm{d}x} \right)}
    }{
    \Big[
    r^2 J_2(r)
    + J_4(r)
    \Big]
    + G_2(r)
    + \cfrac{2\pi^2}{15}
    }\,,
\end{equation}
where $r = x/z$, $ \frac{\mathrm{d}\varrho_{\alpha \alpha}}{\mathrm{d}x}$ and $\frac{\mathrm{d}\bvrho_{\alpha \alpha}}{\mathrm{d}x}$ are the diagonal components of the QKEs for neutrinos and antineutrinos. The remaining functions appearing in the equation are defined as
\begin{eqnarray}
J_a(r)
&=&
\frac{1}{\pi^2}
\int_0^\infty {\rm d}u \, u^a
\frac{\exp(\sqrt{u^2+r^2})}{\Big[\exp(\sqrt{u^2+r^2})+1\Big]^2}
\label{eq:j}
\,,\\
K_a(r)
&=&
\frac{1}{\pi^2}
\int_0^\infty {\rm d}u \, 
\frac{u^a}{\sqrt{u^2+r^2}}\,
\frac{1}{\exp(\sqrt{u^2+r^2})+1}
\label{eq:k}\,.
\end{eqnarray}
and $G_{1,2}$ functions are expanded as a series of powers of the electron charge $e^2 = 4\pi \alpha$ with $\alpha$ the fine structure constant,
\begin{eqnarray}
G_{1,2}(x,z)
&=&
G_{1,2}^{(2)}(x/z)
+
G_{1,2}^{(2+\ln)}(x,z)
+
G_{1,2}^{(3)}(x/z)
+
\ldots
\,.
\end{eqnarray}
At the $\mathcal{O}(e^2)$ order we have
\begin{eqnarray}
G_1^{(2)}(r)
&=&
2\pi\alpha
\left[
  \frac{1}{r}
  \left(
    \frac{K_2}{3}
    + 2 K_2^2
    -\frac{J_2}{6}
    -K_2J_2
  \right)
  +
  G_a
\right]
\label{eq:g1}
\,,\\
G_2^{(2)}(r)
&=&
-8\pi\alpha
\left(
  \frac{K_2}{6}
  +\frac{J_2}{6}
  -\frac{1}{2}K_2^2
  +K_2J_2
\right)
+
2\pi\alpha r
G_a
\label{eq:g2}
\,, \\
G_a(r)
&=&
\frac{K_2'}{6}
-K_2K_2'
+\frac{J_2'}{6}
+K_2'J_2
+K_2J_2'
\,,
\end{eqnarray}
where the prime denotes derivative with respect to $r$ and the explicit dependence on $r$ for the $G$ functions are dropped for simplicity.
An additional log term at the $\mathcal{O}(e^2)$ order is
\begin{eqnarray}
G_1^{(2+\ln)}(x,z)
&=&
\frac{e^2 x}{16\pi^4 z^3}
\int\!\!\!\int_0^\infty
{\rm d}y\,
{\rm d}k\,
\frac{y\,k}{E_y E_k}
\ln\left|\frac{y+k}{y-k}\right|
\Bigg\{
-x\bigg[
  z\Big(
    \partial_x\mathcal{N}_y\partial_z\mathcal{N}_k
    +
    \mathcal{N}_y\partial_x\partial_z\mathcal{N}_k
  \Big)
  -\mathcal{N}_y\partial_x\mathcal{N}_k
\bigg]
\nonumber\\
&&
\qquad\qquad\qquad\qquad
-\mathcal{N}_y\mathcal{N}_k
-z\mathcal{N}_y\partial_z\mathcal{N}_k
+\frac{x^2 (E_y^2+E_k^2)}{2 E_y^2 E_k^2}
\Big(
2z \mathcal{N}_y\partial_z \mathcal{N}_k
-\mathcal{N}_y \mathcal{N}_k
\Big)
\Bigg\}
\,,\\
G_2^{(2+\ln)}(x,z)
&=&
\frac{e^2 x^2}{16\pi^4 z^2}
\int\!\!\!\int_0^\infty
{\rm d}y\,
{\rm d}k\,
\frac{y\,k}{E_y E_k}
\ln\left|\frac{y+k}{y-k}\right|
\partial_z
\Big(
\mathcal{N}_y
\partial_z
\mathcal{N}_k
\Big)
\,,
\end{eqnarray}
where
\begin{eqnarray}
\mathcal{N}_p
&=&
\frac{2}{e^{E_p/z}+1}
\,,\\
\partial_x\mathcal{N}_p
&=&
-\frac{x\,e^{E_p/z}\,\mathcal{N}_p^2}{2z\,E_p}
\,,\\
\partial_z\mathcal{N}_p
&=&
\frac{
 e^{E_p/z}\,
 E_p\,
 \mathcal{N}_p^2
}{2 z^2}
\,,\\
\partial_x\partial_z\mathcal{N}_p
&=&
\frac{
 x\,
 e^{E_p/z}\,
 \mathcal{N}_p^2
}{2z^3}
\left(
 1
 -e^{E_p/z}\,\mathcal{N}_p
 +\frac{z}{E_p}
\right)
\,.
\end{eqnarray}
At the $\mathcal{O}(e^3)$ order we have
\begin{eqnarray}
G_1^{(3)}(r)
&=&
\frac{e^3}{4\pi} G_b
\left\{
 \frac{1}{r}\Big(2J_2-4K_2\Big)
 -2J_2'
 -r^2J_0'
 -r\Big(2K_0+J_0\Big) \right. \nonumber\\
 && \left.-G_c\Big[r(K_0-J_0)+K_2'\Big]
\right\}
\,,\\
G_2^{(3)}(r)
&=&
\frac{e^3}{4\pi} G_b
\left[
 G_c \Big(2J_2 + r^2 J_0\Big)
 -\frac{2}{r}J_4'
 -r\Big(3J_2'+r^2J_0'\Big)
\right]
\,, 
\end{eqnarray}
with
\begin{eqnarray}
G_b(r)
&=&
\sqrt{K_2 + r^2 \frac{K_0}{2}}
\,,\\
G_c(r)
&=&
\frac{2J_2 + r^2 J_0}{2\Big(2K_2 + r^2 K_0\Big)}
\,.
\label{eq:g3}
\end{eqnarray}

\subsection{FTQED corrections to the total energies and the pressures}

The total pressure and energy density of the universe are also modified by the FTQED corrections:
\begin{eqnarray}
P_{\rm tot}
&=&
\sum_{i=\gamma,{\nu_i},e}P_i
+
\delta P(x,z)
\,,\\
\label{eq:total_p}
\rho_{\rm tot}
&=&
\sum_{i=\gamma,{\nu_i},e}\rho_i
+
\delta\rho(x,z)
\,,
\label{eq:total_rho}
\end{eqnarray}
where $\delta P$ and $\delta\rho$ can also be expanded as a series of powers of the electron charge, so that we have~\cite{Bennett:2019ewm}
\begin{eqnarray}
\delta P(x,z)
&=&
\delta P^{(2)}(x/z)
+
\delta P^{(2+\ln)}(x,z)
+
\delta P^{(3)}(x/z)
+\ldots
\,,
\\
\delta P^{(2)}(r)
&=&
-
e^2 z^4\,K_2
\left(
 \frac{1}{6}
 +\frac{K_2}{2}
\right)
\,,\\
\delta P^{(2+\ln)}(x,z)
&=&
\frac{e^2 x^2}{16\pi^4}
\int\!\!\!\int_0^\infty
{\rm d}y\,
{\rm d}k\,
\frac{y\,k}{E_y E_k}
\ln
\left|
\frac{y+k}{y-k}
\right|
\mathcal{N}_y\,
\mathcal{N}_k
\,,\\
\delta P^{(3)}(r)
&=&
\frac{2e^3z^4}{3\pi}
\left(
K_2
+
\frac{r^2}{2} K_0
\right)^{3/2}
\,,
\end{eqnarray}
for the pressure, while
\begin{eqnarray}
\delta\rho(x,z)
&=&
\delta\rho^{(2)}(x/z)
+
\delta\rho^{(2+\ln)}(x,z)
+
\delta\rho^{(3)}(x/z)
+\ldots
\,,
\\
\delta\rho^{(2)}(r)
&=&
e^2 z^4
\left(
\frac{K_2^2}{2}
-\frac{K_2+J_2}{6}
-K_2 J_2
\right)
\,,\\
\delta\rho^{(2+\ln)}(x,z)
&=&
\frac{e^2 x^2}{16\pi^4}
\int\!\!\!\int_0^\infty
{\rm d}y\,
{\rm d}k\,
\frac{y\,k}{E_y E_k}
\ln
\left|
\frac{y+k}{y-k}
\right|
\mathcal{N}_y
\Big(
2z \partial_z \mathcal{N}_k
- \mathcal{N}_k
\Big)
\,,\\
\delta\rho^{(3)}(r)
&=&
\frac{e^3z^4}{\pi}
\left(
K_2
+
\frac{r^2}{2} K_0
\right)^{1/2}
\left(
J_2
+
\frac{r^2}{2} J_0
\right)
\,.
\end{eqnarray}
for the energy densities.

\subsection{FTQED correction to the electron mass}
In addition, the mass of electron/positrons are also modified by the FTQED correction, which 
 is mainly used in the collision terms.
In the comoving coordinates, the additional FTQED contribution to the electron mass can be written as \cite{Fornengo:1997wa,Mangano:2001iu,Bennett:2019ewm}:
\begin{equation}
\delta m_e^2(x, y, z)
=
\frac{2\pi\alpha z^2}{3}
+
\frac{4\alpha}{\pi}
\int_0^\infty{\rm d}k\,
\frac{k^2}{E_k}
\frac{1}{e^{E_k/z}+1}
-
\frac{x^2\alpha}{\pi y}
\int_0^\infty{\rm d}k\,
\frac{k}{E_k}
\log\left|\frac{y+k}{y-k}\right|
\mathcal{N}_k
\,,
\end{equation}
so that the comoving electron mass in the collision terms is replaced using $x^2\rightarrow x^2+\delta m_e^2$.
Finally, we note that the the log term that depends on $y$ in the FTQED corrections to the continuity equation, the total energy densities and the pressure, as long as the electron mass are ignored in the computation, as it is shown that their contribution is negligible~\cite{Froustey:2020mcq,Bennett:2020zkv}.

\bibliographystyle{JHEP}
\bibliography{ref}

\providecommand{\href}[2]{#2}\begingroup\raggedright\begin{thebibliography}{100}

\bibitem{Lesgourgues:2013sjj}
J.~Lesgourgues, G.~Mangano, G.~Miele and S.~Pastor, \emph{{Neutrino Cosmology}}. Cambridge University Press, 2, 2013.

\bibitem{Pitrou:2018cgg}
C.~Pitrou, A.~Coc, J.-P. Uzan and E.~Vangioni, \emph{{Precision big bang nucleosynthesis with improved Helium-4 predictions}}, \href{https://doi.org/10.1016/j.physrep.2018.04.005}{\emph{Phys. Rept.} {\bfseries 754} (2018) 1} [\href{https://arxiv.org/abs/1801.08023}{{\ttfamily 1801.08023}}].

\bibitem{Planck:2018vyg}
{\scshape Planck} collaboration, \emph{{Planck 2018 results. VI. Cosmological parameters}}, \href{https://doi.org/10.1051/0004-6361/201833910}{\emph{Astron. Astrophys.} {\bfseries 641} (2020) A6} [\href{https://arxiv.org/abs/1807.06209}{{\ttfamily 1807.06209}}].

\bibitem{empress}
A.~Matsumoto et~al., \emph{{EMPRESS. VIII. A New Determination of Primordial He Abundance with Extremely Metal-poor Galaxies: A Suggestion of the Lepton Asymmetry and Implications for the Hubble Tension}}, \href{https://doi.org/10.3847/1538-4357/ac9ea1}{\emph{Astrophys. J.} {\bfseries 941} (2022) 167} [\href{https://arxiv.org/abs/2203.09617}{{\ttfamily 2203.09617}}].

\bibitem{DESI:2024mwx}
{\scshape DESI} collaboration, \emph{{DESI 2024 VI: cosmological constraints from the measurements of baryon acoustic oscillations}}, \href{https://doi.org/10.1088/1475-7516/2025/02/021}{\emph{JCAP} {\bfseries 02} (2025) 021} [\href{https://arxiv.org/abs/2404.03002}{{\ttfamily 2404.03002}}].

\bibitem{Hannestad:1995rs}
S.~Hannestad and J.~Madsen, \emph{{Neutrino decoupling in the early universe}}, \href{https://doi.org/10.1103/PhysRevD.52.1764}{\emph{Phys. Rev. D} {\bfseries 52} (1995) 1764} [\href{https://arxiv.org/abs/astro-ph/9506015}{{\ttfamily astro-ph/9506015}}].

\bibitem{Dolgov:1997mb}
A.~D. Dolgov, S.~H. Hansen and D.~V. Semikoz, \emph{{Nonequilibrium corrections to the spectra of massless neutrinos in the early universe}}, \href{https://doi.org/10.1016/S0550-3213(97)00479-3}{\emph{Nucl. Phys. B} {\bfseries 503} (1997) 426} [\href{https://arxiv.org/abs/hep-ph/9703315}{{\ttfamily hep-ph/9703315}}].

\bibitem{Mangano:2001iu}
G.~Mangano, G.~Miele, S.~Pastor and M.~Peloso, \emph{{A Precision calculation of the effective number of cosmological neutrinos}}, \href{https://doi.org/10.1016/S0370-2693(02)01622-2}{\emph{Phys. Lett. B} {\bfseries 534} (2002) 8} [\href{https://arxiv.org/abs/astro-ph/0111408}{{\ttfamily astro-ph/0111408}}].

\bibitem{Mangano:2005cc}
G.~Mangano, G.~Miele, S.~Pastor, T.~Pinto, O.~Pisanti and P.~D. Serpico, \emph{{Relic neutrino decoupling including flavor oscillations}}, \href{https://doi.org/10.1016/j.nuclphysb.2005.09.041}{\emph{Nucl. Phys. B} {\bfseries 729} (2005) 221} [\href{https://arxiv.org/abs/hep-ph/0506164}{{\ttfamily hep-ph/0506164}}].

\bibitem{Grohs:2015tfy}
E.~Grohs, G.~M. Fuller, C.~T. Kishimoto, M.~W. Paris and A.~Vlasenko, \emph{{Neutrino energy transport in weak decoupling and big bang nucleosynthesis}}, \href{https://doi.org/10.1103/PhysRevD.93.083522}{\emph{Phys. Rev. D} {\bfseries 93} (2016) 083522} [\href{https://arxiv.org/abs/1512.02205}{{\ttfamily 1512.02205}}].

\bibitem{deSalas:2016ztq}
P.~F. de~Salas and S.~Pastor, \emph{{Relic neutrino decoupling with flavour oscillations revisited}}, \href{https://doi.org/10.1088/1475-7516/2016/07/051}{\emph{JCAP} {\bfseries 07} (2016) 051} [\href{https://arxiv.org/abs/1606.06986}{{\ttfamily 1606.06986}}].

\bibitem{Akita:2020szl}
K.~Akita and M.~Yamaguchi, \emph{{A precision calculation of relic neutrino decoupling}}, \href{https://doi.org/10.1088/1475-7516/2020/08/012}{\emph{JCAP} {\bfseries 08} (2020) 012} [\href{https://arxiv.org/abs/2005.07047}{{\ttfamily 2005.07047}}].

\bibitem{Froustey:2019owm}
J.~Froustey and C.~Pitrou, \emph{{Incomplete neutrino decoupling effect on big bang nucleosynthesis}}, \href{https://doi.org/10.1103/PhysRevD.101.043524}{\emph{Phys. Rev. D} {\bfseries 101} (2020) 043524} [\href{https://arxiv.org/abs/1912.09378}{{\ttfamily 1912.09378}}].

\bibitem{EscuderoAbenza:2020cmq}
M.~Escudero~Abenza, \emph{{Precision early universe thermodynamics made simple: $N_{\rm eff}$ and neutrino decoupling in the Standard Model and beyond}}, \href{https://doi.org/10.1088/1475-7516/2020/05/048}{\emph{JCAP} {\bfseries 05} (2020) 048} [\href{https://arxiv.org/abs/2001.04466}{{\ttfamily 2001.04466}}].

\bibitem{Froustey:2020mcq}
J.~Froustey, C.~Pitrou and M.~C. Volpe, \emph{{Neutrino decoupling including flavour oscillations and primordial nucleosynthesis}}, \href{https://doi.org/10.1088/1475-7516/2020/12/015}{\emph{JCAP} {\bfseries 12} (2020) 015} [\href{https://arxiv.org/abs/2008.01074}{{\ttfamily 2008.01074}}].

\bibitem{Bennett:2020zkv}
J.~J. Bennett, G.~Buldgen, P.~F. De~Salas, M.~Drewes, S.~Gariazzo, S.~Pastor et~al., \emph{{Towards a precision calculation of $N_{\rm eff}$ in the Standard Model II: Neutrino decoupling in the presence of flavour oscillations and finite-temperature QED}}, \href{https://doi.org/10.1088/1475-7516/2021/04/073}{\emph{JCAP} {\bfseries 04} (2021) 073} [\href{https://arxiv.org/abs/2012.02726}{{\ttfamily 2012.02726}}].

\bibitem{Barenboim:2016lxv}
G.~Barenboim, W.~H. Kinney and W.-I. Park, \emph{{Flavor versus mass eigenstates in neutrino asymmetries: implications for cosmology}}, \href{https://doi.org/10.1140/epjc/s10052-017-5147-4}{\emph{Eur. Phys. J. C} {\bfseries 77} (2017) 590} [\href{https://arxiv.org/abs/1609.03200}{{\ttfamily 1609.03200}}].

\bibitem{Yeung:2020zde}
S.~Yeung, K.~Lau and M.~C. Chu, \emph{{Relic Neutrino Degeneracies and Their Impact on Cosmological Parameters}}, \href{https://doi.org/10.1088/1475-7516/2021/04/024}{\emph{JCAP} {\bfseries 04} (2021) 024} [\href{https://arxiv.org/abs/2010.01696}{{\ttfamily 2010.01696}}].

\bibitem{Seto:2021tad}
O.~Seto and Y.~Toda, \emph{{Hubble tension in lepton asymmetric cosmology with an extra radiation}}, \href{https://doi.org/10.1103/PhysRevD.104.063019}{\emph{Phys. Rev. D} {\bfseries 104} (2021) 063019} [\href{https://arxiv.org/abs/2104.04381}{{\ttfamily 2104.04381}}].

\bibitem{Kumar:2022vee}
S.~Kumar, R.~C. Nunes and P.~Yadav, \emph{{Updating non-standard neutrinos properties with Planck-CMB data and full-shape analysis of BOSS and eBOSS galaxies}}, \href{https://doi.org/10.1088/1475-7516/2022/09/060}{\emph{JCAP} {\bfseries 09} (2022) 060} [\href{https://arxiv.org/abs/2205.04292}{{\ttfamily 2205.04292}}].

\bibitem{Yeung:2024krv}
S.~Yeung, W.~Zhang and M.-c. Chu, \emph{{Resolving the $H_0$ and $S_8$ tensions with neutrino mass and chemical potential}},  \href{https://arxiv.org/abs/2403.11499}{{\ttfamily 2403.11499}}.

\bibitem{Kuzmin:1985mm}
V.~A. Kuzmin, V.~A. Rubakov and M.~E. Shaposhnikov, \emph{{On the Anomalous Electroweak Baryon Number Nonconservation in the Early Universe}}, \href{https://doi.org/10.1016/0370-2693(85)91028-7}{\emph{Phys. Lett. B} {\bfseries 155} (1985) 36}.

\bibitem{Khlebnikov:1988sr}
S.~Y. Khlebnikov and M.~E. Shaposhnikov, \emph{{The Statistical Theory of Anomalous Fermion Number Nonconservation}}, \href{https://doi.org/10.1016/0550-3213(88)90133-2}{\emph{Nucl. Phys. B} {\bfseries 308} (1988) 885}.

\bibitem{Harvey:1990qw}
J.~A. Harvey and M.~S. Turner, \emph{{Cosmological baryon and lepton number in the presence of electroweak fermion number violation}}, \href{https://doi.org/10.1103/PhysRevD.42.3344}{\emph{Phys. Rev. D} {\bfseries 42} (1990) 3344}.

\bibitem{Dreiner:1992vm}
H.~K. Dreiner and G.~G. Ross, \emph{{Sphaleron erasure of primordial baryogenesis}}, \href{https://doi.org/10.1016/0550-3213(93)90579-E}{\emph{Nucl. Phys. B} {\bfseries 410} (1993) 188} [\href{https://arxiv.org/abs/hep-ph/9207221}{{\ttfamily hep-ph/9207221}}].

\bibitem{Casas:1997gx}
A.~Casas, W.~Y. Cheng and G.~Gelmini, \emph{{Generation of large lepton asymmetries}}, \href{https://doi.org/10.1016/S0550-3213(98)00606-3}{\emph{Nucl. Phys. B} {\bfseries 538} (1999) 297} [\href{https://arxiv.org/abs/hep-ph/9709289}{{\ttfamily hep-ph/9709289}}].

\bibitem{Dolgov:1989us}
A.~D. Dolgov and D.~P. Kirilova, \emph{{ON PARTICLE CREATION BY A TIME DEPENDENT SCALAR FIELD}}, {\emph{Sov. J. Nucl. Phys.} {\bfseries 51} (1990) 172}.

\bibitem{Bajc:1997ky}
B.~Bajc, A.~Riotto and G.~Senjanovic, \emph{{Large lepton number of the universe and the fate of topological defects}}, \href{https://doi.org/10.1103/PhysRevLett.81.1355}{\emph{Phys. Rev. Lett.} {\bfseries 81} (1998) 1355} [\href{https://arxiv.org/abs/hep-ph/9710415}{{\ttfamily hep-ph/9710415}}].

\bibitem{Asaka:2005pn}
T.~Asaka and M.~Shaposhnikov, \emph{{The $\nu$MSM, dark matter and baryon asymmetry of the universe}}, \href{https://doi.org/10.1016/j.physletb.2005.06.020}{\emph{Phys. Lett. B} {\bfseries 620} (2005) 17} [\href{https://arxiv.org/abs/hep-ph/0505013}{{\ttfamily hep-ph/0505013}}].

\bibitem{Asaka:2005an}
T.~Asaka, S.~Blanchet and M.~Shaposhnikov, \emph{{The nuMSM, dark matter and neutrino masses}}, \href{https://doi.org/10.1016/j.physletb.2005.09.070}{\emph{Phys. Lett. B} {\bfseries 631} (2005) 151} [\href{https://arxiv.org/abs/hep-ph/0503065}{{\ttfamily hep-ph/0503065}}].

\bibitem{Pilaftsis:2003gt}
A.~Pilaftsis and T.~E.~J. Underwood, \emph{{Resonant leptogenesis}}, \href{https://doi.org/10.1016/j.nuclphysb.2004.05.029}{\emph{Nucl. Phys. B} {\bfseries 692} (2004) 303} [\href{https://arxiv.org/abs/hep-ph/0309342}{{\ttfamily hep-ph/0309342}}].

\bibitem{Borah:2022uos}
D.~Borah and A.~Dasgupta, \emph{{Large neutrino asymmetry from TeV scale leptogenesis}}, \href{https://doi.org/10.1103/PhysRevD.108.035015}{\emph{Phys. Rev. D} {\bfseries 108} (2023) 035015} [\href{https://arxiv.org/abs/2206.14722}{{\ttfamily 2206.14722}}].

\bibitem{Kawasaki:2002hq}
M.~Kawasaki, F.~Takahashi and M.~Yamaguchi, \emph{{Large lepton asymmetry from Q balls}}, \href{https://doi.org/10.1103/PhysRevD.66.043516}{\emph{Phys. Rev. D} {\bfseries 66} (2002) 043516} [\href{https://arxiv.org/abs/hep-ph/0205101}{{\ttfamily hep-ph/0205101}}].

\bibitem{Kawasaki:2022hvx}
M.~Kawasaki and K.~Murai, \emph{{Lepton asymmetric universe}}, \href{https://doi.org/10.1088/1475-7516/2022/08/041}{\emph{JCAP} {\bfseries 08} (2022) 041} [\href{https://arxiv.org/abs/2203.09713}{{\ttfamily 2203.09713}}].

\bibitem{Grohs:2016cuu}
E.~Grohs, G.~M. Fuller, C.~T. Kishimoto and M.~W. Paris, \emph{{Lepton asymmetry, neutrino spectral distortions, and big bang nucleosynthesis}}, \href{https://doi.org/10.1103/PhysRevD.95.063503}{\emph{Phys. Rev. D} {\bfseries 95} (2017) 063503} [\href{https://arxiv.org/abs/1612.01986}{{\ttfamily 1612.01986}}].

\bibitem{Bell:1998ds}
N.~F. Bell, R.~R. Volkas and Y.~Y.~Y. Wong, \emph{{Relic neutrino asymmetry evolution from first principles}}, \href{https://doi.org/10.1103/PhysRevD.59.113001}{\emph{Phys. Rev. D} {\bfseries 59} (1999) 113001} [\href{https://arxiv.org/abs/hep-ph/9809363}{{\ttfamily hep-ph/9809363}}].

\bibitem{Pastor:2001iu}
S.~Pastor, G.~G. Raffelt and D.~V. Semikoz, \emph{{Physics of synchronized neutrino oscillations caused by selfinteractions}}, \href{https://doi.org/10.1103/PhysRevD.65.053011}{\emph{Phys. Rev. D} {\bfseries 65} (2002) 053011} [\href{https://arxiv.org/abs/hep-ph/0109035}{{\ttfamily hep-ph/0109035}}].

\bibitem{Abazajian:2002qx}
K.~N. Abazajian, J.~F. Beacom and N.~F. Bell, \emph{{Stringent Constraints on Cosmological Neutrino Antineutrino Asymmetries from Synchronized Flavor Transformation}}, \href{https://doi.org/10.1103/PhysRevD.66.013008}{\emph{Phys. Rev. D} {\bfseries 66} (2002) 013008} [\href{https://arxiv.org/abs/astro-ph/0203442}{{\ttfamily astro-ph/0203442}}].

\bibitem{Wong:2002fa}
Y.~Y.~Y. Wong, \emph{{Analytical treatment of neutrino asymmetry equilibration from flavor oscillations in the early universe}}, \href{https://doi.org/10.1103/PhysRevD.66.025015}{\emph{Phys. Rev. D} {\bfseries 66} (2002) 025015} [\href{https://arxiv.org/abs/hep-ph/0203180}{{\ttfamily hep-ph/0203180}}].

\bibitem{Dolgov:2002ab}
A.~D. Dolgov, S.~H. Hansen, S.~Pastor, S.~T. Petcov, G.~G. Raffelt and D.~V. Semikoz, \emph{{Cosmological bounds on neutrino degeneracy improved by flavor oscillations}}, \href{https://doi.org/10.1016/S0550-3213(02)00274-2}{\emph{Nucl. Phys. B} {\bfseries 632} (2002) 363} [\href{https://arxiv.org/abs/hep-ph/0201287}{{\ttfamily hep-ph/0201287}}].

\bibitem{Pastor:2008ti}
S.~Pastor, T.~Pinto and G.~G. Raffelt, \emph{{Relic density of neutrinos with primordial asymmetries}}, \href{https://doi.org/10.1103/PhysRevLett.102.241302}{\emph{Phys. Rev. Lett.} {\bfseries 102} (2009) 241302} [\href{https://arxiv.org/abs/0808.3137}{{\ttfamily 0808.3137}}].

\bibitem{Mangano:2010ei}
G.~Mangano, G.~Miele, S.~Pastor, O.~Pisanti and S.~Sarikas, \emph{{Constraining the cosmic radiation density due to lepton number with Big Bang Nucleosynthesis}}, \href{https://doi.org/10.1088/1475-7516/2011/03/035}{\emph{JCAP} {\bfseries 03} (2011) 035} [\href{https://arxiv.org/abs/1011.0916}{{\ttfamily 1011.0916}}].

\bibitem{Mangano:2011ip}
G.~Mangano, G.~Miele, S.~Pastor, O.~Pisanti and S.~Sarikas, \emph{{Updated BBN bounds on the cosmological lepton asymmetry for non-zero $\theta_{13}$}}, \href{https://doi.org/10.1016/j.physletb.2012.01.015}{\emph{Phys. Lett. B} {\bfseries 708} (2012) 1} [\href{https://arxiv.org/abs/1110.4335}{{\ttfamily 1110.4335}}].

\bibitem{Castorina:2012md}
E.~Castorina, U.~Franca, M.~Lattanzi, J.~Lesgourgues, G.~Mangano, A.~Melchiorri et~al., \emph{{Cosmological lepton asymmetry with a nonzero mixing angle $\theta_{13}$}}, \href{https://doi.org/10.1103/PhysRevD.86.023517}{\emph{Phys. Rev. D} {\bfseries 86} (2012) 023517} [\href{https://arxiv.org/abs/1204.2510}{{\ttfamily 1204.2510}}].

\bibitem{Barenboim:2016shh}
G.~Barenboim, W.~H. Kinney and W.-I. Park, \emph{{Resurrection of large lepton number asymmetries from neutrino flavor oscillations}}, \href{https://doi.org/10.1103/PhysRevD.95.043506}{\emph{Phys. Rev. D} {\bfseries 95} (2017) 043506} [\href{https://arxiv.org/abs/1609.01584}{{\ttfamily 1609.01584}}].

\bibitem{Johns:2016enc}
L.~Johns, M.~Mina, V.~Cirigliano, M.~W. Paris and G.~M. Fuller, \emph{{Neutrino flavor transformation in the lepton-asymmetric universe}}, \href{https://doi.org/10.1103/PhysRevD.94.083505}{\emph{Phys. Rev. D} {\bfseries 94} (2016) 083505} [\href{https://arxiv.org/abs/1608.01336}{{\ttfamily 1608.01336}}].

\bibitem{Froustey:2021azz}
J.~Froustey and C.~Pitrou, \emph{{Primordial neutrino asymmetry evolution with full mean-field effects and collisions}}, \href{https://doi.org/10.1088/1475-7516/2022/03/065}{\emph{JCAP} {\bfseries 03} (2022) 065} [\href{https://arxiv.org/abs/2110.11889}{{\ttfamily 2110.11889}}].

\bibitem{Sarkar:1995dd}
S.~Sarkar, \emph{{Big bang nucleosynthesis and physics beyond the standard model}}, \href{https://doi.org/10.1088/0034-4885/59/12/001}{\emph{Rept. Prog. Phys.} {\bfseries 59} (1996) 1493} [\href{https://arxiv.org/abs/hep-ph/9602260}{{\ttfamily hep-ph/9602260}}].

\bibitem{Iocco:2008va}
F.~Iocco, G.~Mangano, G.~Miele, O.~Pisanti and P.~D. Serpico, \emph{{Primordial Nucleosynthesis: from precision cosmology to fundamental physics}}, \href{https://doi.org/10.1016/j.physrep.2009.02.002}{\emph{Phys. Rept.} {\bfseries 472} (2009) 1} [\href{https://arxiv.org/abs/0809.0631}{{\ttfamily 0809.0631}}].

\bibitem{Serpico:2005bc}
P.~D. Serpico and G.~G. Raffelt, \emph{{Lepton asymmetry and primordial nucleosynthesis in the era of precision cosmology}}, \href{https://doi.org/10.1103/PhysRevD.71.127301}{\emph{Phys. Rev. D} {\bfseries 71} (2005) 127301} [\href{https://arxiv.org/abs/astro-ph/0506162}{{\ttfamily astro-ph/0506162}}].

\bibitem{Chu:2006ua}
Y.-Z. Chu and M.~Cirelli, \emph{{Sterile neutrinos, lepton asymmetries, primordial elements: How much of each?}}, \href{https://doi.org/10.1103/PhysRevD.74.085015}{\emph{Phys. Rev. D} {\bfseries 74} (2006) 085015} [\href{https://arxiv.org/abs/astro-ph/0608206}{{\ttfamily astro-ph/0608206}}].

\bibitem{Simha:2008mt}
V.~Simha and G.~Steigman, \emph{{Constraining The Universal Lepton Asymmetry}}, \href{https://doi.org/10.1088/1475-7516/2008/08/011}{\emph{JCAP} {\bfseries 08} (2008) 011} [\href{https://arxiv.org/abs/0806.0179}{{\ttfamily 0806.0179}}].

\bibitem{Saviano:2013ktj}
N.~Saviano, A.~Mirizzi, O.~Pisanti, P.~D. Serpico, G.~Mangano and G.~Miele, \emph{{Multi-momentum and multi-flavour active-sterile neutrino oscillations in the early universe: role of neutrino asymmetries and effects on nucleosynthesis}}, \href{https://doi.org/10.1103/PhysRevD.87.073006}{\emph{Phys. Rev. D} {\bfseries 87} (2013) 073006} [\href{https://arxiv.org/abs/1302.1200}{{\ttfamily 1302.1200}}].

\bibitem{Burns:2023sgx}
A.-K. Burns, T.~M.~P. Tait and M.~Valli, \emph{{PRyMordial: the first three minutes, within and beyond the standard model}}, \href{https://doi.org/10.1140/epjc/s10052-024-12442-0}{\emph{Eur. Phys. J. C} {\bfseries 84} (2024) 86} [\href{https://arxiv.org/abs/2307.07061}{{\ttfamily 2307.07061}}].

\bibitem{Oldengott:2017tzj}
I.~M. Oldengott and D.~J. Schwarz, \emph{{Improved constraints on lepton asymmetry from the cosmic microwave background}}, \href{https://doi.org/10.1209/0295-5075/119/29001}{\emph{EPL} {\bfseries 119} (2017) 29001} [\href{https://arxiv.org/abs/1706.01705}{{\ttfamily 1706.01705}}].

\bibitem{Matsumoto:2022tlr}
A.~Matsumoto et~al., \emph{{EMPRESS. VIII. A New Determination of Primordial He Abundance with Extremely Metal-poor Galaxies: A Suggestion of the Lepton Asymmetry and Implications for the Hubble Tension}}, \href{https://doi.org/10.3847/1538-4357/ac9ea1}{\emph{Astrophys. J.} {\bfseries 941} (2022) 167} [\href{https://arxiv.org/abs/2203.09617}{{\ttfamily 2203.09617}}].

\bibitem{Burns:2022hkq}
A.-K. Burns, T.~M.~P. Tait and M.~Valli, \emph{{Indications for a Nonzero Lepton Asymmetry from Extremely Metal-Poor Galaxies}}, \href{https://doi.org/10.1103/PhysRevLett.130.131001}{\emph{Phys. Rev. Lett.} {\bfseries 130} (2023) 131001} [\href{https://arxiv.org/abs/2206.00693}{{\ttfamily 2206.00693}}].

\bibitem{Escudero:2022okz}
M.~Escudero, A.~Ibarra and V.~Maura, \emph{{Primordial lepton asymmetries in the precision cosmology era: Current status and future sensitivities from BBN and the CMB}}, \href{https://doi.org/10.1103/PhysRevD.107.035024}{\emph{Phys. Rev. D} {\bfseries 107} (2023) 035024} [\href{https://arxiv.org/abs/2208.03201}{{\ttfamily 2208.03201}}].

\bibitem{Froustey:2024mgf}
J.~Froustey and C.~Pitrou, \emph{{Constraints on primordial lepton asymmetries with full neutrino transport}}, \href{https://doi.org/10.1103/PhysRevD.110.103551}{\emph{Phys. Rev. D} {\bfseries 110} (2024) 103551} [\href{https://arxiv.org/abs/2405.06509}{{\ttfamily 2405.06509}}].

\bibitem{Sigl:1993ctk}
G.~Sigl and G.~Raffelt, \emph{{General kinetic description of relativistic mixed neutrinos}}, \href{https://doi.org/10.1016/0550-3213(93)90175-O}{\emph{Nucl. Phys. B} {\bfseries 406} (1993) 423}.

\bibitem{Volpe:2013uxl}
C.~Volpe, D.~V\"a\"an\"anen and C.~Espinoza, \emph{{Extended evolution equations for neutrino propagation in astrophysical and cosmological environments}}, \href{https://doi.org/10.1103/PhysRevD.87.113010}{\emph{Phys. Rev. D} {\bfseries 87} (2013) 113010} [\href{https://arxiv.org/abs/1302.2374}{{\ttfamily 1302.2374}}].

\bibitem{Volpe:2015rla}
C.~Volpe, \emph{{Neutrino Quantum Kinetic Equations}}, \href{https://doi.org/10.1142/S0218301315410098}{\emph{Int. J. Mod. Phys. E} {\bfseries 24} (2015) 1541009} [\href{https://arxiv.org/abs/1506.06222}{{\ttfamily 1506.06222}}].

\bibitem{Vlasenko:2013fja}
A.~Vlasenko, G.~M. Fuller and V.~Cirigliano, \emph{{Neutrino Quantum Kinetics}}, \href{https://doi.org/10.1103/PhysRevD.89.105004}{\emph{Phys. Rev. D} {\bfseries 89} (2014) 105004} [\href{https://arxiv.org/abs/1309.2628}{{\ttfamily 1309.2628}}].

\bibitem{Blaschke:2016xxt}
D.~N. Blaschke and V.~Cirigliano, \emph{{Neutrino Quantum Kinetic Equations: The Collision Term}}, \href{https://doi.org/10.1103/PhysRevD.94.033009}{\emph{Phys. Rev. D} {\bfseries 94} (2016) 033009} [\href{https://arxiv.org/abs/1605.09383}{{\ttfamily 1605.09383}}].

\bibitem{Kainulainen:2023ocv}
K.~Kainulainen and H.~Parkkinen, \emph{{Quantum transport theory for neutrinos with flavor and particle-antiparticle mixing}}, \href{https://doi.org/10.1007/JHEP02(2024)217}{\emph{JHEP} {\bfseries 02} (2024) 217} [\href{https://arxiv.org/abs/2309.00881}{{\ttfamily 2309.00881}}].

\bibitem{Kainulainen:2024fdg}
K.~Kainulainen and H.~Parkkinen, \emph{{Coherent collision integrals for neutrino transport equations}}, \href{https://doi.org/10.1007/JHEP12(2024)169}{\emph{JHEP} {\bfseries 12} (2024) 169} [\href{https://arxiv.org/abs/2407.08598}{{\ttfamily 2407.08598}}].

\bibitem{Cielo:2023bqp}
M.~Cielo, M.~Escudero, G.~Mangano and O.~Pisanti, \emph{{Neff in the Standard Model at NLO is 3.043}}, \href{https://doi.org/10.1103/PhysRevD.108.L121301}{\emph{Phys. Rev. D} {\bfseries 108} (2023) L121301} [\href{https://arxiv.org/abs/2306.05460}{{\ttfamily 2306.05460}}].

\bibitem{Jackson:2023zkl}
G.~Jackson and M.~Laine, \emph{{QED corrections to the thermal neutrino interaction rate}}, \href{https://doi.org/10.1007/JHEP05(2024)089}{\emph{JHEP} {\bfseries 05} (2024) 089} [\href{https://arxiv.org/abs/2312.07015}{{\ttfamily 2312.07015}}].

\bibitem{Drewes:2024wbw}
M.~Drewes, Y.~Georis, M.~Klasen, L.~P. Wiggering and Y.~Y.~Y. Wong, \emph{{Towards a precision calculation of N $_{eff}$ in the Standard Model. Part III. Improved estimate of NLO contributions to the collision integral}}, \href{https://doi.org/10.1088/1475-7516/2024/06/032}{\emph{JCAP} {\bfseries 06} (2024) 032} [\href{https://arxiv.org/abs/2402.18481}{{\ttfamily 2402.18481}}].

\bibitem{Berges:2004yj}
J.~Berges, \emph{{Introduction to nonequilibrium quantum field theory}}, \href{https://doi.org/10.1063/1.1843591}{\emph{AIP Conf. Proc.} {\bfseries 739} (2004) 3} [\href{https://arxiv.org/abs/hep-ph/0409233}{{\ttfamily hep-ph/0409233}}].

\bibitem{Giunti:2007ry}
C.~Giunti and C.~W. Kim, \emph{{Fundamentals of Neutrino Physics and Astrophysics}}. 2007.

\bibitem{Gariazzo:2019gyi}
S.~Gariazzo, P.~F. de~Salas and S.~Pastor, \emph{{Thermalisation of sterile neutrinos in the early Universe in the 3+1 scheme with full mixing matrix}}, \href{https://doi.org/10.1088/1475-7516/2019/07/014}{\emph{JCAP} {\bfseries 07} (2019) 014} [\href{https://arxiv.org/abs/1905.11290}{{\ttfamily 1905.11290}}].

\bibitem{Fornengo:1997wa}
N.~Fornengo, C.~W. Kim and J.~Song, \emph{{Finite temperature effects on the neutrino decoupling in the early universe}}, \href{https://doi.org/10.1103/PhysRevD.56.5123}{\emph{Phys. Rev. D} {\bfseries 56} (1997) 5123} [\href{https://arxiv.org/abs/hep-ph/9702324}{{\ttfamily hep-ph/9702324}}].

\bibitem{Bennett:2019ewm}
J.~J. Bennett, G.~Buldgen, M.~Drewes and Y.~Y.~Y. Wong, \emph{{Towards a precision calculation of the effective number of neutrinos $N_{\rm eff}$ in the Standard Model I: the QED equation of state}}, \href{https://doi.org/10.1088/1475-7516/2020/03/003}{\emph{JCAP} {\bfseries 03} (2020) 003} [\href{https://arxiv.org/abs/1911.04504}{{\ttfamily 1911.04504}}].

\bibitem{pdg}
{\scshape Particle Data Group} collaboration, \emph{{Review of Particle Physics}}, \href{https://doi.org/10.1093/ptep/ptac097}{\emph{PTEP} {\bfseries 2022} (2022) 083C01}.

\bibitem{hindmarsh1982odepack}
A.~Hindmarsh and L.~L. Laboratory, \emph{{ODEPACK, a Systematized Collection of ODE Solvers}}. Lawrence Livermore National Laboratory, 1982.

\bibitem{PTOLEMY:2018jst}
{\scshape PTOLEMY} collaboration, \emph{{PTOLEMY: A Proposal for Thermal Relic Detection of Massive Neutrinos and Directional Detection of MeV Dark Matter}},  \href{https://arxiv.org/abs/1808.01892}{{\ttfamily 1808.01892}}.

\bibitem{Betti:2018bjv}
M.~G. Betti et~al., \emph{{A design for an electromagnetic filter for precision energy measurements at the tritium endpoint}}, \href{https://doi.org/10.1016/j.ppnp.2019.02.004}{\emph{Prog. Part. Nucl. Phys.} {\bfseries 106} (2019) 120} [\href{https://arxiv.org/abs/1810.06703}{{\ttfamily 1810.06703}}].

\bibitem{PTOLEMY:2019hkd}
{\scshape PTOLEMY} collaboration, \emph{{Neutrino physics with the PTOLEMY project: active neutrino properties and the light sterile case}}, \href{https://doi.org/10.1088/1475-7516/2019/07/047}{\emph{JCAP} {\bfseries 07} (2019) 047} [\href{https://arxiv.org/abs/1902.05508}{{\ttfamily 1902.05508}}].

\bibitem{Long:2014zva}
A.~J. Long, C.~Lunardini and E.~Sabancilar, \emph{{Detecting non-relativistic cosmic neutrinos by capture on tritium: phenomenology and physics potential}}, \href{https://doi.org/10.1088/1475-7516/2014/08/038}{\emph{JCAP} {\bfseries 08} (2014) 038} [\href{https://arxiv.org/abs/1405.7654}{{\ttfamily 1405.7654}}].

\bibitem{Akita:2020jbo}
K.~Akita, S.~Hurwitz and M.~Yamaguchi, \emph{{Precise Capture Rates of Cosmic Neutrinos and Their Implications on Cosmology}}, \href{https://doi.org/10.1140/epjc/s10052-021-09133-5}{\emph{Eur. Phys. J. C} {\bfseries 81} (2021) 344} [\href{https://arxiv.org/abs/2010.04454}{{\ttfamily 2010.04454}}].

\bibitem{Bauer:2022lri}
M.~Bauer and J.~D. Shergold, \emph{{Limits on the cosmic neutrino background}}, \href{https://doi.org/10.1088/1475-7516/2023/01/003}{\emph{JCAP} {\bfseries 01} (2023) 003} [\href{https://arxiv.org/abs/2207.12413}{{\ttfamily 2207.12413}}].

\bibitem{Duda:2001hd}
G.~Duda, G.~Gelmini and S.~Nussinov, \emph{{Expected signals in relic neutrino detectors}}, \href{https://doi.org/10.1103/PhysRevD.64.122001}{\emph{Phys. Rev. D} {\bfseries 64} (2001) 122001} [\href{https://arxiv.org/abs/hep-ph/0107027}{{\ttfamily hep-ph/0107027}}].

\bibitem{Domcke:2017aqj}
V.~Domcke and M.~Spinrath, \emph{{Detection prospects for the Cosmic Neutrino Background using laser interferometers}}, \href{https://doi.org/10.1088/1475-7516/2017/06/055}{\emph{JCAP} {\bfseries 06} (2017) 055} [\href{https://arxiv.org/abs/1703.08629}{{\ttfamily 1703.08629}}].

\bibitem{Descouvemont:2004cw}
P.~Descouvemont, A.~Adahchour, C.~Angulo, A.~Coc and E.~Vangioni-Flam, \emph{{Compilation and R-matrix analysis of Big Bang nuclear reaction rates}}, \href{https://doi.org/10.1016/j.adt.2004.08.001}{\emph{Atom. Data Nucl. Data Tabl.} {\bfseries 88} (2004) 203} [\href{https://arxiv.org/abs/astro-ph/0407101}{{\ttfamily astro-ph/0407101}}].

\bibitem{Iliadis:2016vkw}
C.~Iliadis, K.~Anderson, A.~Coc, F.~Timmes and S.~Starrfield, \emph{{Bayesian Estimation of Thermonuclear Reaction Rates}}, \href{https://doi.org/10.3847/0004-637X/831/1/107}{\emph{Astrophys. J.} {\bfseries 831} (2016) 107} [\href{https://arxiv.org/abs/1608.05853}{{\ttfamily 1608.05853}}].

\bibitem{InestaGomez:2017eya}
A.~I\~nesta G\'omez, C.~Iliadis and A.~Coc, \emph{{Bayesian estimation of thermonuclear reaction rates for deuterium+deuterium reactions}}, \href{https://doi.org/10.3847/1538-4357/aa9025}{\emph{Astrophys. J.} {\bfseries 849} (2017) 134} [\href{https://arxiv.org/abs/1710.01647}{{\ttfamily 1710.01647}}].

\bibitem{Xu:2013fha}
Y.~Xu, K.~Takahashi, S.~Goriely, M.~Arnould, M.~Ohta and H.~Utsunomiya, \emph{{NACRE II: an update of the NACRE compilation of charged-particle-induced thermonuclear reaction rates for nuclei with mass number $A < 16$}}, \href{https://doi.org/10.1016/j.nuclphysa.2013.09.007}{\emph{Nucl. Phys. A} {\bfseries 918} (2013) 61} [\href{https://arxiv.org/abs/1310.7099}{{\ttfamily 1310.7099}}].

\bibitem{Angulo:1999zz}
C.~Angulo et~al., \emph{{A compilation of charged-particle induced thermonuclear reaction rates}}, \href{https://doi.org/10.1016/S0375-9474(99)00030-5}{\emph{Nucl. Phys. A} {\bfseries 656} (1999) 3}.

\bibitem{Li:2025rjr}
Y.-Z. Li and J.-H. Yu, \emph{{Neutrinophilic $\mathbf{\Lambda}$CDM Extension for EMPRESS, DESI and Hubble Tension}},  \href{https://arxiv.org/abs/2501.13153}{{\ttfamily 2501.13153}}.

\bibitem{Lesgourgues:2011re}
J.~Lesgourgues, \emph{{The Cosmic Linear Anisotropy Solving System (CLASS) I: Overview}},  \href{https://arxiv.org/abs/1104.2932}{{\ttfamily 1104.2932}}.

\bibitem{Lesgourgues:2011rh}
J.~Lesgourgues and T.~Tram, \emph{{The Cosmic Linear Anisotropy Solving System (CLASS) IV: efficient implementation of non-cold relics}}, \href{https://doi.org/10.1088/1475-7516/2011/09/032}{\emph{JCAP} {\bfseries 09} (2011) 032} [\href{https://arxiv.org/abs/1104.2935}{{\ttfamily 1104.2935}}].

\bibitem{Ma:1995ey}
C.-P. Ma and E.~Bertschinger, \emph{{Cosmological perturbation theory in the synchronous and conformal Newtonian gauges}}, \href{https://doi.org/10.1086/176550}{\emph{Astrophys. J.} {\bfseries 455} (1995) 7} [\href{https://arxiv.org/abs/astro-ph/9506072}{{\ttfamily astro-ph/9506072}}].

\bibitem{Peebles:1970ag}
P.~J.~E. Peebles and J.~T. Yu, \emph{{Primeval adiabatic perturbation in an expanding universe}}, \href{https://doi.org/10.1086/150713}{\emph{Astrophys. J.} {\bfseries 162} (1970) 815}.

\bibitem{Sunyaev:1970bma}
R.~A. Sunyaev and Y.~B. Zeldovich, \emph{{Small-scale fluctuations of relic radiation}}, \href{https://doi.org/10.1007/BF00653471}{\emph{Astrophys. Space Sci.} {\bfseries 7} (1970) 3}.

\bibitem{Cooke:2017cwo}
R.~J. Cooke, M.~Pettini and C.~C. Steidel, \emph{{One Percent Determination of the Primordial Deuterium Abundance}}, \href{https://doi.org/10.3847/1538-4357/aaab53}{\emph{Astrophys. J.} {\bfseries 855} (2018) 102} [\href{https://arxiv.org/abs/1710.11129}{{\ttfamily 1710.11129}}].

\bibitem{Planck:2018lbu}
{\scshape Planck} collaboration, \emph{{Planck 2018 results. VIII. Gravitational lensing}}, \href{https://doi.org/10.1051/0004-6361/201833886}{\emph{Astron. Astrophys.} {\bfseries 641} (2020) A8} [\href{https://arxiv.org/abs/1807.06210}{{\ttfamily 1807.06210}}].

\bibitem{Planck:2019nip}
{\scshape Planck} collaboration, \emph{{Planck 2018 results. V. CMB power spectra and likelihoods}}, \href{https://doi.org/10.1051/0004-6361/201936386}{\emph{Astron. Astrophys.} {\bfseries 641} (2020) A5} [\href{https://arxiv.org/abs/1907.12875}{{\ttfamily 1907.12875}}].

\bibitem{BOSS:2016wmc}
{\scshape BOSS} collaboration, \emph{{The clustering of galaxies in the completed SDSS-III Baryon Oscillation Spectroscopic Survey: cosmological analysis of the DR12 galaxy sample}}, \href{https://doi.org/10.1093/mnras/stx721}{\emph{Mon. Not. Roy. Astron. Soc.} {\bfseries 470} (2017) 2617} [\href{https://arxiv.org/abs/1607.03155}{{\ttfamily 1607.03155}}].

\bibitem{Beutler:2011hx}
F.~Beutler, C.~Blake, M.~Colless, D.~H. Jones, L.~Staveley-Smith, L.~Campbell et~al., \emph{{The 6dF Galaxy Survey: Baryon Acoustic Oscillations and the Local Hubble Constant}}, \href{https://doi.org/10.1111/j.1365-2966.2011.19250.x}{\emph{Mon. Not. Roy. Astron. Soc.} {\bfseries 416} (2011) 3017} [\href{https://arxiv.org/abs/1106.3366}{{\ttfamily 1106.3366}}].

\bibitem{Ross:2014qpa}
A.~J. Ross, L.~Samushia, C.~Howlett, W.~J. Percival, A.~Burden and M.~Manera, \emph{{The clustering of the SDSS DR7 main Galaxy sample \textendash{} I. A 4 per cent distance measure at $z = 0.15$}}, \href{https://doi.org/10.1093/mnras/stv154}{\emph{Mon. Not. Roy. Astron. Soc.} {\bfseries 449} (2015) 835} [\href{https://arxiv.org/abs/1409.3242}{{\ttfamily 1409.3242}}].

\bibitem{Audren:2012wb}
B.~Audren, J.~Lesgourgues, K.~Benabed and S.~Prunet, \emph{{Conservative Constraints on Early Cosmology: an illustration of the Monte Python cosmological parameter inference code}}, \href{https://doi.org/10.1088/1475-7516/2013/02/001}{\emph{JCAP} {\bfseries 02} (2013) 001} [\href{https://arxiv.org/abs/1210.7183}{{\ttfamily 1210.7183}}].

\bibitem{Brinckmann:2018cvx}
T.~Brinckmann and J.~Lesgourgues, \emph{{MontePython 3: boosted MCMC sampler and other features}}, \href{https://doi.org/10.1016/j.dark.2018.100260}{\emph{Phys. Dark Univ.} {\bfseries 24} (2019) 100260} [\href{https://arxiv.org/abs/1804.07261}{{\ttfamily 1804.07261}}].

\bibitem{Kusenko:1997si}
A.~Kusenko and M.~E. Shaposhnikov, \emph{{Supersymmetric Q balls as dark matter}}, \href{https://doi.org/10.1016/S0370-2693(97)01375-0}{\emph{Phys. Lett. B} {\bfseries 418} (1998) 46} [\href{https://arxiv.org/abs/hep-ph/9709492}{{\ttfamily hep-ph/9709492}}].

\bibitem{Enqvist:1997si}
K.~Enqvist and J.~McDonald, \emph{{Q balls and baryogenesis in the MSSM}}, \href{https://doi.org/10.1016/S0370-2693(98)00271-8}{\emph{Phys. Lett. B} {\bfseries 425} (1998) 309} [\href{https://arxiv.org/abs/hep-ph/9711514}{{\ttfamily hep-ph/9711514}}].

\bibitem{Kasuya:1999wu}
S.~Kasuya and M.~Kawasaki, \emph{{Q ball formation through Affleck-Dine mechanism}}, \href{https://doi.org/10.1103/PhysRevD.61.041301}{\emph{Phys. Rev. D} {\bfseries 61} (2000) 041301} [\href{https://arxiv.org/abs/hep-ph/9909509}{{\ttfamily hep-ph/9909509}}].

\bibitem{Gelmini:2020ekg}
G.~B. Gelmini, M.~Kawasaki, A.~Kusenko, K.~Murai and V.~Takhistov, \emph{{Big Bang Nucleosynthesis constraints on sterile neutrino and lepton asymmetry of the Universe}}, \href{https://doi.org/10.1088/1475-7516/2020/09/051}{\emph{JCAP} {\bfseries 09} (2020) 051} [\href{https://arxiv.org/abs/2005.06721}{{\ttfamily 2005.06721}}].

\bibitem{Du:2021idh}
Y.~Du and J.-H. Yu, \emph{{Neutrino non-standard interactions meet precision measurements of N$_{eff}$}}, \href{https://doi.org/10.1007/JHEP05(2021)058}{\emph{JHEP} {\bfseries 05} (2021) 058} [\href{https://arxiv.org/abs/2101.10475}{{\ttfamily 2101.10475}}].

\bibitem{Du:2023upj}
Y.~Du, \emph{{Neff as a new physics probe in the precision era of cosmology}}, \href{https://doi.org/10.1103/PhysRevD.110.055030}{\emph{Phys. Rev. D} {\bfseries 110} (2024) 055030} [\href{https://arxiv.org/abs/2310.10034}{{\ttfamily 2310.10034}}].

\end{thebibliography}\endgroup

\end{document}